\documentclass[usenatbib,usegraphicx,useAMS,usenatbib]{mn2e}
\pdfoutput=1
\usepackage{color}
\usepackage{graphicx}
\usepackage[caption=true]{subfig}
\usepackage{amsmath}

\usepackage{rotating}
\usepackage{lscape}
\usepackage{color}
\usepackage[labelfont=bf,justification=justified,format=plain,singlelinecheck=false,compatibility=false]{caption}
\usepackage[export]{adjustbox}

\newcommand{\aj}{AJ}			
\newcommand{\araa}{ARA\&A}		
\newcommand{\apj}{ApJ}			
\newcommand{\apjl}{ApJLett}		
\newcommand{\apjs}{ApJS}		
\newcommand{\aap}{A\&A}			
\newcommand{\aaps}{A\&AS}		
\newcommand{\mnras}{MNRAS}		
\newcommand{\pasp}{PASP}		
\newcommand{\procspie}{Proc.~SPIE}   

\title[VIPS CSOs]{Compact Symmetric Objects and Supermassive Binary Black Holes in the VLBA Imaging and Polarimetry Survey}
\author[Tremblay et al.]{S.E. Tremblay$^{1,2,3}$\thanks{E-mail:
steve.tremblay@curtin.edu.au}, G.B. Taylor$^{1,4}$,  A.A. Ortiz$^{1}$, C.D. Tremblay$^{2}$, J.F. Helmboldt$^{5}$, \and R.W. Romani$^{6}$
\\
$^{1}$Department of Physics and Astronomy, University of New Mexico, Albuquerque, NM 87131\\
$^{2}$ International Centre for Radio Astronomy Research (ICRAR), Curtin University, Bentley, WA 6102, Australia\\
$^{3}$ARC Centre of Excellence for All-sky Astrophysics (CAASTRO), Sydney, Australia\\
$^{4}$National Radio Astronomy Observatory, Socorro NM 87801\\
$^{5}$Naval Research Laboratory, Code 7213, Washington, DC 20375\\
$^{6}$Department of Physics, Stanford University, Stanford, CA 94305
}

\begin{document}



\maketitle



\begin{abstract}
We present multi-frequency Very Long Baseline Array (VLBA) follow-up observations of VLBA Imaging and Polarimetry Survey sources identified as likely compact symmetric objects (CSOs) or super-massive binary black holes (SBBHs). We also present new spectroscopic redshifts for 11 sources observed with the Hobby-Eberly Telescope. While no new SBBHs can be confirmed from these observations, we have identified 24 CSOs in the sample, 15 of which are newly designated, and refuted 52 candidates leaving 33 unconfirmed candidates. This is the first large uniform sample of CSOs which can be used to elicit some of the general properties of these sources, including morphological evolution and environmental interaction. We have detected polarised emission from two of these CSOs  the properties of which are consistent with Active Galactic Nuclei unification schemes.

\end{abstract}



\section{Introduction}
In 2006, the Very Long Baseline Array (VLBA) performed a large full polarisation survey of radio sources, the VLBA Imaging and Polarimetry Survey (VIPS; Helmboldt et al. 2007). This survey was comprised of 198 hours of 5 GHz observations, targeting a subset of 1127 sources from the cosmic lens all-sky survey (CLASS; Myers et al. 2003). While the primary goal of the survey was to identify and characterise potential \textit{Fermi Gamma-ray Space Telescope} sources \citep{2011ApJ...726...16L}, two concurrent objectives were to identify candidate members of two classes of rare astronomical objects, compact symmetric objects (CSOs) and supermassive binary black holes (SBBHs). 

CSOs can broadly be described as dual-lobed sources whose extent is less than a kiloparsec and that are oriented close to the plane of the sky \citep{1994ApJ...432L..87W}. Similar to larger radio galaxies, the morphology we observe consists of relativistic jets emitted from opposing sides of the core, presumably where the central supermassive black hole resides. These jets then terminate at hotspots, where the fast moving jet material interacts with the surrounding media, intergalactic medium for large radio galaxies and interstellar medium for small CSOs,  and inflate lobes. The emission of CSOs is typically dominated by these hot spots \citep{1996ApJ...460..634R}, typically leading to edge brightening. Multi-epoch observations show the distance separating the furthest extents of the lobes typically increasing at $\sim$0.1 c \citep{2000ApJ...541..112T} which, when extracted backwards to estimate the ages of these sources, implies that CSOs are young, ranging from $\sim20-2000$ years old \citep{2005ApJ...622..136G}. The small size and orientation of these sources make CSOs ideal for studying both galactic evolution and the properties of the  media surrounding the AGN.  Coupling this with the small number of heretofore confirmed CSOs, 50-90 \citep{2006MNRAS.368.1411A,2009AN....330..190A} justifies the further study of these objects. Furthermore, the confirmed CSOs to date have mostly resulted from targeted observations aimed at a high rate of confirmation resulting in a lack of understanding in how these objects fit into our general understanding of AGN, which a large sample of CSOs found within a uniform sample could provide.

The combined theories of galactic growth via mergers and the residence of supermassive black holes in the centres of most galaxies necessitate supermassive black hole interactions if both are correct \citep{2005LRR.....8....8M}. Binary black holes, which themselves end either in a merger event or ejection, are a naturally expected outcome of these interactions. The observed correlation between central supermassive black hole mass and the mass of the stellar bulge in galaxies \citep{1998AJ....115.2285M, 2013ARA&A..51..511K, 2015ApJ...798...54G} further strengthens the expectation of binary systems. To date, the number of close separation ($\leq 10$  pc) SBBH candidates is small with only 0402+379 having morphological verification \citep{2006ApJ...646...49R,2009ApJ...697...37R}. One of the goals for this survey of over 1000 galaxies was to identify, and potentially confirm, new SBBH candidates.

This paper explores the follow-up observations performed on sources identified as either CSO or SBBH candidates. More detailed analysis, in particular on the individual sources comprising the VIPS CSO sample, will be carried out in a subsequent paper. Throughout this paper, we assume H$_{0}$=73 km s$^{-1}$
Mpc$^{-1}$, $\Omega_m$ = 0.27, $\Omega_\Lambda$ = 0.73, unless noted otherwise. This paper makes use of Ned Wright's Javascript Cosmology Calculator \citep{2006PASP..118...1711W}.


\section{Sample Selection, Observations and Data Reduction}
\subsection{Sample Selection}
\label{selection}

The sample selection is described in detail in Helmboldt et al. (2007). Here we outline the VIPS classification process to better understand the followup sample being studied in this paper. 

After imaging, an automated morphological classification was performed on all 1127 VIPS sources. This started with the AIPS \citep{2003ASSL..285..109G} task SAD being used to generate multi-component Gaussian models for each source. The sources were then categorised as follows:

\begin{itemize}
\item Sources with a single Gaussian component containing $\ge$ 95\% of the total flux density were flagged as single component sources
\item Sources where the two brightest Gaussian components contained $\ge$ 95\% of the total flux density were flagged as double sources
\item Sources that didn't fit these criteria were flagged as multiple component sources
\end{itemize}

These preliminary categories were then refined as follows:

\begin{itemize}
\item Single component sources with minor/major axis ratios of $>$ 0.6 were classified as point sources (PSs), otherwise they were flagged as core-jets (0.6 was chosen since this was the axis ratio of the restoring beam used in the VIPS images)
\item Double sources where the flux densities of the two brightest components match within a factor of 2.5 were classified as CSO candidates (CSOs)
\item Multiple components sources where the components containing $\ge$ 95\% of the total flux density were located along a single line were flagged as core-jets, otherwise they were classified as complex (CPLX)
\item Core-jets longer than 6 mas were classified as long jets (LJETs), otherwise they were placed in the category of short jets (SJETs)
\item LJETs longer than 12 mas whose brightest Gaussian component was positioned within 3 mas of the centre of the structure were reclassified as CSO candidates (CSOs)
\end{itemize}

The two separate ways sources could be classified as CSO candidates were instituted since the the jets connecting the hotspots and lobes of CSOs back to the core are sometimes, but not always, detectable.
The automatic source classifications were then verified visually and false classifications corrected (87\% CSOs and 71\% CPLX sources were correct). This yielded 103 CSOs and 17 CPLX sources needing follow-up observations. 

The list of SBBH candidates was generated by visually inspecting the 5 GHz images and looking for sources that appeared to have two distinct axes of emission, possibly indicating the existence of two separate bi-directional jet flows from two cores. The CPLX sources were included in the follow-up both because they were a good place to look for SBBHs and since non-archetypical CSOs can easily be classified as such.

\subsection{Observations, Calibration and Imaging}
\label{observations}

\subsubsection{Radio Data}

Two subsets of multi-frequency observations of the candidates were performed.  A series of 4 separate observing runs from September 2006 to February 2007 (BT088) and another 5 observing runs were performed from June 2007 to June 2008 (BT094) all with the VLBA. Each of these observations consisted of four 8 MHz wide IFs in the C, X and U bands with full polarisation centred at: 4605.5 , 4675.5, 4990.5, 5091.5, 8106.0, 8176.0, 8491.0, 8590.0, 14902.5, 14910.5, 15356.5 and 15364.5 MHz at an aggregate bit rate of 256 Mbps to maximise ($u$,$v$) coverage and sensitivity. When the data in each band were combined, the three central frequencies were: 4844.7, 8344.7, and 15137.5 MHz. The BT088 observations had typical times on source of $\sim$38 minutes at C and X band and $\sim$116 minutes at U Band. For BT094 typical time on source was $\sim$10 minutes for C and X band and $\sim$32 minutes for U Band. The integrations for each source were spread out over approximately 10 hours to maximise ($u$,$v$) coverage for the observations. 

Most of the calibration and initial imaging of the data were carried out by automated AIPS and Difmap \citep{1997ASPC..125...77S} scripts similar to those used in reducing the VIPS 5 GHz survey data \citep{2007ApJ...658..203H, 2005ApJS..159...27T}. To summarise, flagging of bad data and calibration were performed using the VLBA data calibration pipeline \citep{2005ASPC..340..613S}, while imaging was performed using Difmap scripts described in \citet{2005ApJS..159...27T}. Final images were inspected and manually improved using the Difmap program to attain the best fidelity possible. Part of the final imaging scripts  for C and U bands involved running the Difmap command selfcal on the data starting with the clean-component model of the X band data as the input model to help in aligning the source across all three frequencies. This creates a phase `correction' to shift the entire map in a way that maximises the alignment of the bright components across the map. This is the same method described in \citet{1997ApJ...490..291B}, where they align the maps for different frequencies for spectral index analysis.


The four IFs for each observing band were treated separately in order to maximise ($u$,$v$) coverage. The average root-mean-square scatter of the baseline visibility amplitudes in the calibrated data were $\pm$18 mJy, $\pm$16 mJy, and $\pm$14 mJy for the C, X, and U bands respectively for BT088 and $\pm$25 mJy, $\pm$23 mJy , and $\pm$8 mJy for the C, X, and U bands respectively of BT094. The BT094 15 GHz scatter is notably lower because we combined the four IFs after fringe fitting and calibration (at the expense of ($u$,$v$) coverage) in order to compensate for the lower flux levels resulting from the steep spectrum of many of the sources. For the polarimetric analysis, the lower two and upper two IF pairs in each band were combined and imaged to maximise the sensitivity while retaining some `in band' frequency information.

Polarisation calibration was carried out by first aligning the polarisation angles of all four IFs, within each band, to the pixel brightest in polarisation for 3C279. Then, the mean polarisation angles of J0854+2006 and J1310+3220 were compared to the values recorded in the \textit{VLA/VLBA Polarisation Calibration Page}\footnote{http://www.vla.nrao.edu/astro/calib/polar/} applying any necessary interpolations in either time or frequency. Typical errors in the electric vector polarisation angle (EVPA) were 7$^{\circ}$, 8$^{\circ}$, and 13$^{\circ}$ for the 5, 8, and 15 GHz maps respectively. This was determined using the deviation of the integrated EVPA across all three EVPA calibrators for each source.

\subsubsection{Optical Data}

For a subset of our candidates (11 sources) which didn't have published redshifts, spectroscopic observations were performed with the 9.2 m Hobby-Eberly Telescope (HET) at McDonald Observatory to obtain spectral redshifts. The HET observes in the declination range $-11^\circ < \delta < 73^\circ$ and is fully queue scheduled \citep{2007PASP..119..556S}. We used the Marcario Low-Resolution Spectrograph (LRS; Hill et al. 1998), with grism G1 ($300$ lines mm$^{-1}$), a 2''   slit, and a Schott GG385 long-pass filter for a resolution of R $\sim$ 500 between 4150 \AA \ and 10500 \AA. Typical exposures were 2 x 900 s with the slit placed along the parallactic angle.

\begin{figure}
        \includegraphics[width=0.5355\textwidth,center]{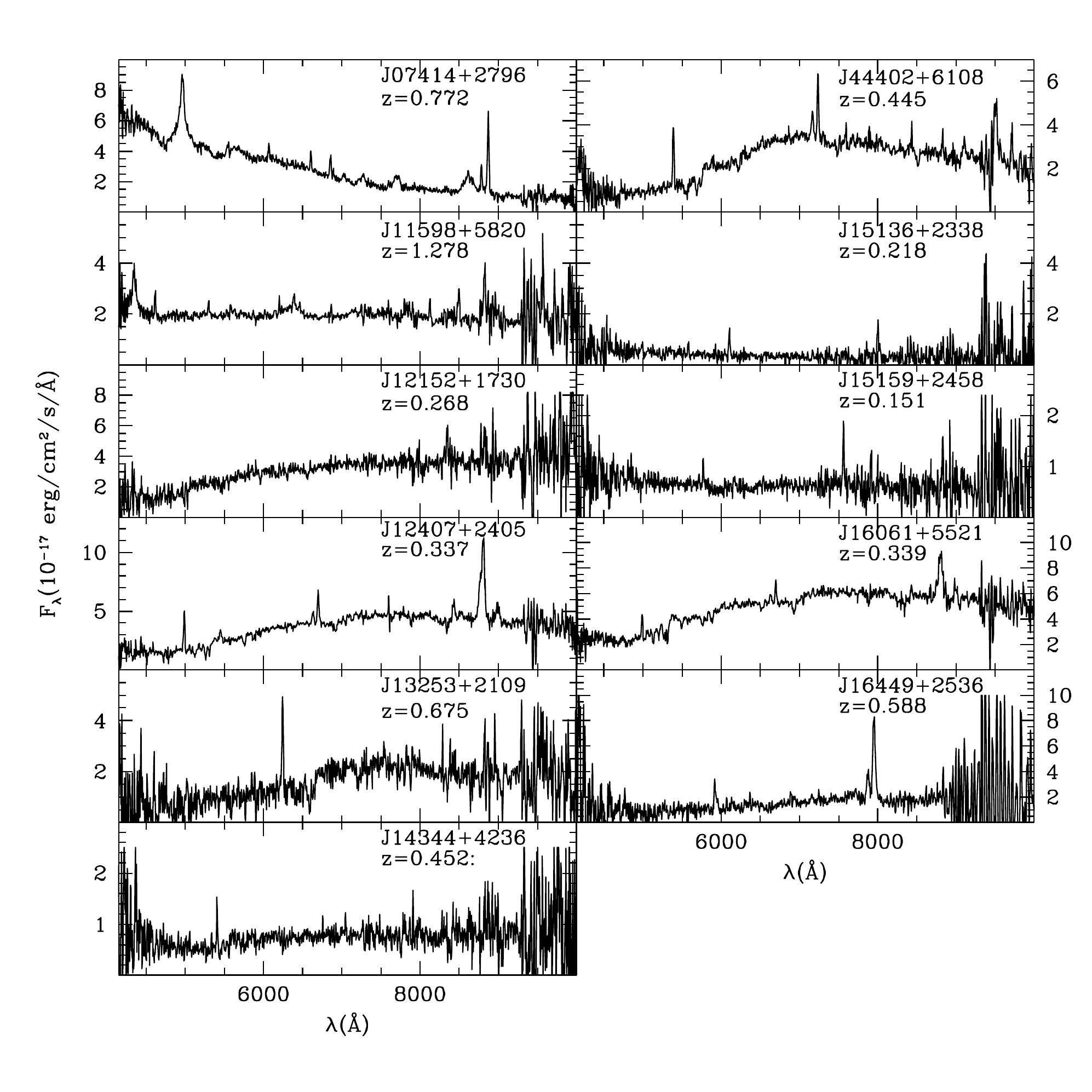}
        \caption{Optical spectra of CSOs with new redshifts from observing at the HET. All redshifts are confirmed from emission lines, although there remains some uncertainty in the redshift of J14344+4236. If the single strong line at 5406\,\AA \ towards J14344+4236 is identified as OII (3727\,\AA), then the redshift is z=0.5405. This identification is supported by a substantial 4000\,\AA \ break, as well as weak detections of Ca H \& K and Balmer features.}
        \label{fig:spectra}
\end{figure}

Data reduction was performed with the IRAF package \citep{1986SPIE..627..733T} using standard techniques. Wavelength calibration was performed with a neon-argon lamp. We employed an optimal extraction algorithm \citep{1992ASPC...25..398V} to maximise the signal to noise ratio (S/N). We performed spectrophotometric calibration using standard stars from \cite{1990AJ.....99.1621O}. Spectra were corrected for telluric absorptions and visually cleaned of cosmic rays. Multiple exposures on a single target were combined into a single spectrum, weighted by S/N. The resulting spectra can be seen in Fig. \ref{fig:spectra}.

\section{Results}
\label{results}

\subsection{Spectral Index Distribution}
\label{spectral_index}

\begin{figure*}
	\centering
	 {%
	 \includegraphics[width=0.32273\textwidth]{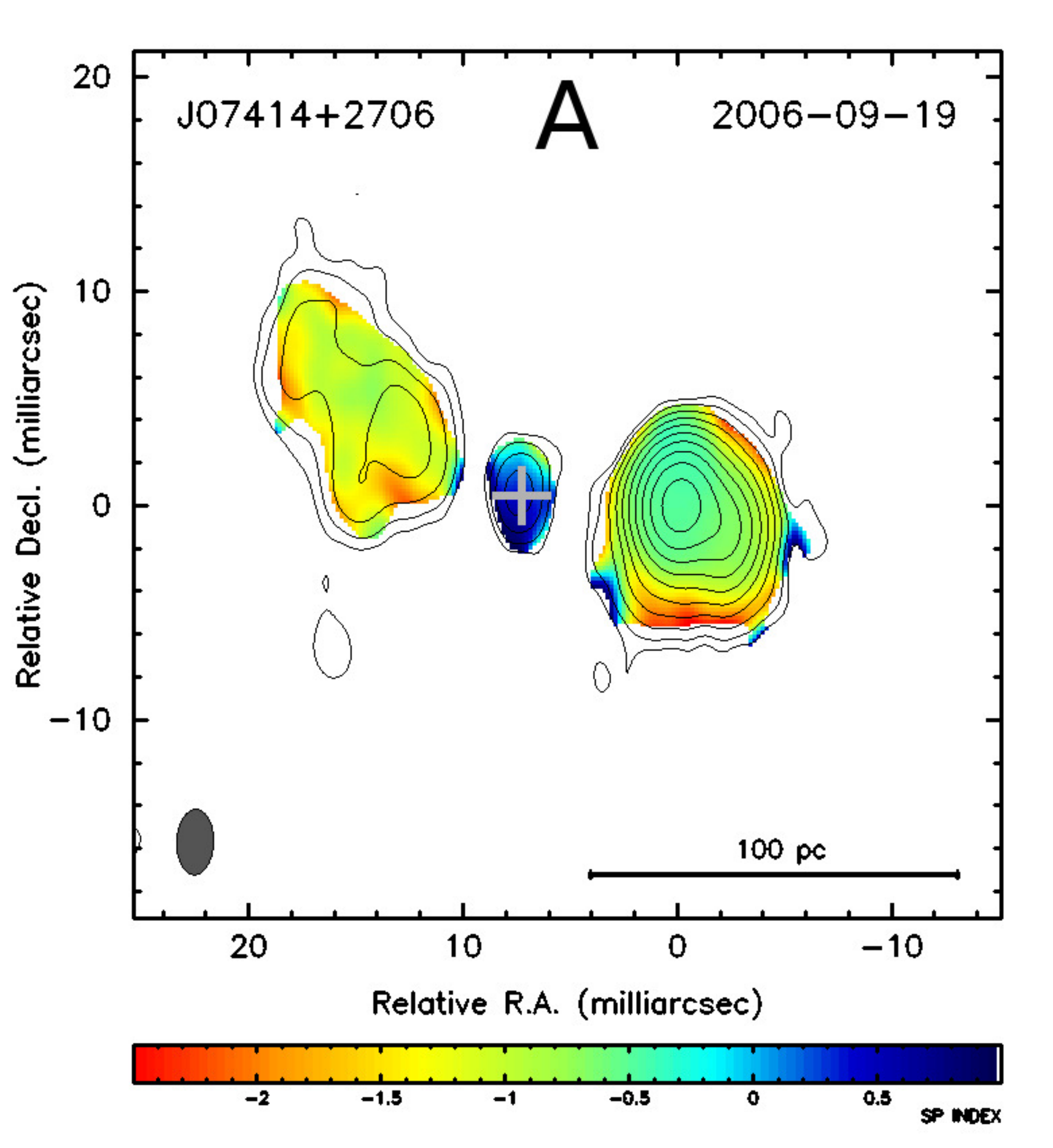}
	}
~\hfill
	\centering
	{%
	\includegraphics[width=0.32273\textwidth]{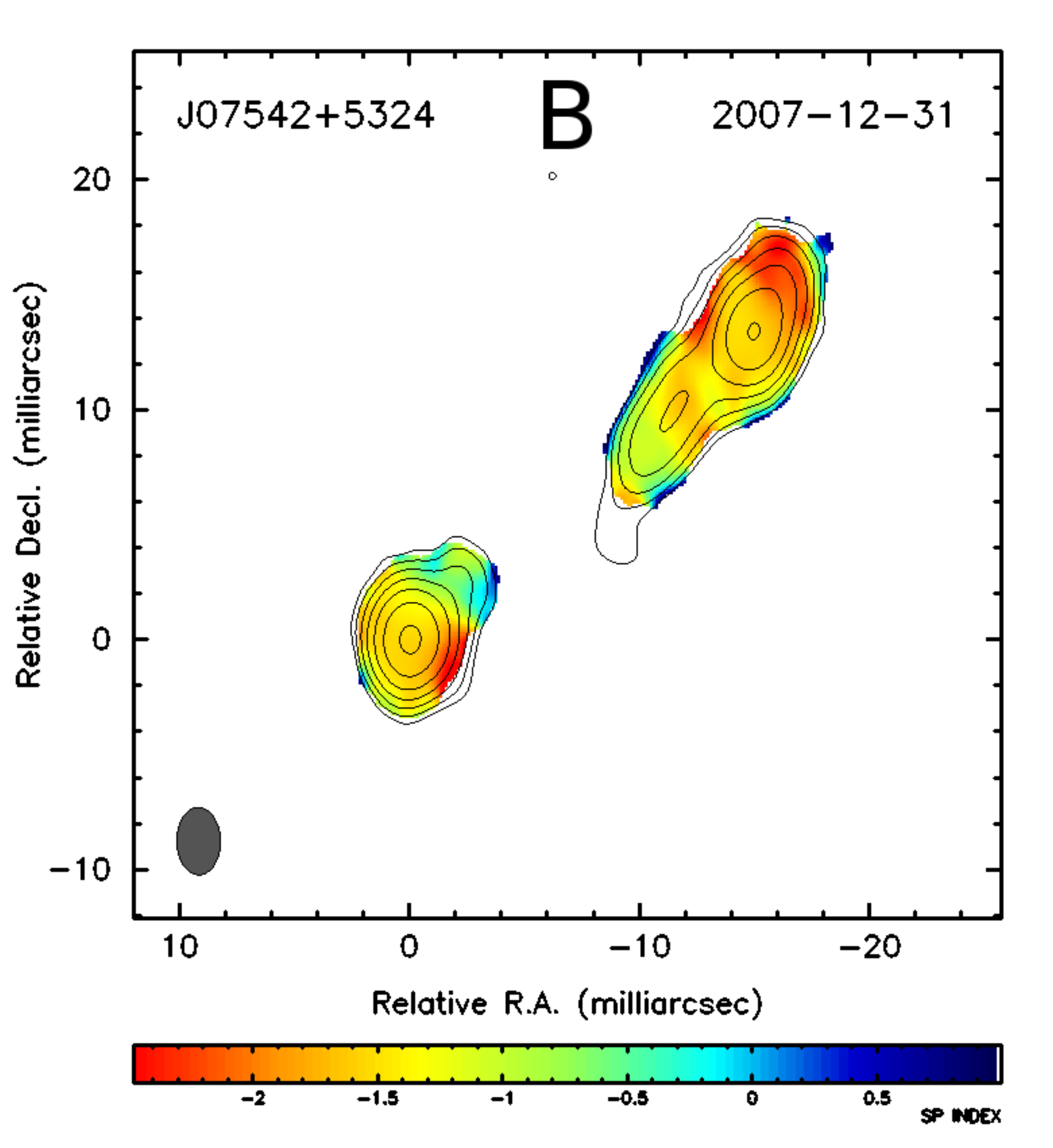}
	}
~\hfill
	\centering
	{%
	\includegraphics[width=0.32273\textwidth]{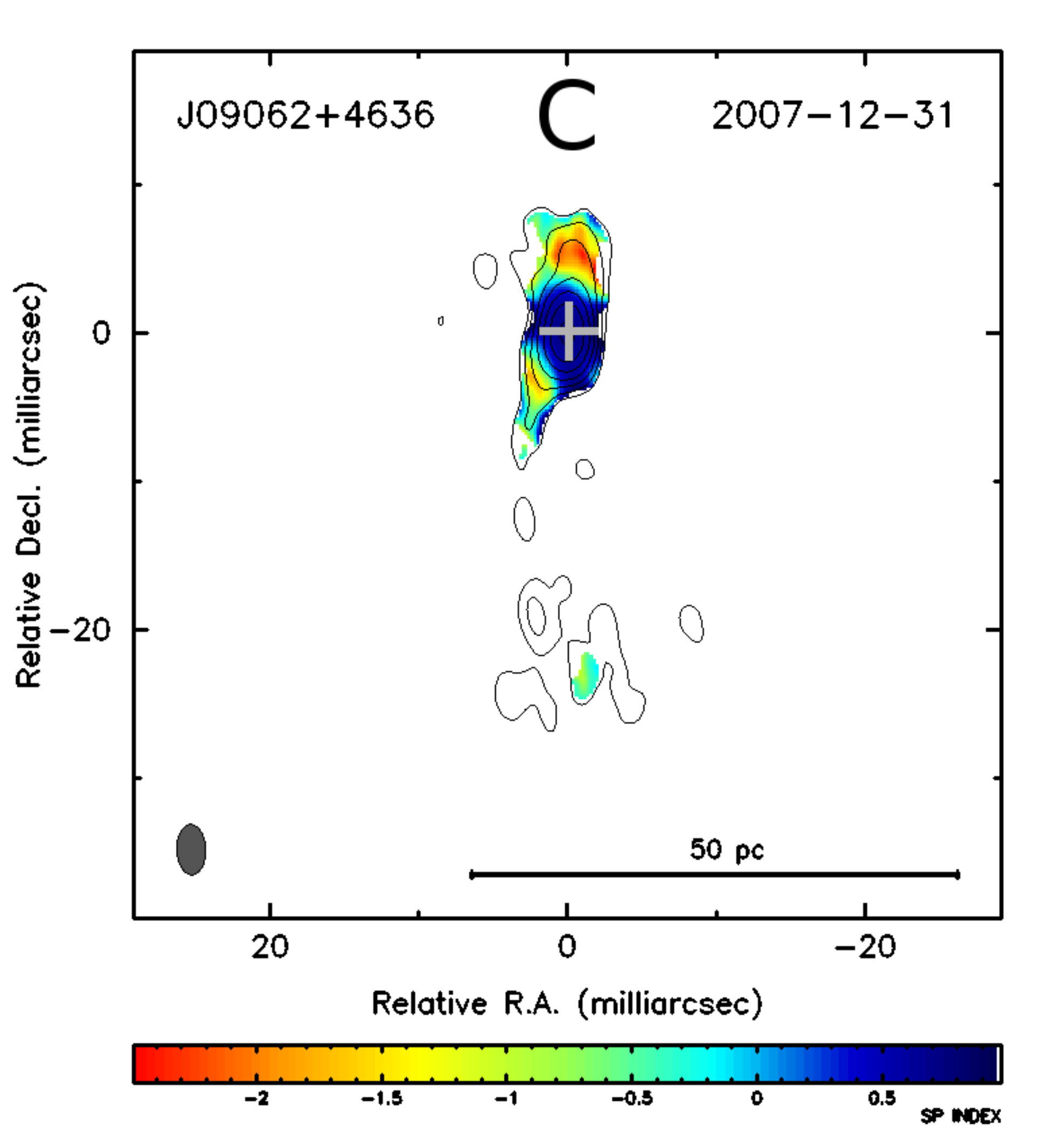}
	}	
~\hfill
	\centering
	{%
	\includegraphics[width=0.32273\textwidth]{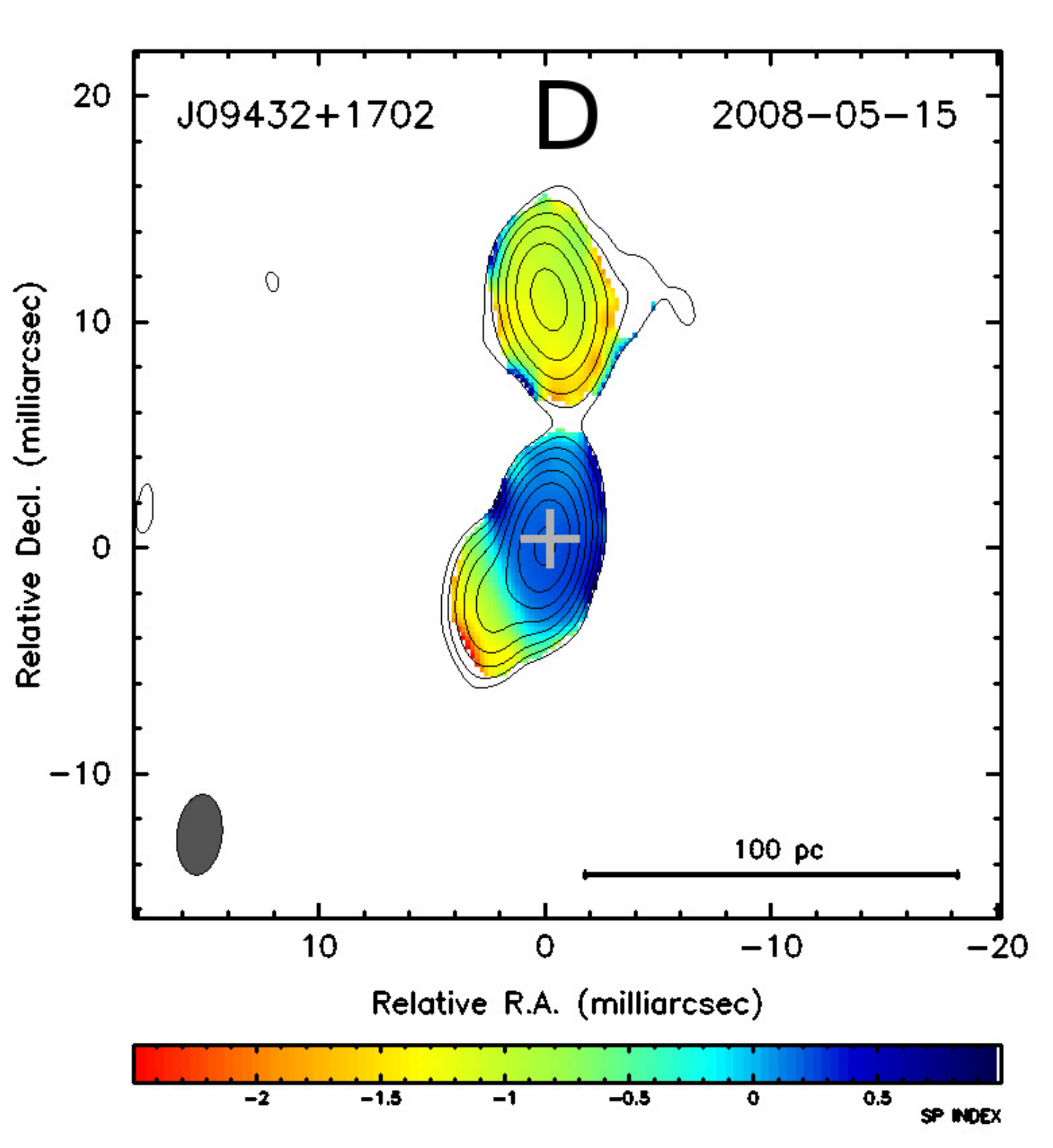}
	}	
~\hfill
	\centering
	{%
	\includegraphics[width=0.32273\textwidth]{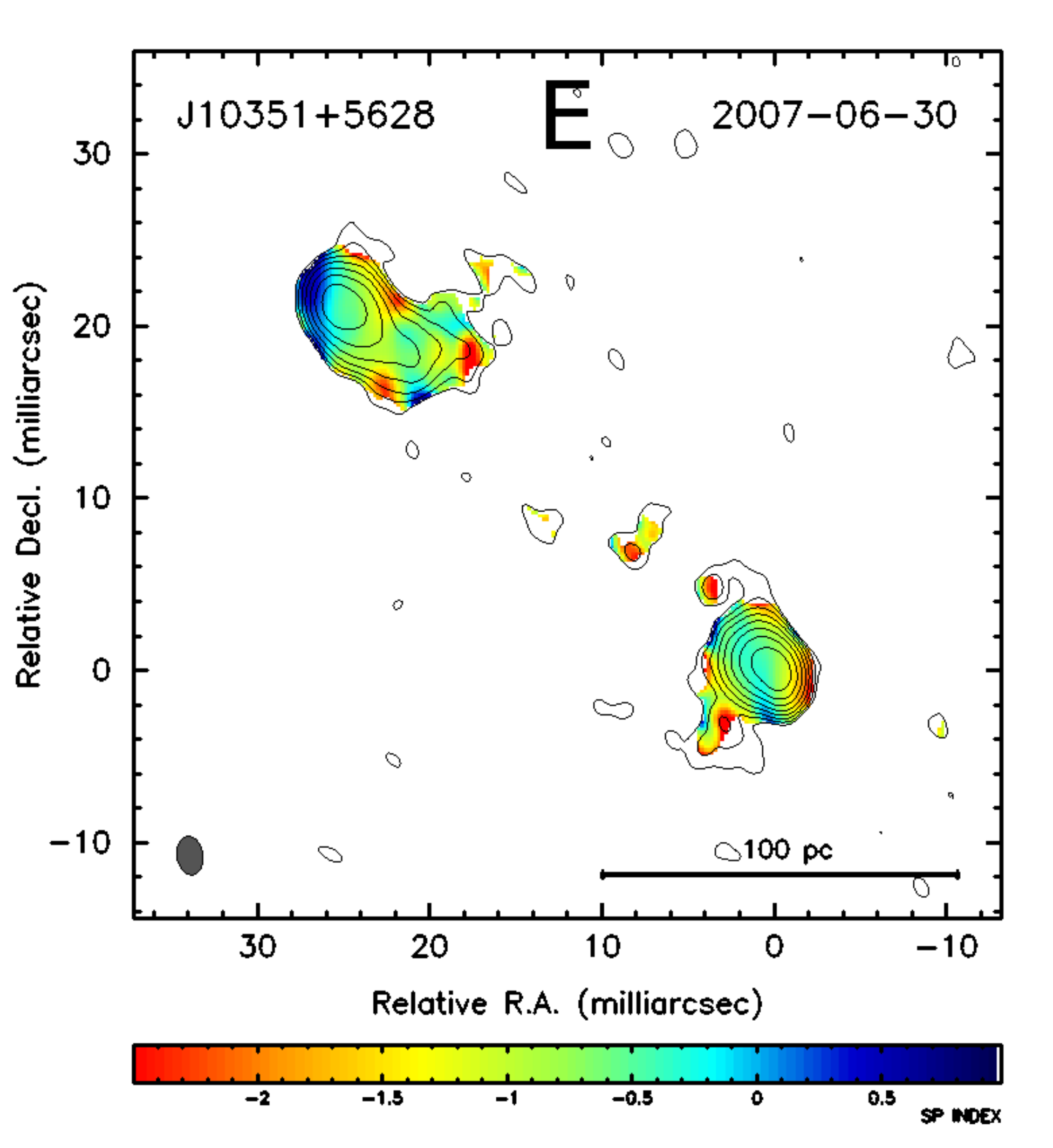}
	}	
~\hfill
	\centering
	{%
	\includegraphics[width=0.32273\textwidth]{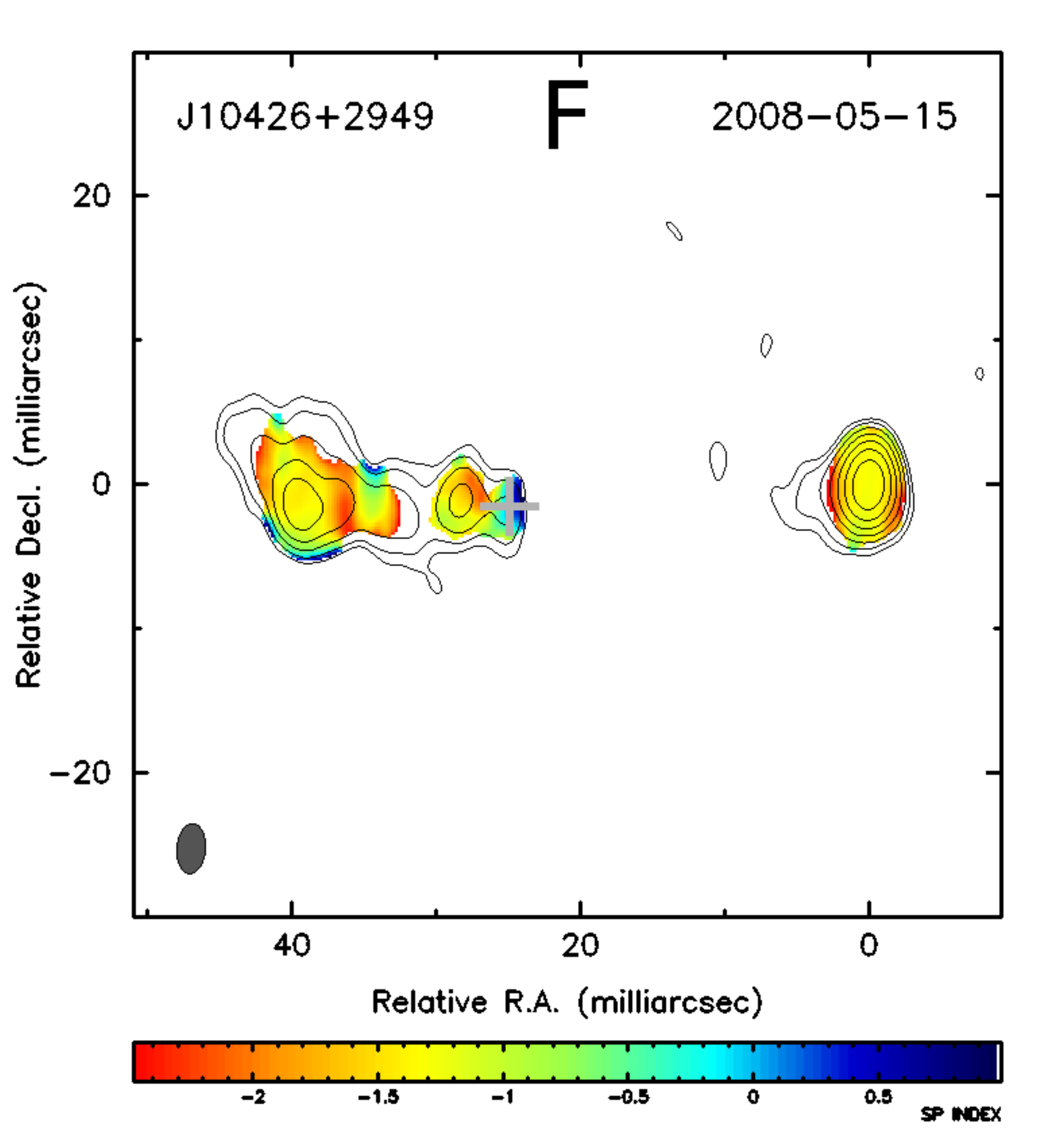}
	}	
~\hfill
	\centering
	{%
	\includegraphics[width=0.32273\textwidth]{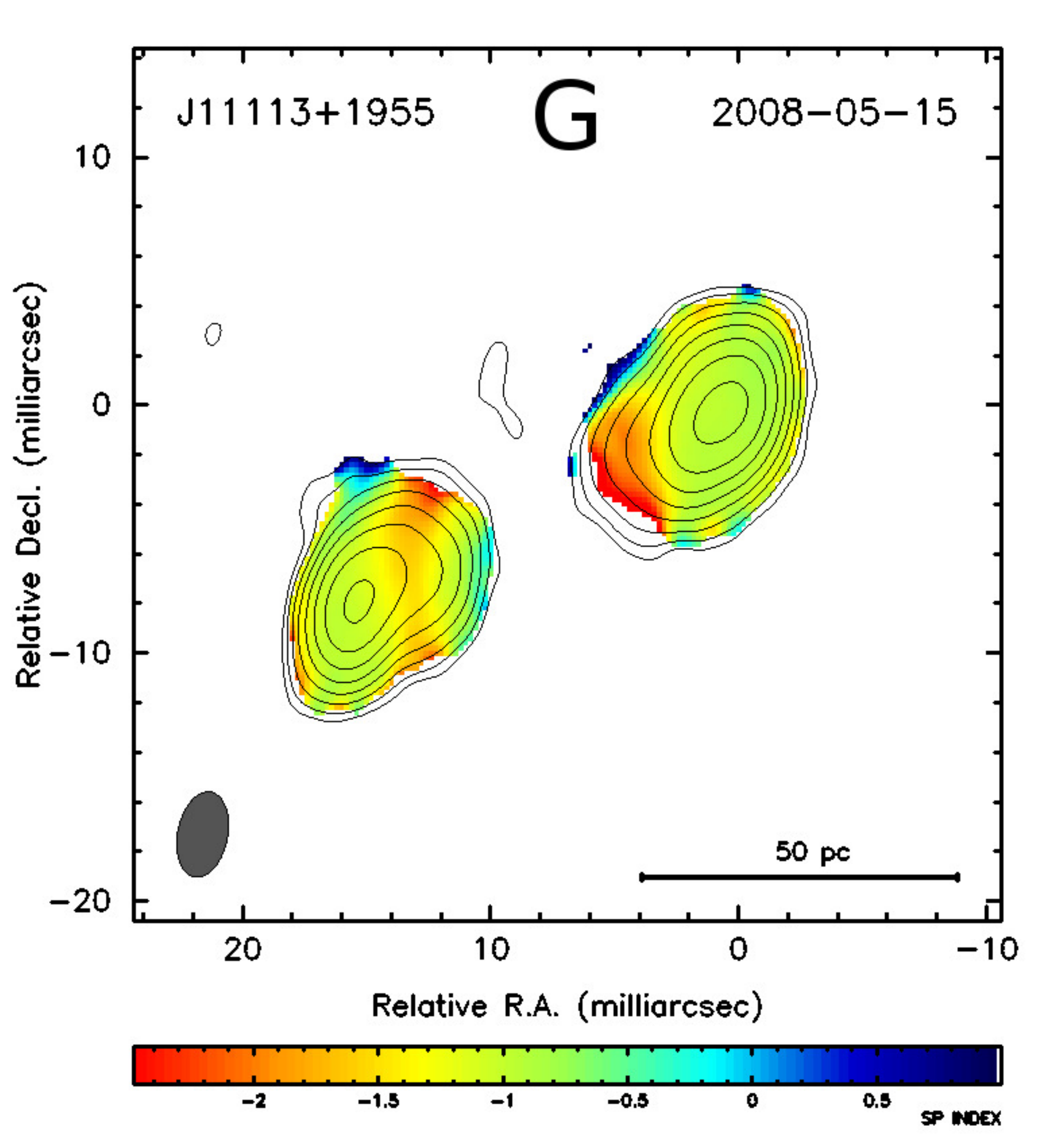}
	}
~\hfill
	\centering
	{%
	\includegraphics[width=0.32273\textwidth]{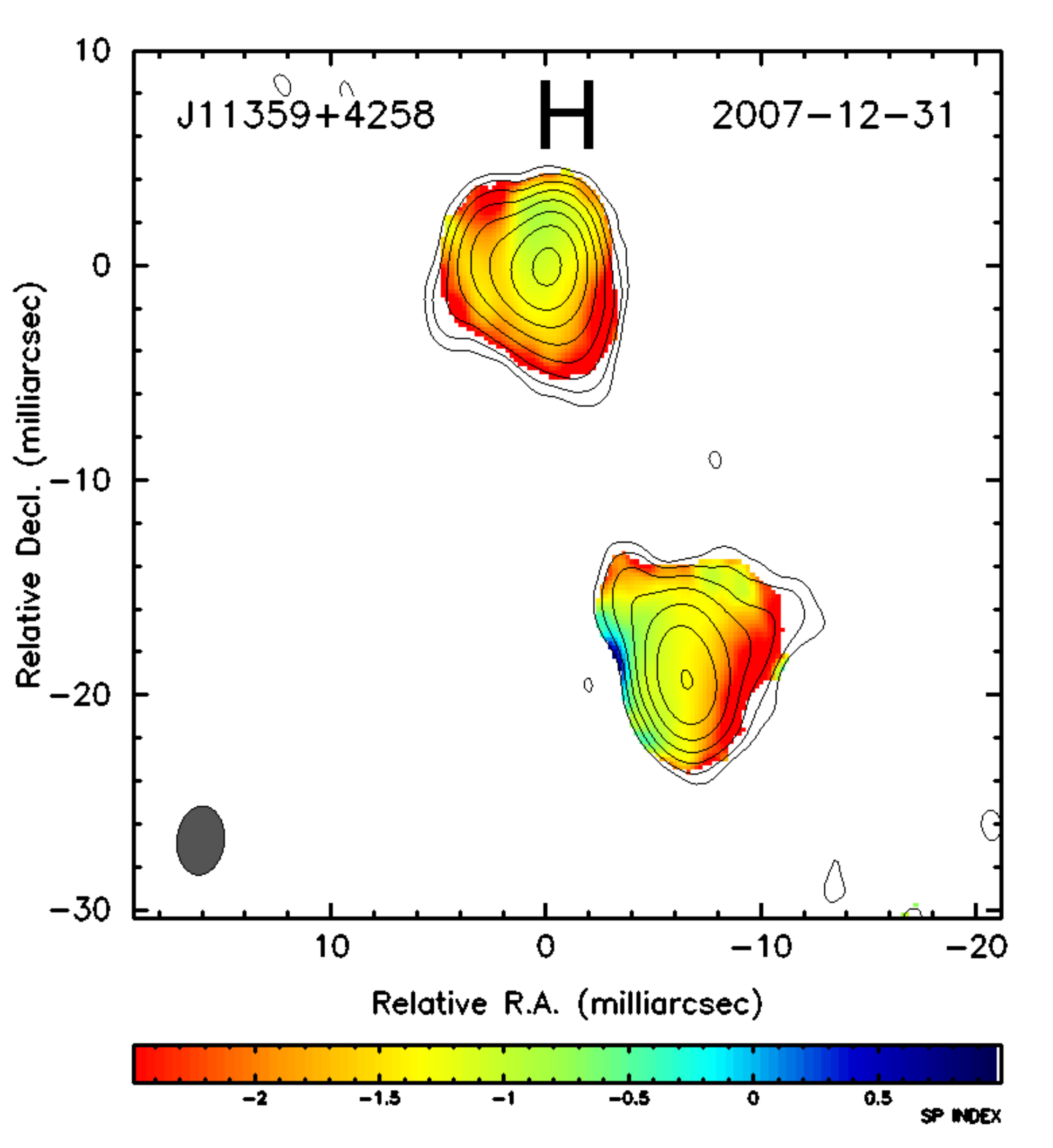}
	}		
~\hfill
	\centering
	{%
	\includegraphics[width=0.32273\textwidth]{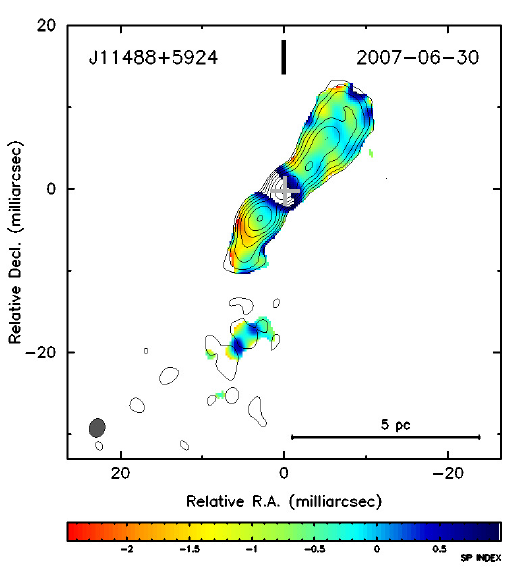}
	}	
	\caption{5 GHz contour maps of VIPS CSOs with 5-8 GHz spectral index map overlays. The contour levels begin at thrice the theoretical noise (typically $\sim$ 0.4 mJy for BT088 and 1.0 mJy for BT094) and increases by powers of 2. The colour scale is fixed from -2.5 to 1 to facilitate comparison. When detected, a grey cross is placed over the core to guide the reader's eye. Sources with confirmed spectral redshifts have associated distance bars for linear scale.}
\label{fig:cso_cxspix}
\end{figure*}

\begin{figure*}
\ContinuedFloat
 \captionsetup{list=off}
	\centering
	 {%
	 \includegraphics[width=0.32273\textwidth]{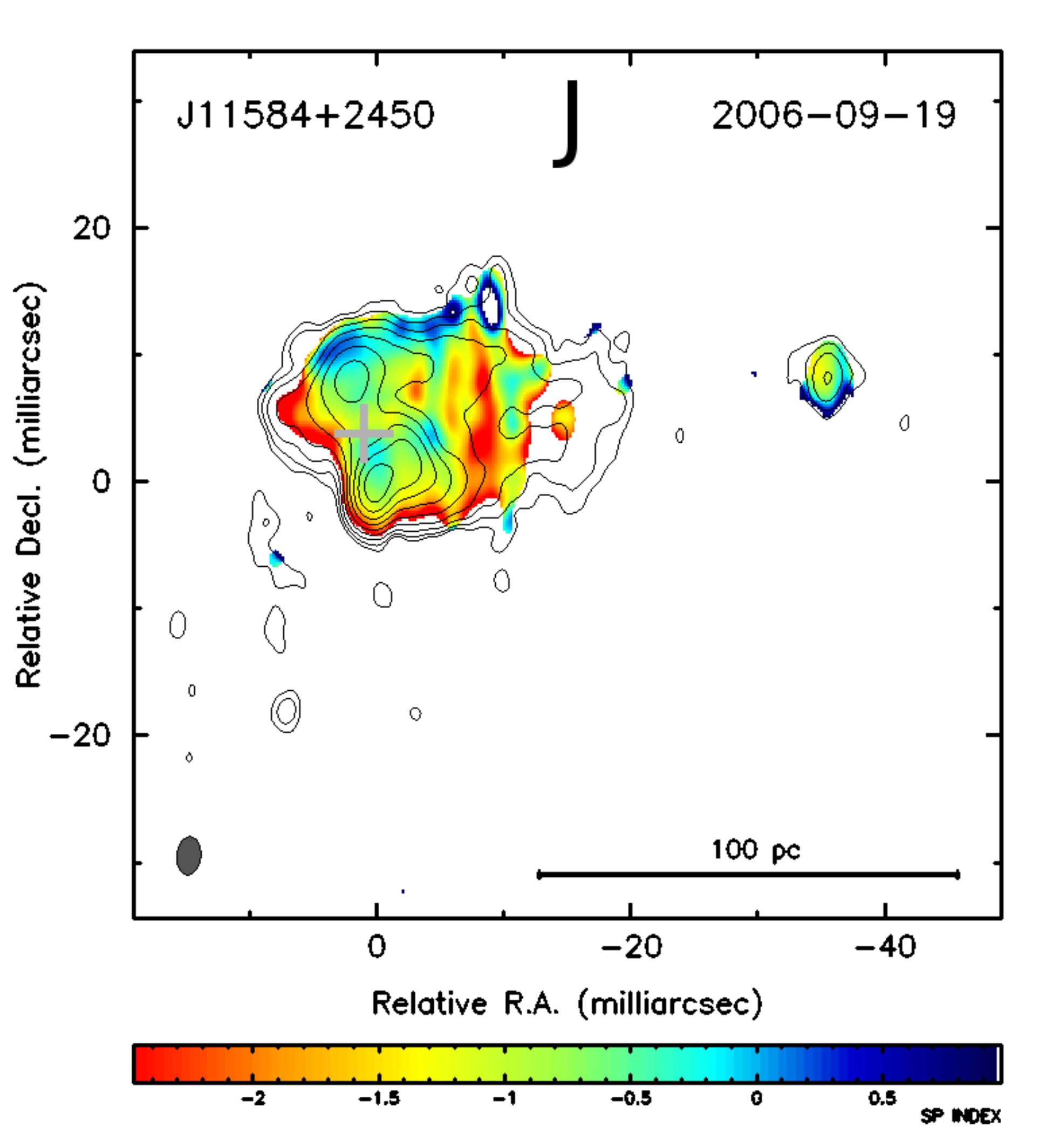}
	}
~\hfill
	\centering
	{%
	\includegraphics[width=0.32273\textwidth]{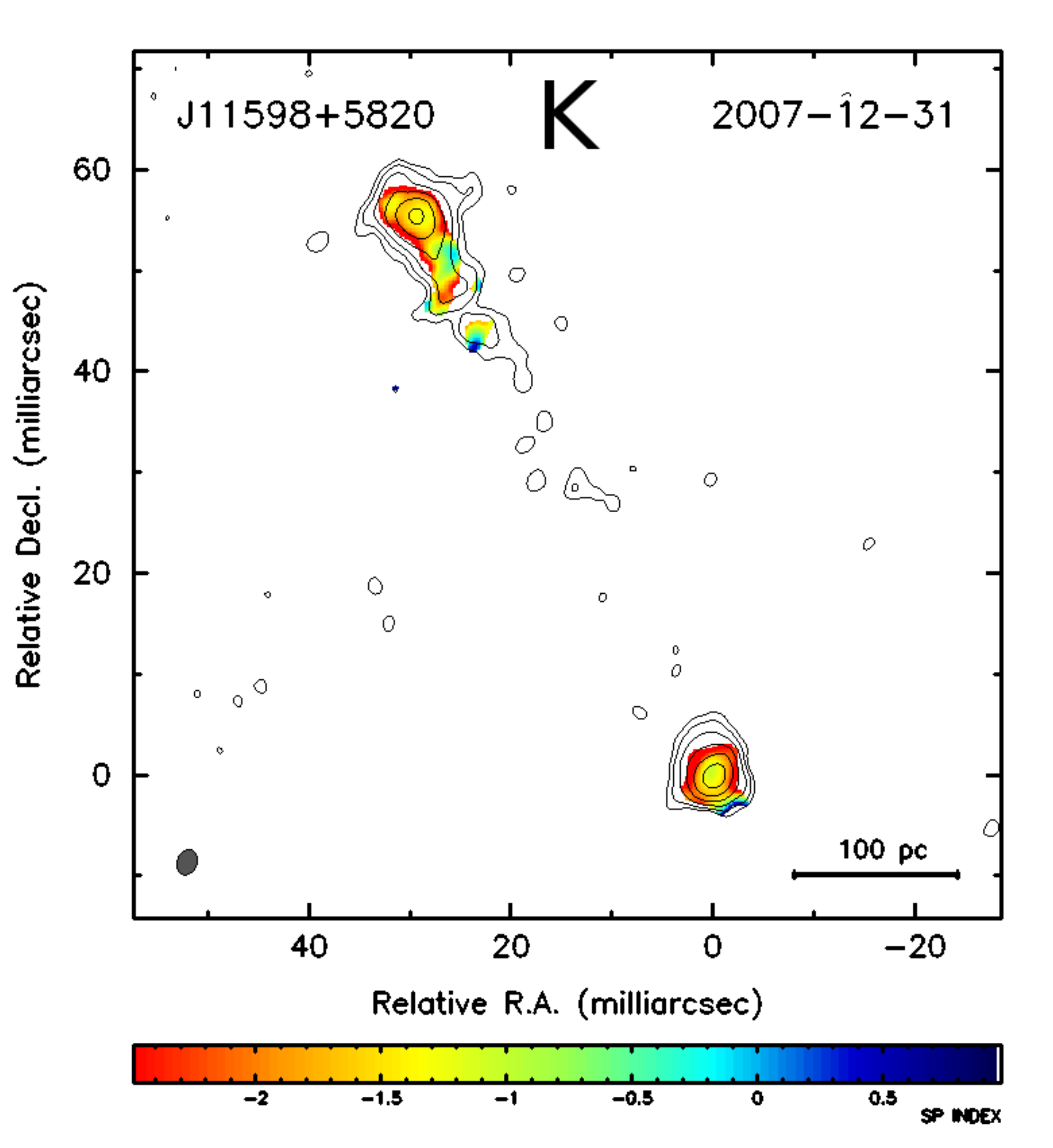}
	}
~\hfill
	\centering
	{%
	\includegraphics[width=0.32273\textwidth]{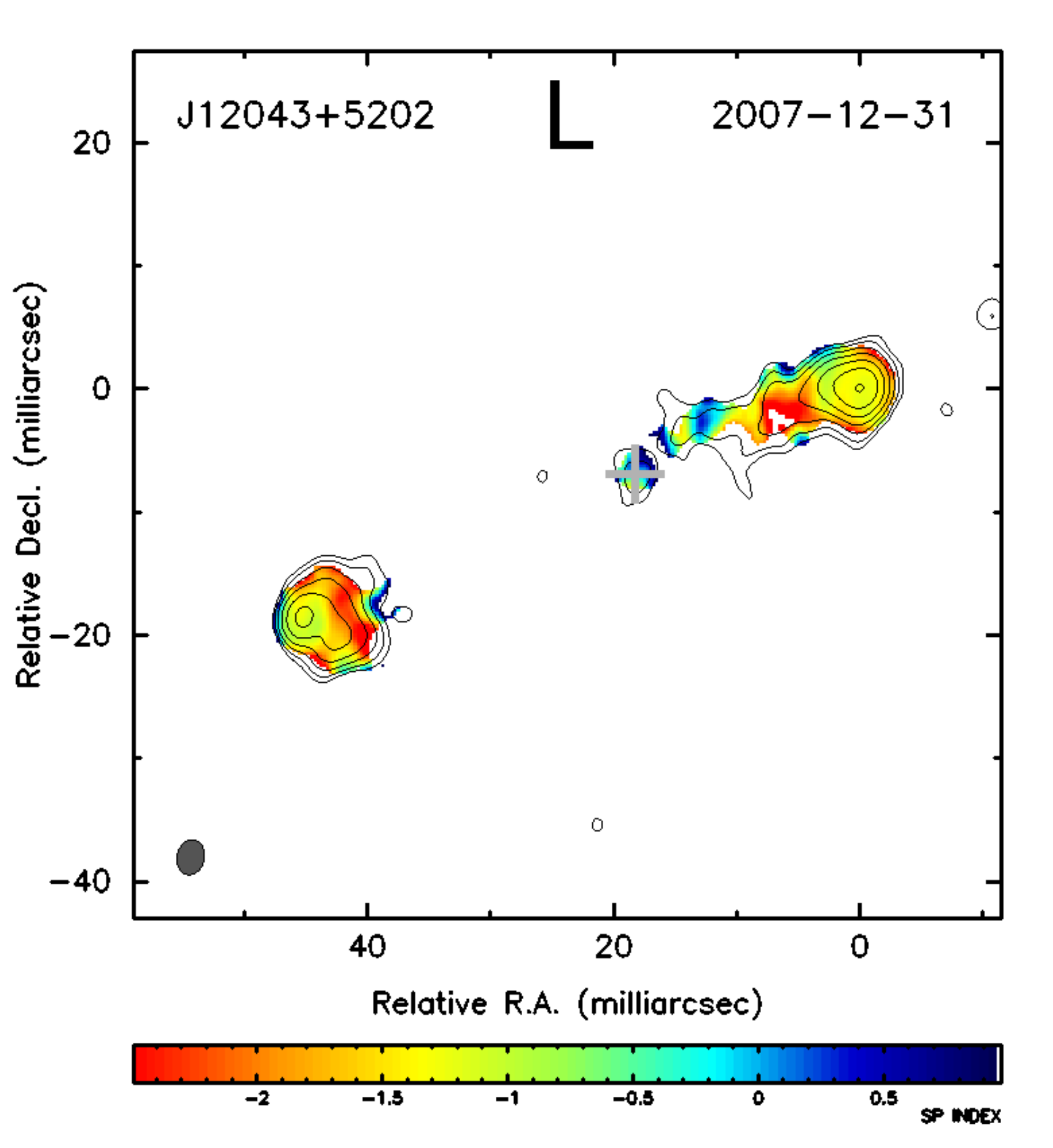}
	}	
~\hfill
	\centering
	{%
	\includegraphics[width=0.32273\textwidth]{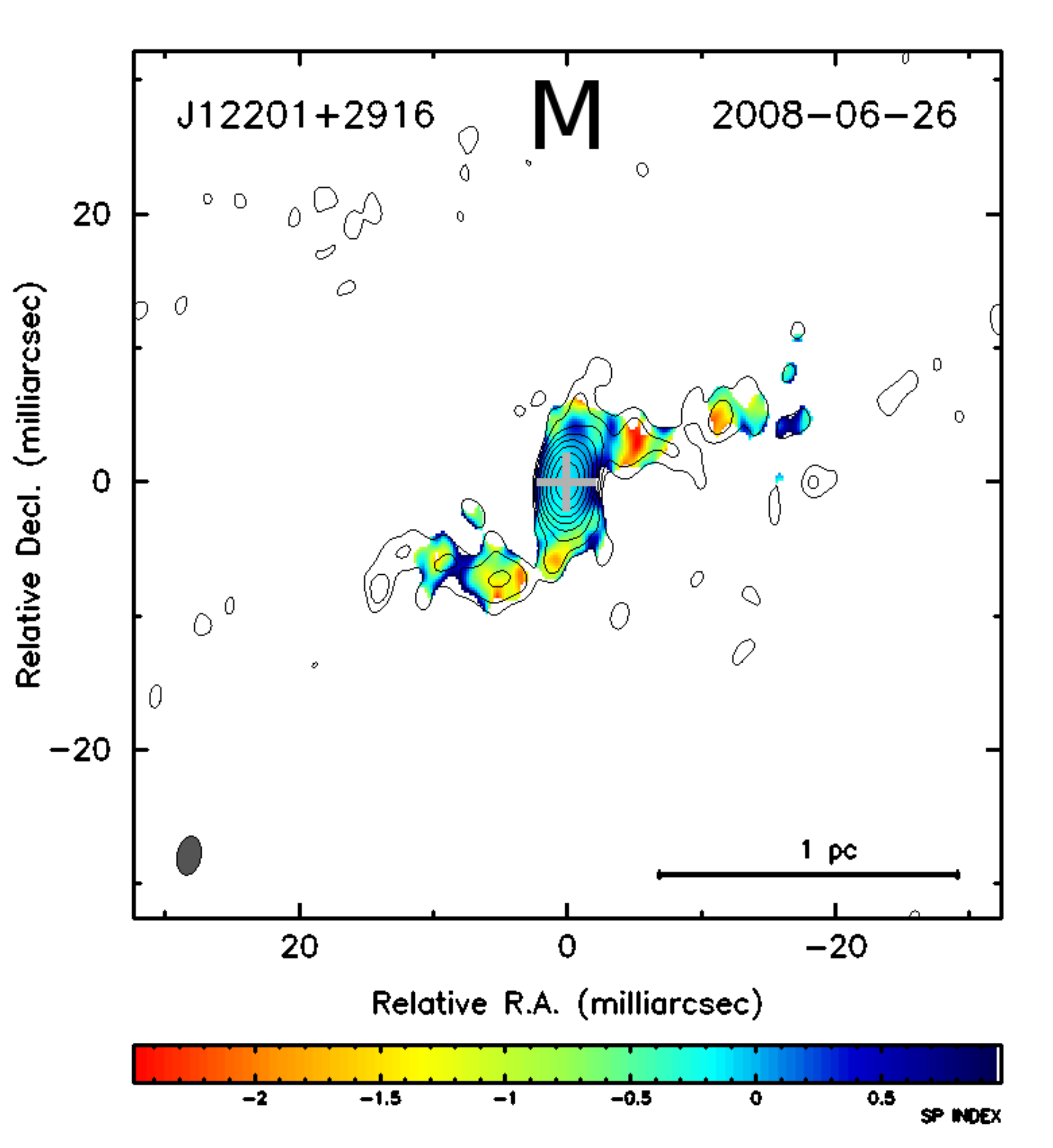}
	}	
~\hfill
	\centering
	{%
	\includegraphics[width=0.32273\textwidth]{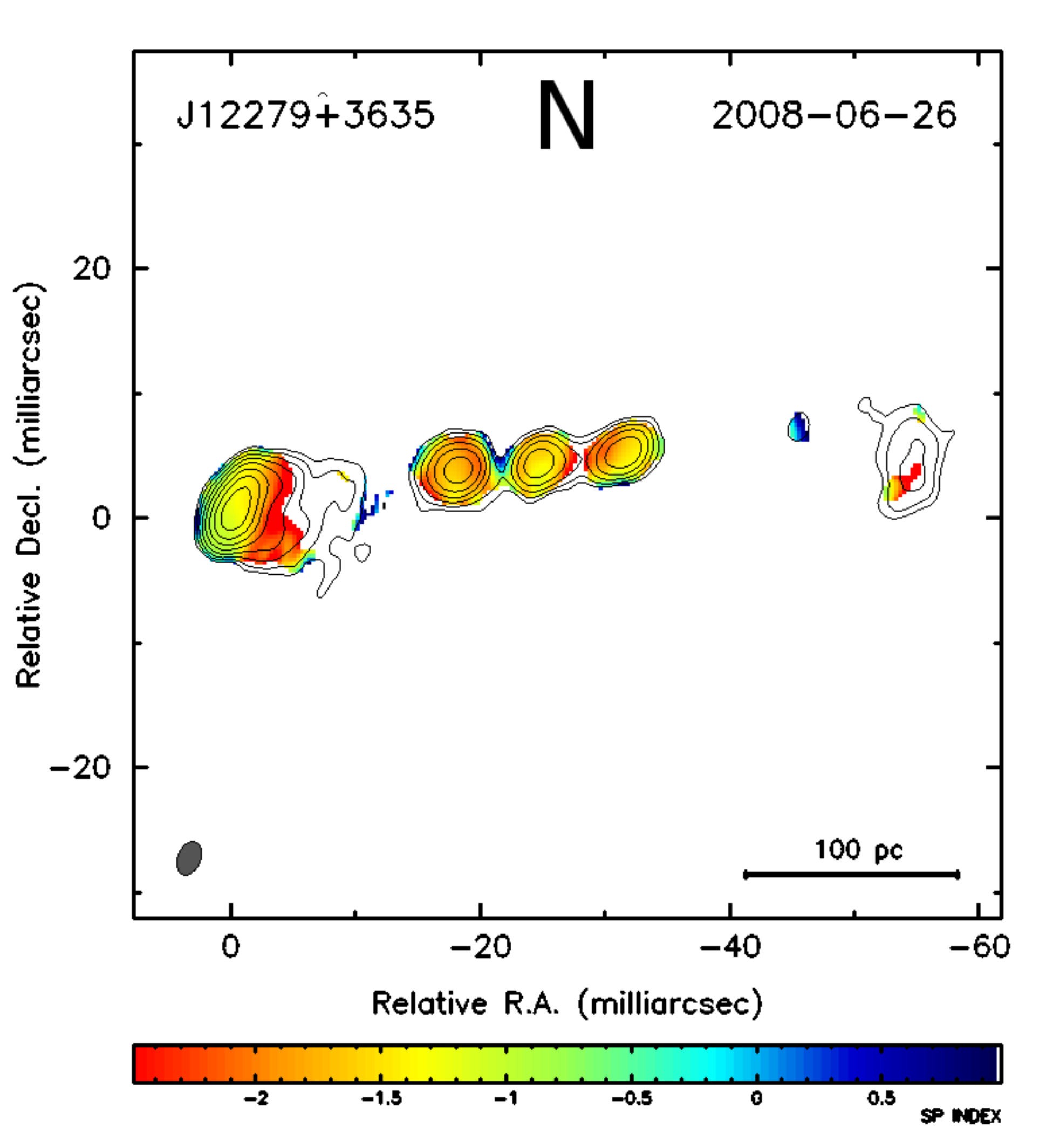}
	}	
~\hfill
	\centering
	{%
	\includegraphics[width=0.32273\textwidth]{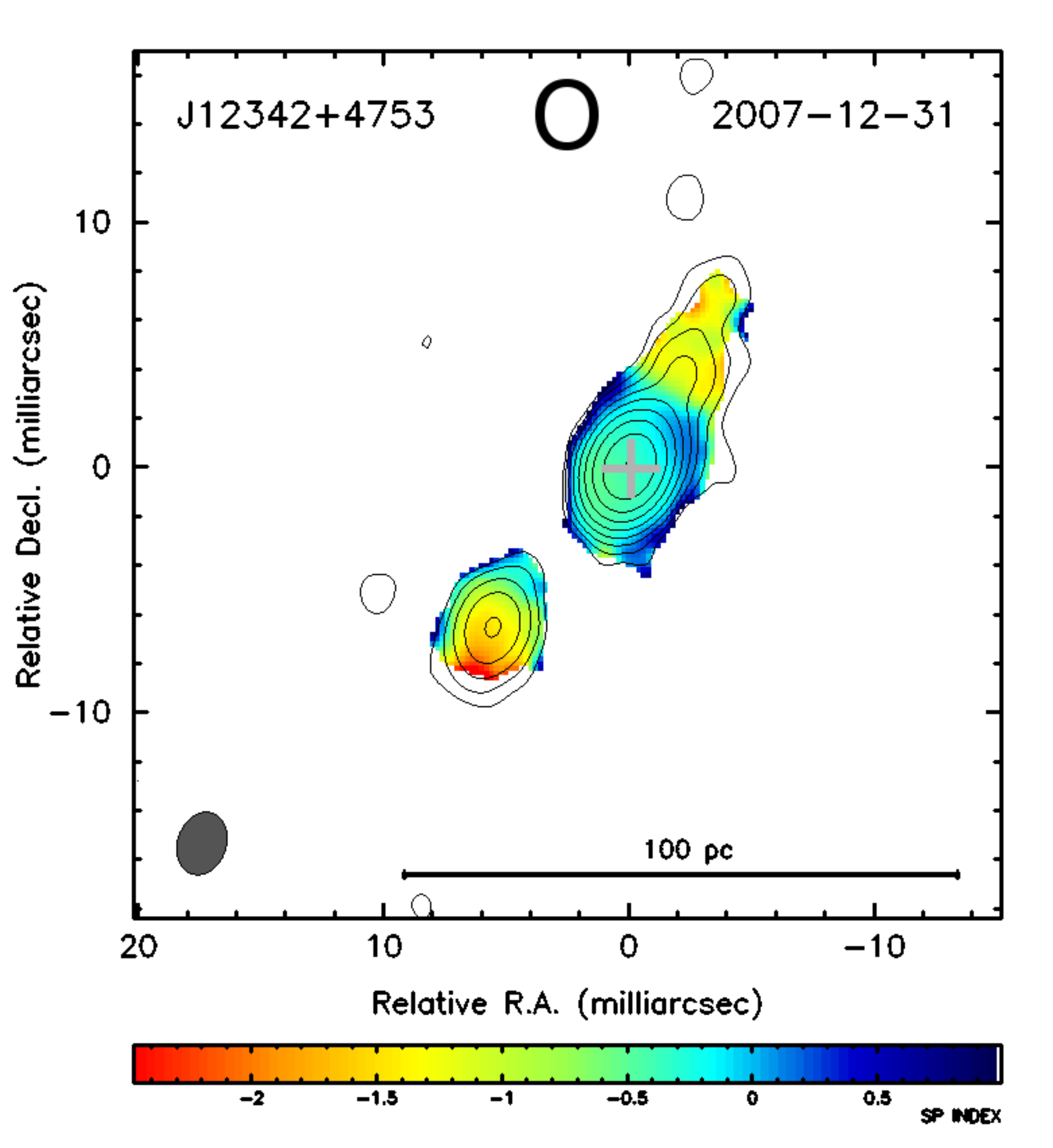}
	}	
~\hfill
	\centering
	{%
	\includegraphics[width=0.32273\textwidth]{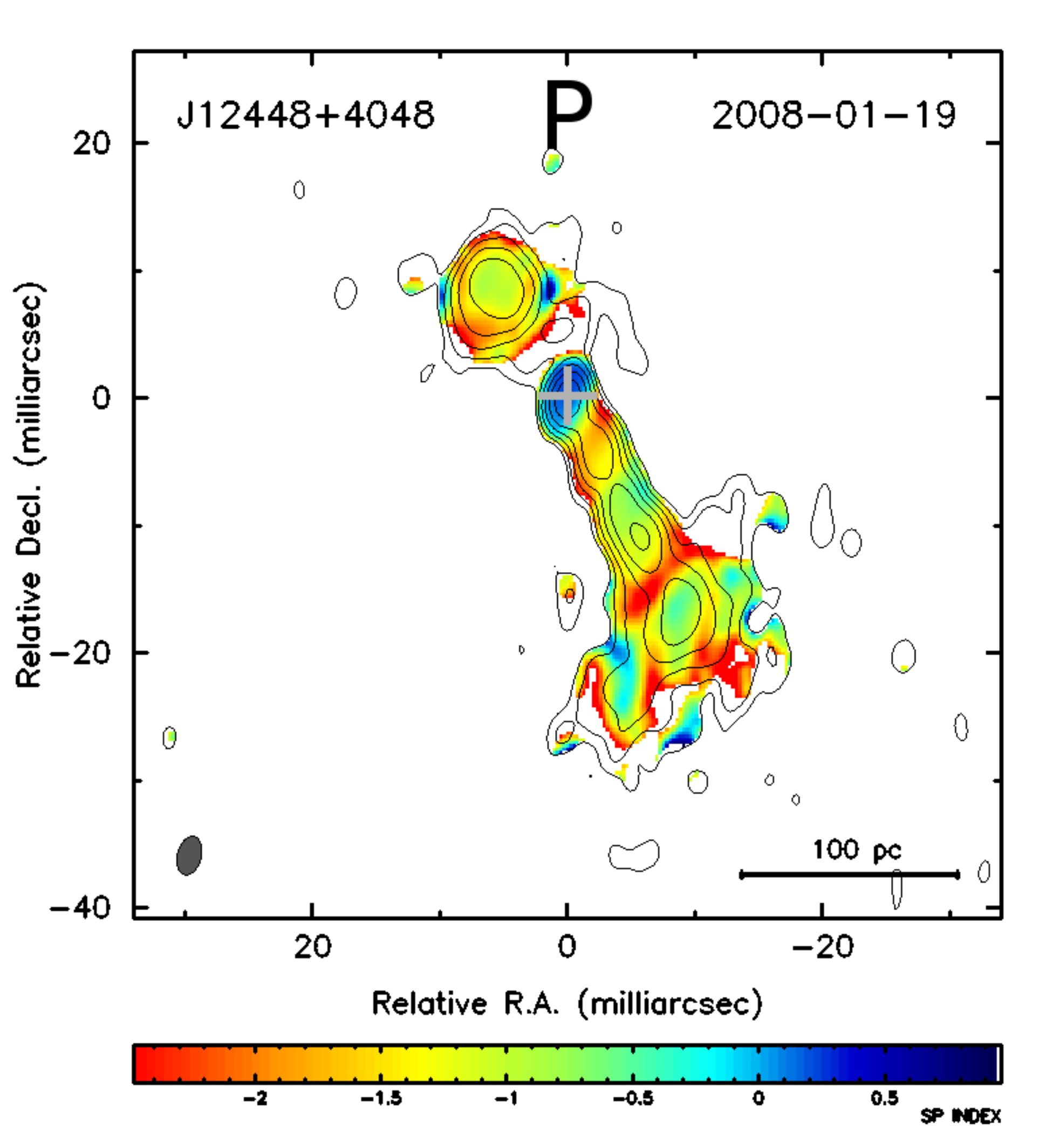}
	}
~\hfill
	\centering
	{%
	\includegraphics[width=0.32273\textwidth]{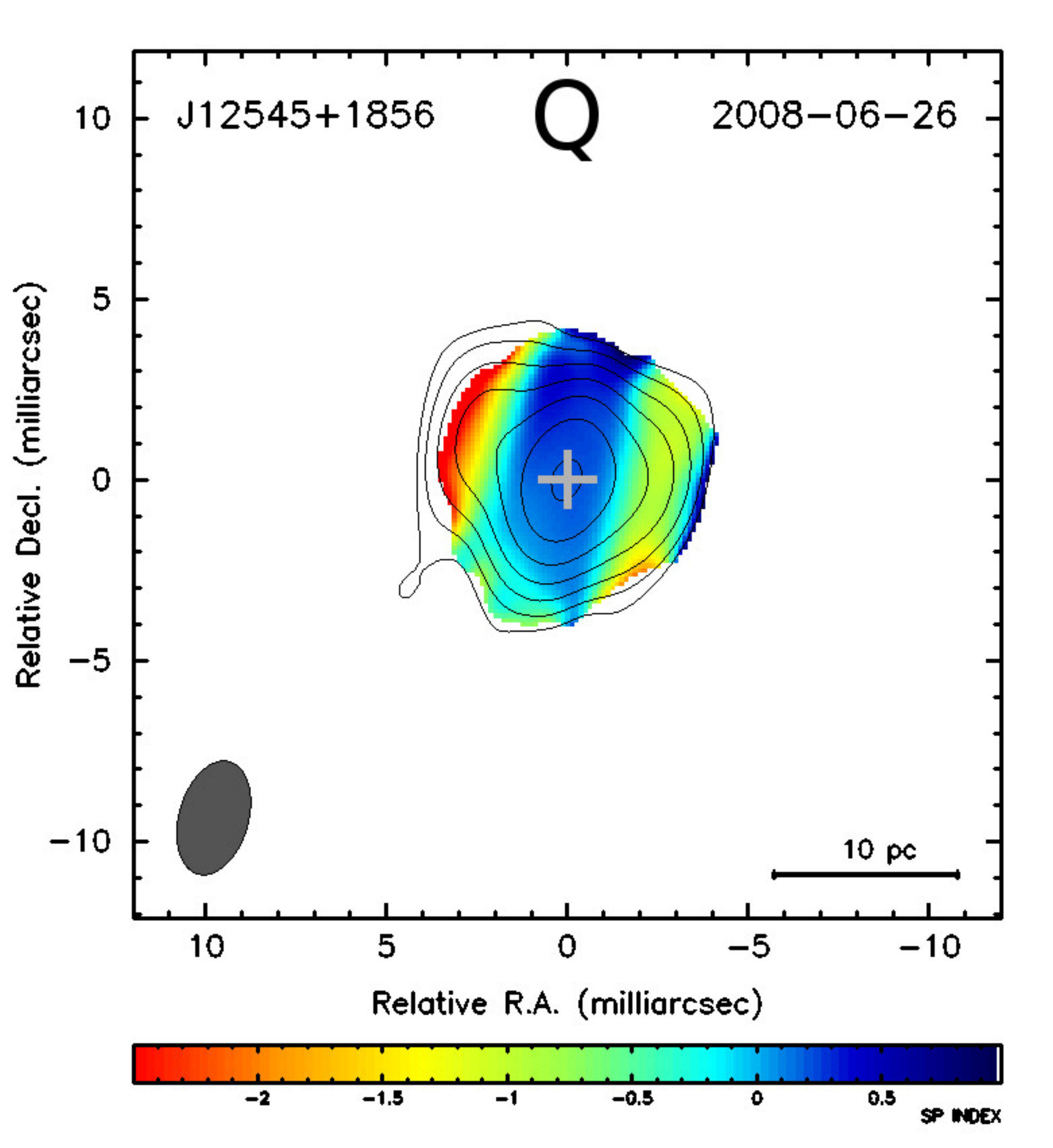}
	}		
~\hfill
	\centering
	{%
	\includegraphics[width=0.32273\textwidth]{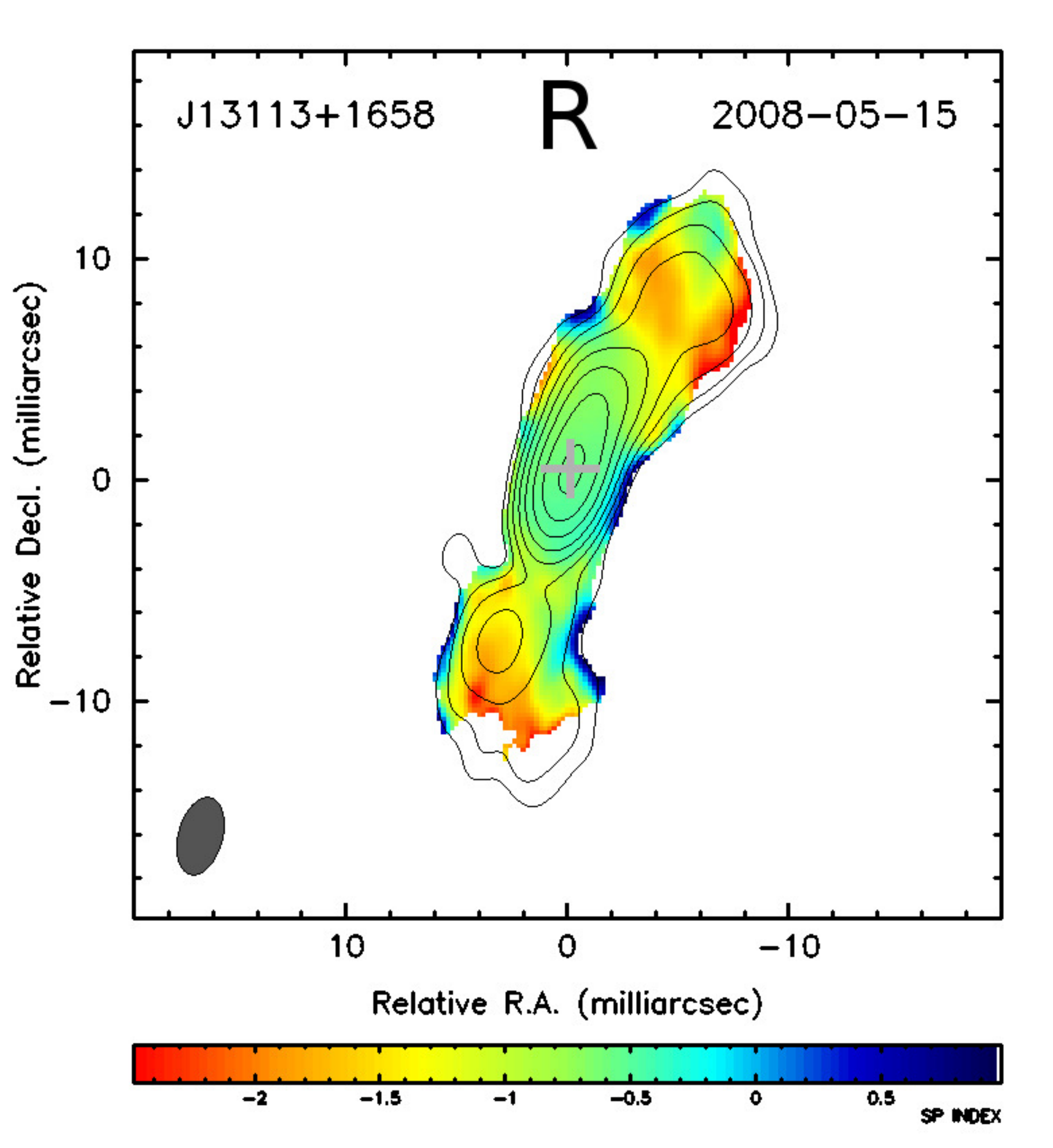}
	}	
	\caption{\textit{\textbf{- Continued}} 5 GHz contour maps of VIPS CSOs with 5-8 GHz spectral index map overlays. The contour levels begin at thrice the theoretical noise (typically $\sim$ 0.4 mJy for BT088 and 1.0 mJy for BT094) and increases by powers of 2. The colour scale is fixed from -2.5 to 1 to facilitate comparison. When detected, a grey cross is placed over the core to guide the reader's eye. Sources with confirmed spectral redshifts have associated distance bars for linear scale.}
\label{fig:cso_cxspix}
\end{figure*}

\begin{figure*}
\ContinuedFloat
 \captionsetup{list=off}
	\centering
	 {%
	 \includegraphics[width=0.32273\textwidth]{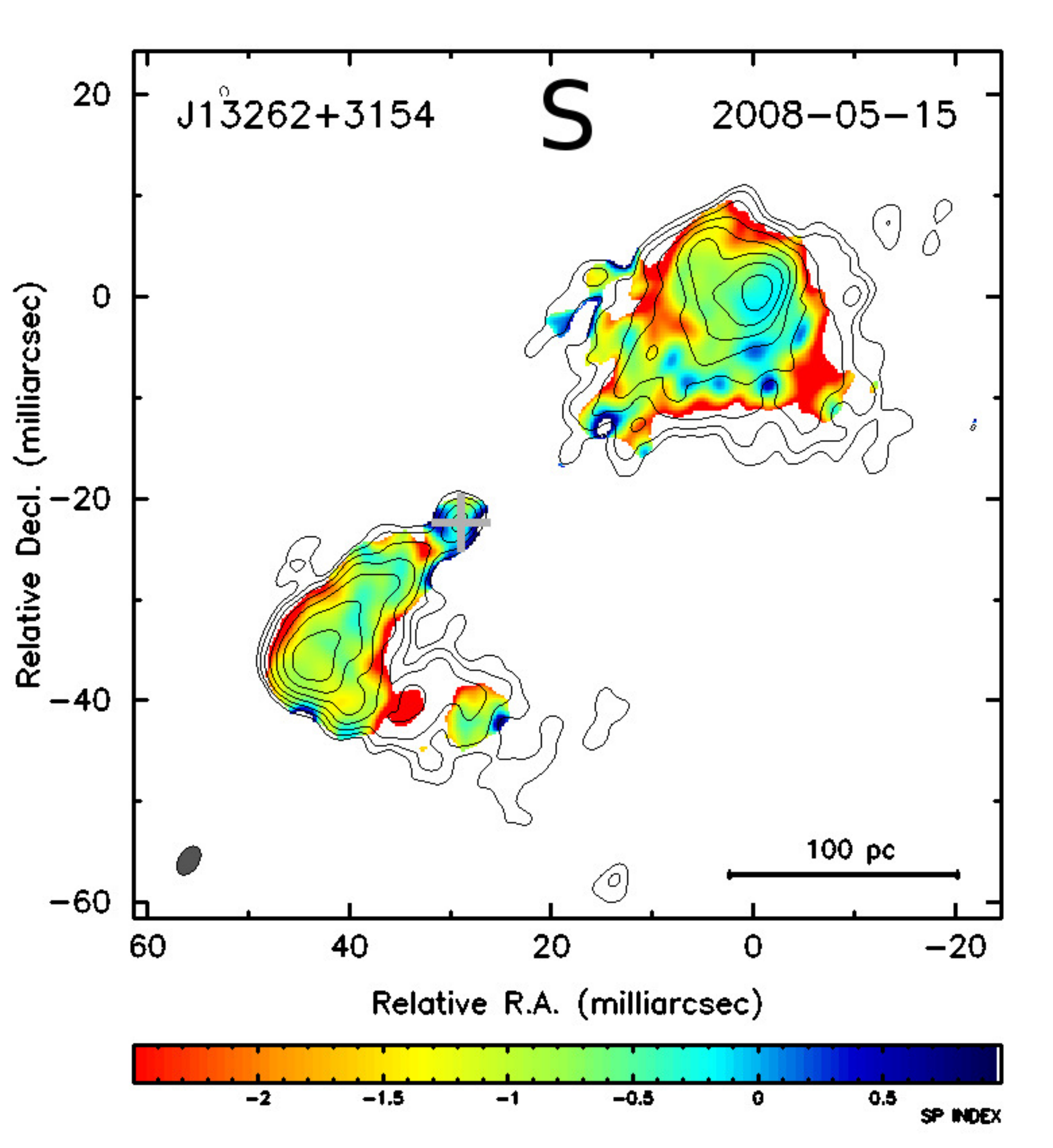}
	}
~\hfill
	\centering
	{%
	\includegraphics[width=0.32273\textwidth]{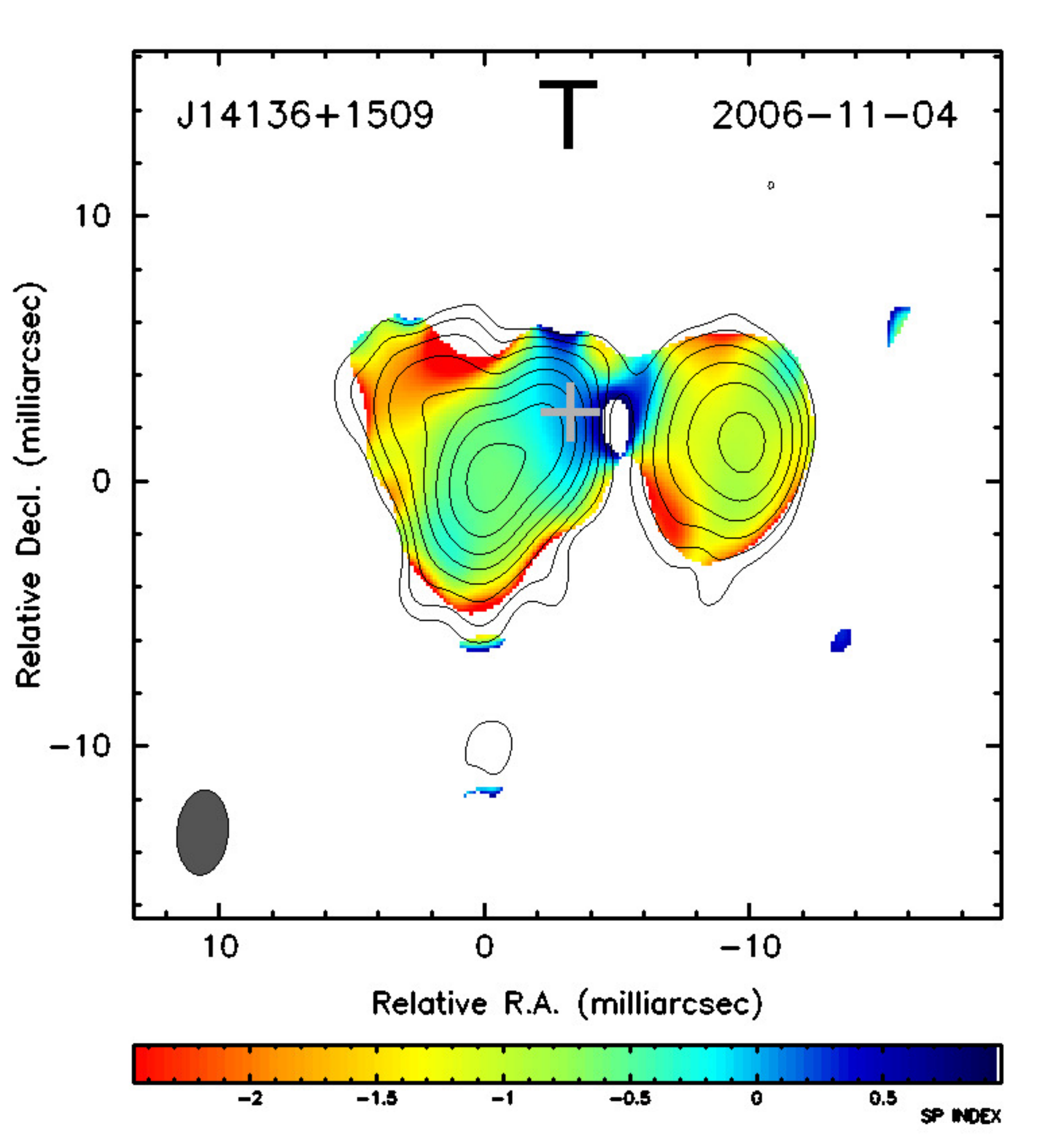}
	}
~\hfill
	\centering
	{%
	\includegraphics[width=0.32273\textwidth]{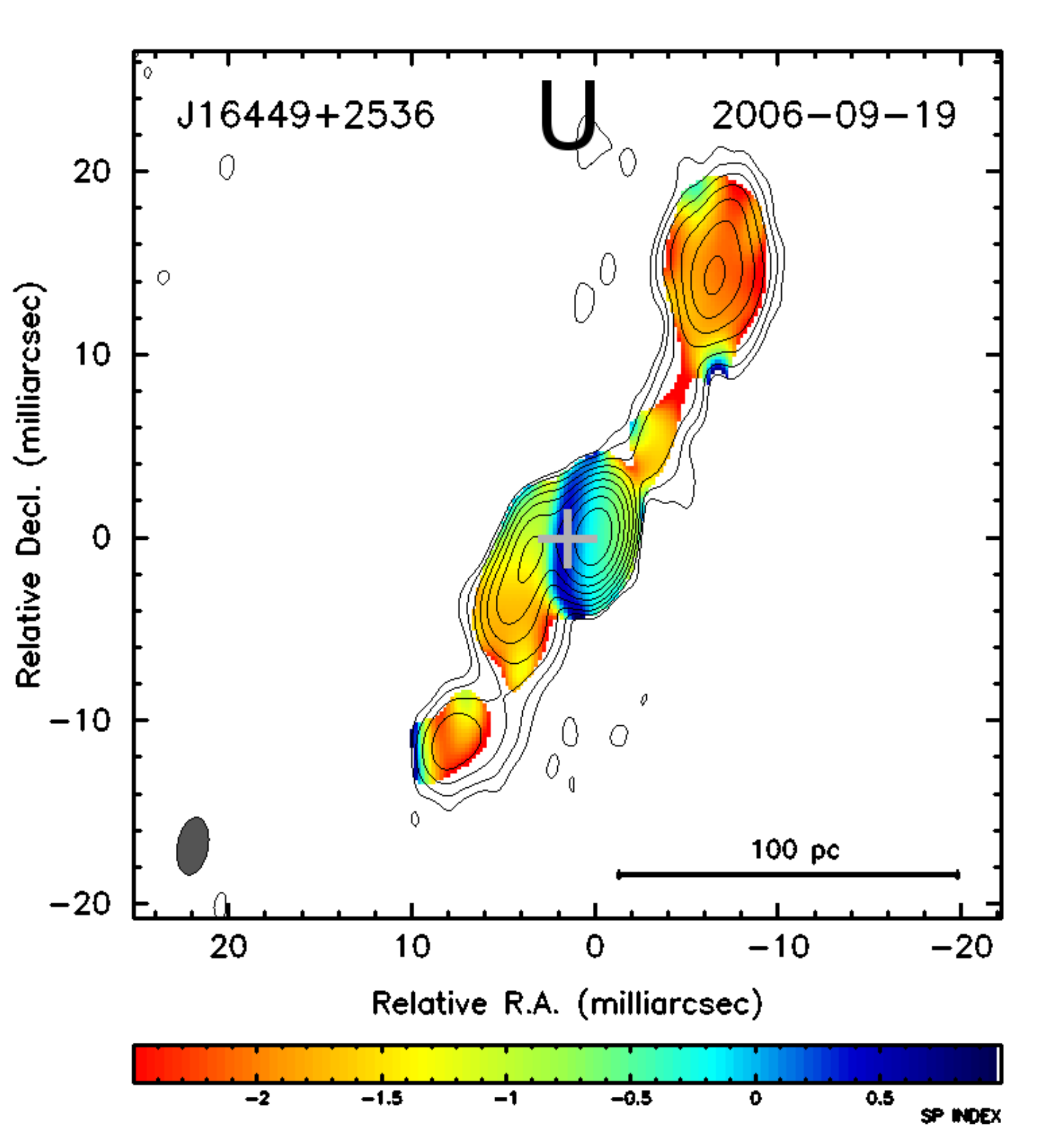}
	}	
~\hfill
	\centering
	{%
	\includegraphics[width=0.32273\textwidth]{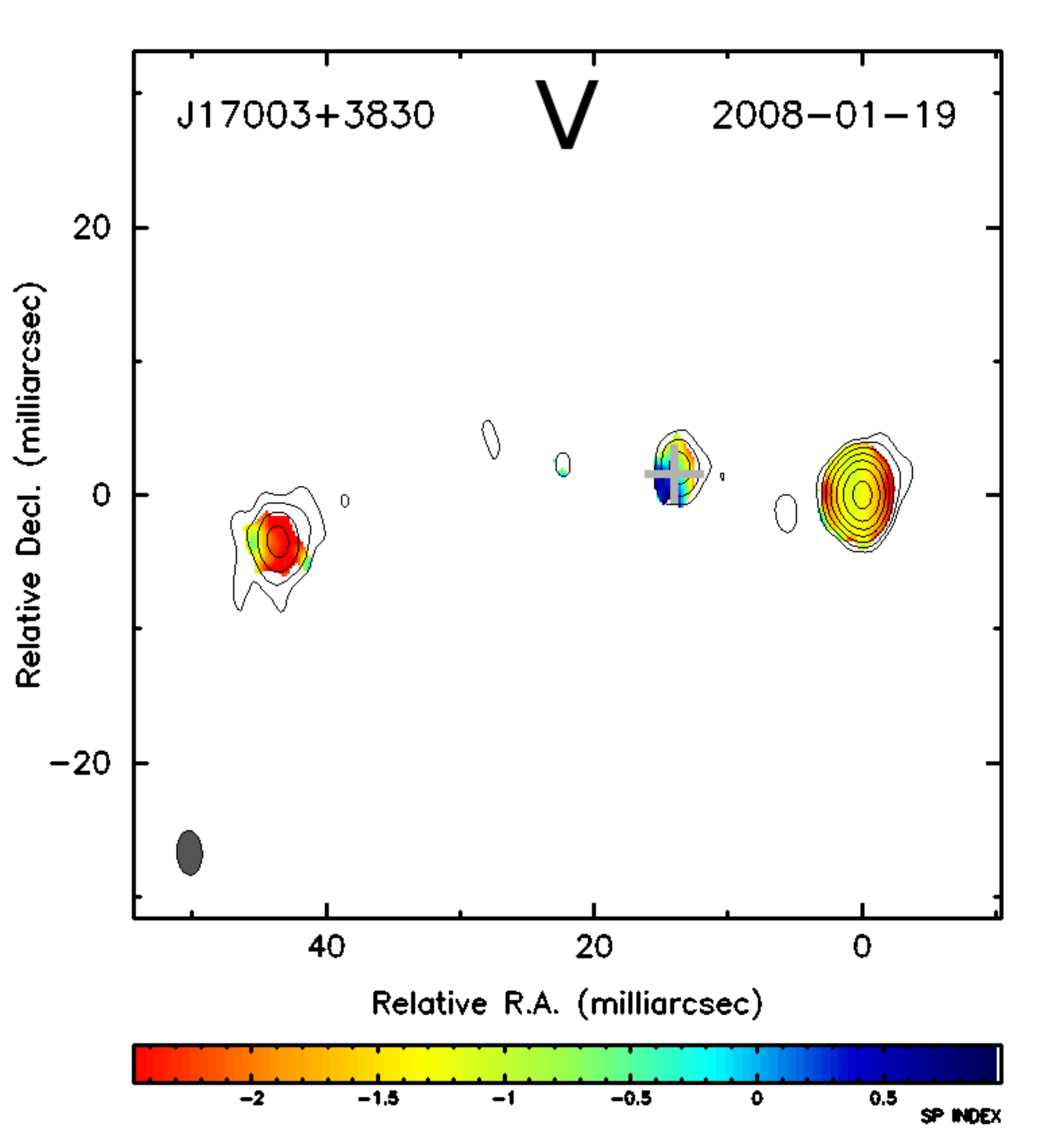}
	}
	\caption{\textit{\textbf{- Continued}} 5 GHz contour maps of VIPS CSOs with 5-8\,GHz spectral index map overlays. The contour levels begin at thrice the theoretical noise (typically $\sim$ 0.4\,mJy for BT088 and 1.0\,mJy for BT094) and increases by powers of 2. The colour scale is fixed from -2.5 to 1 to facilitate comparison. When detected, a grey cross is placed over the core to guide the reader's eye. Sources with confirmed spectral redshifts have associated distance bars for linear scale.}
\label{fig:cso_cxspix}
\end{figure*}

 \begin{figure*}

	\centering
	 {%
	 \includegraphics[width=0.32273\textwidth]{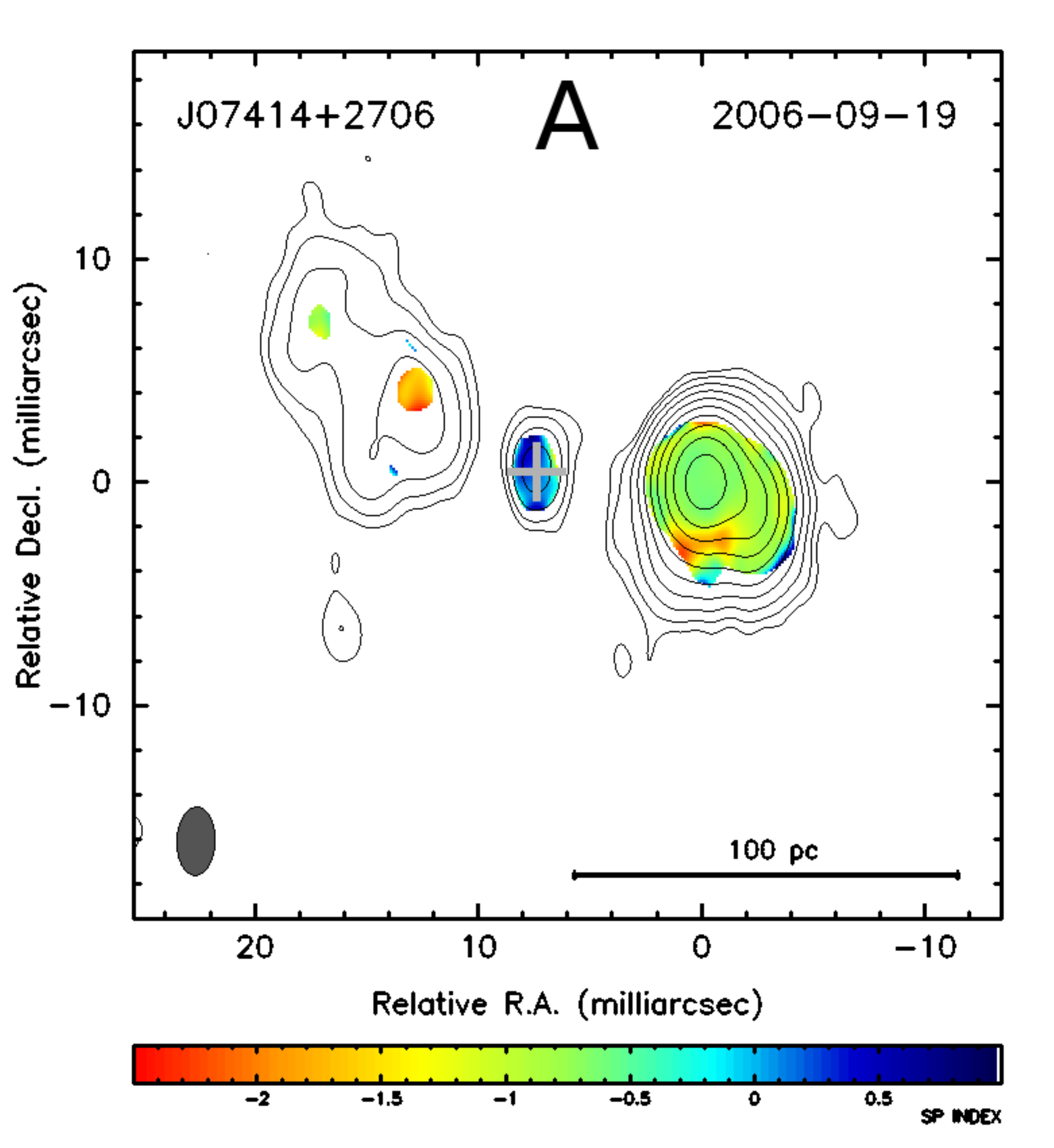}
	}
~\hfill
	\centering
	{%
	\includegraphics[width=0.32273\textwidth]{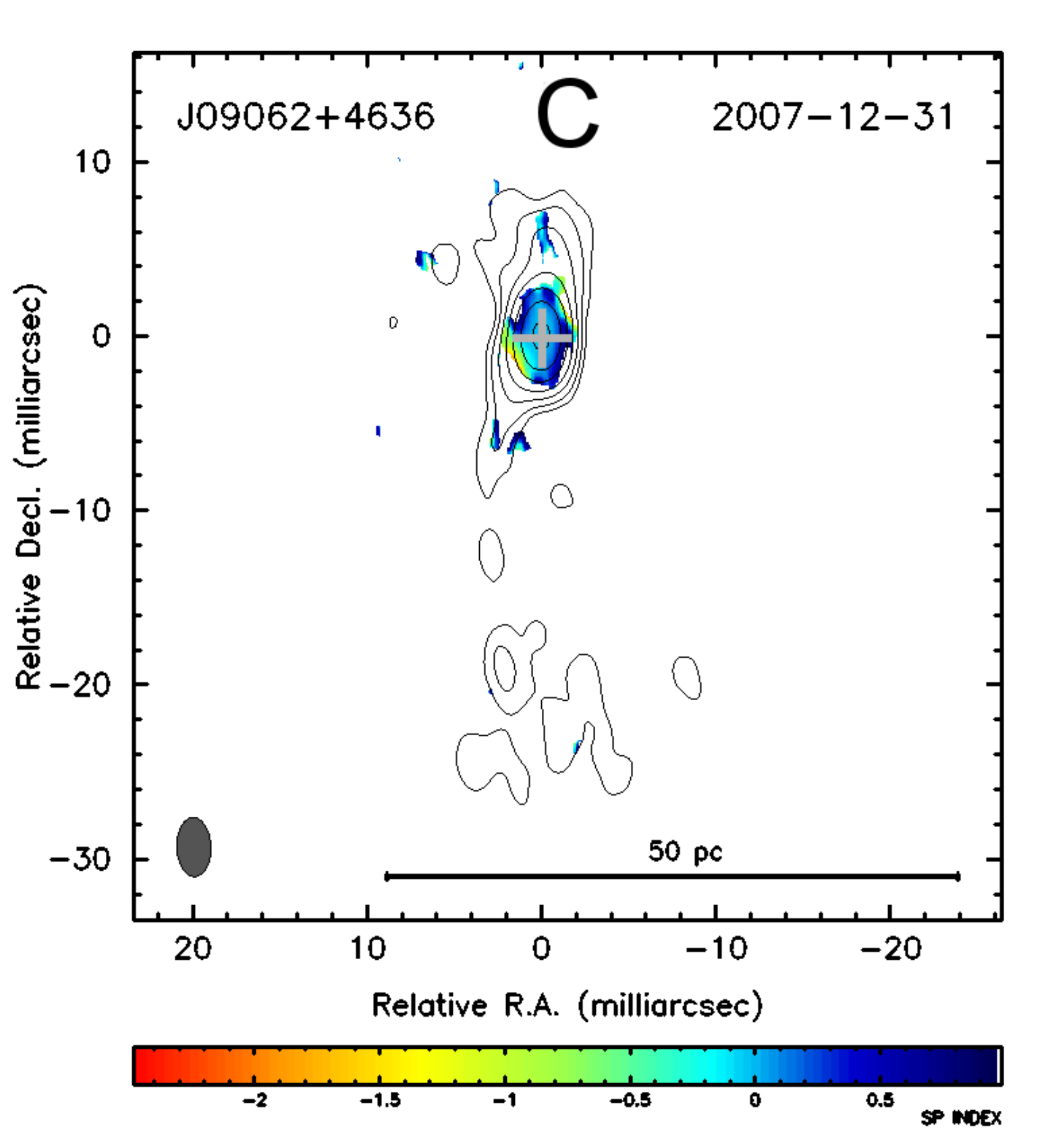}
	}
~\hfill
	\centering
	{%
	\includegraphics[width=0.32273\textwidth]{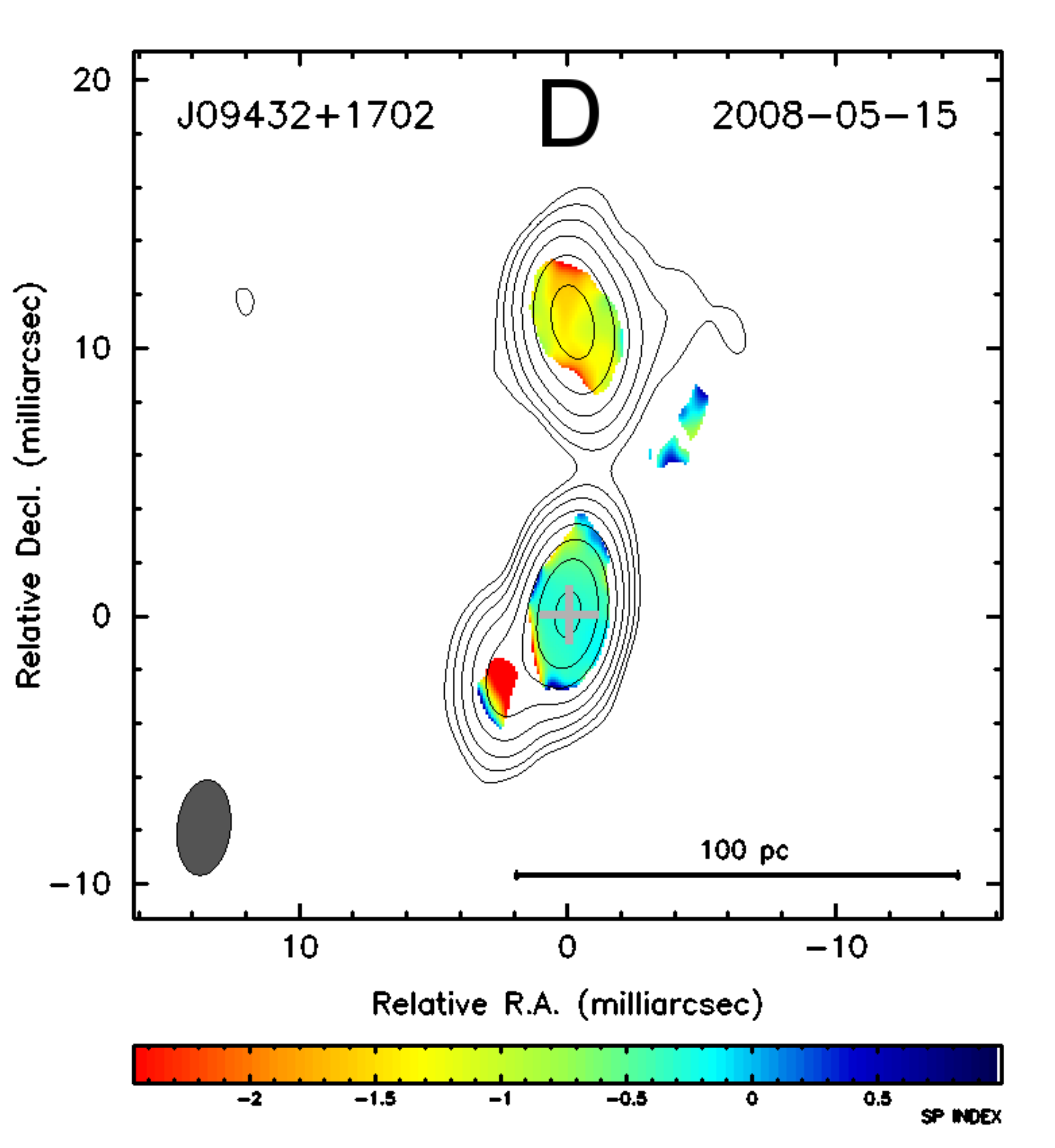}
	}	
~\hfill
	\centering
	{%
	\includegraphics[width=0.32273\textwidth]{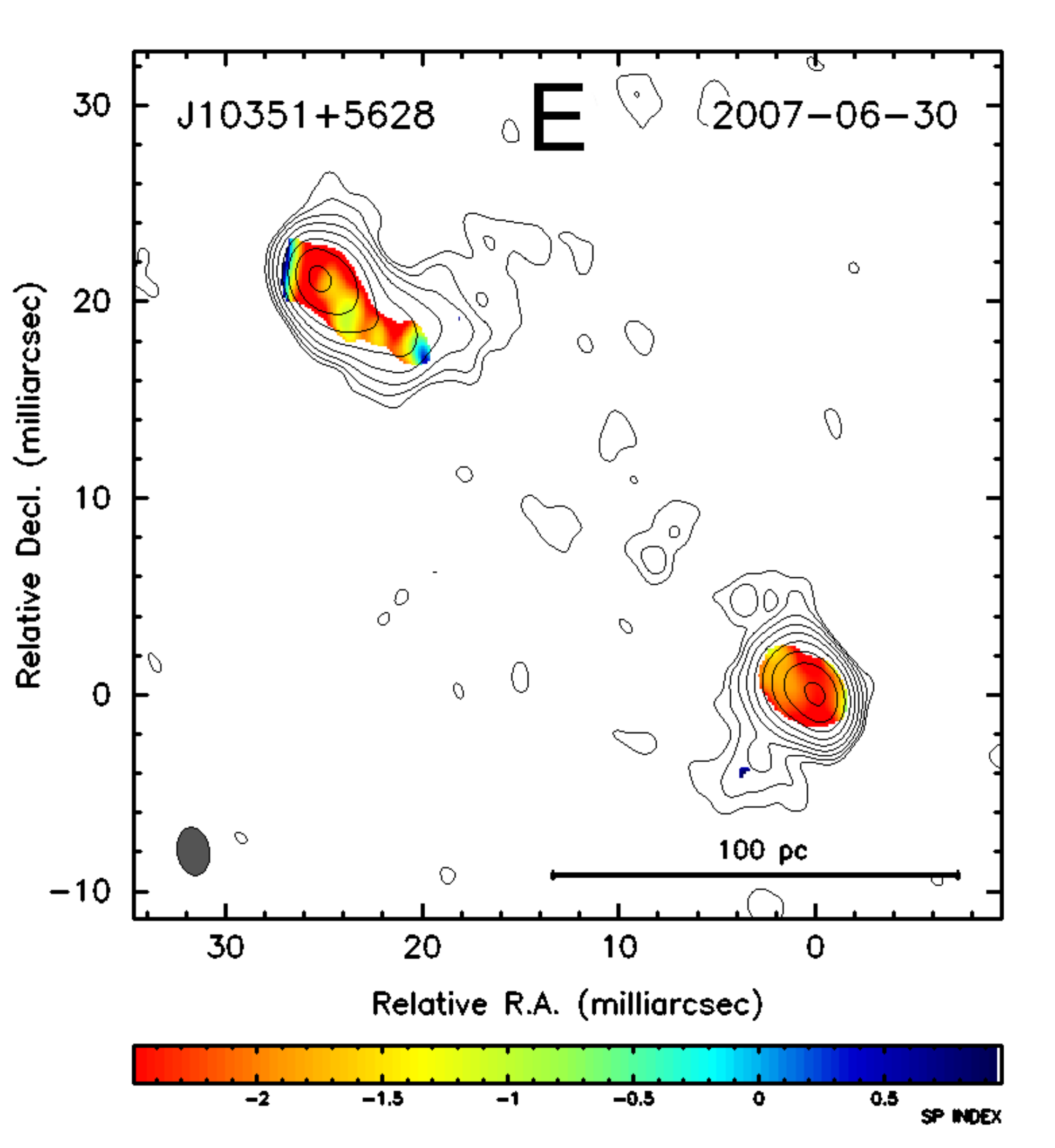}
	}	
~\hfill
	\centering
	{%
	\includegraphics[width=0.32273\textwidth]{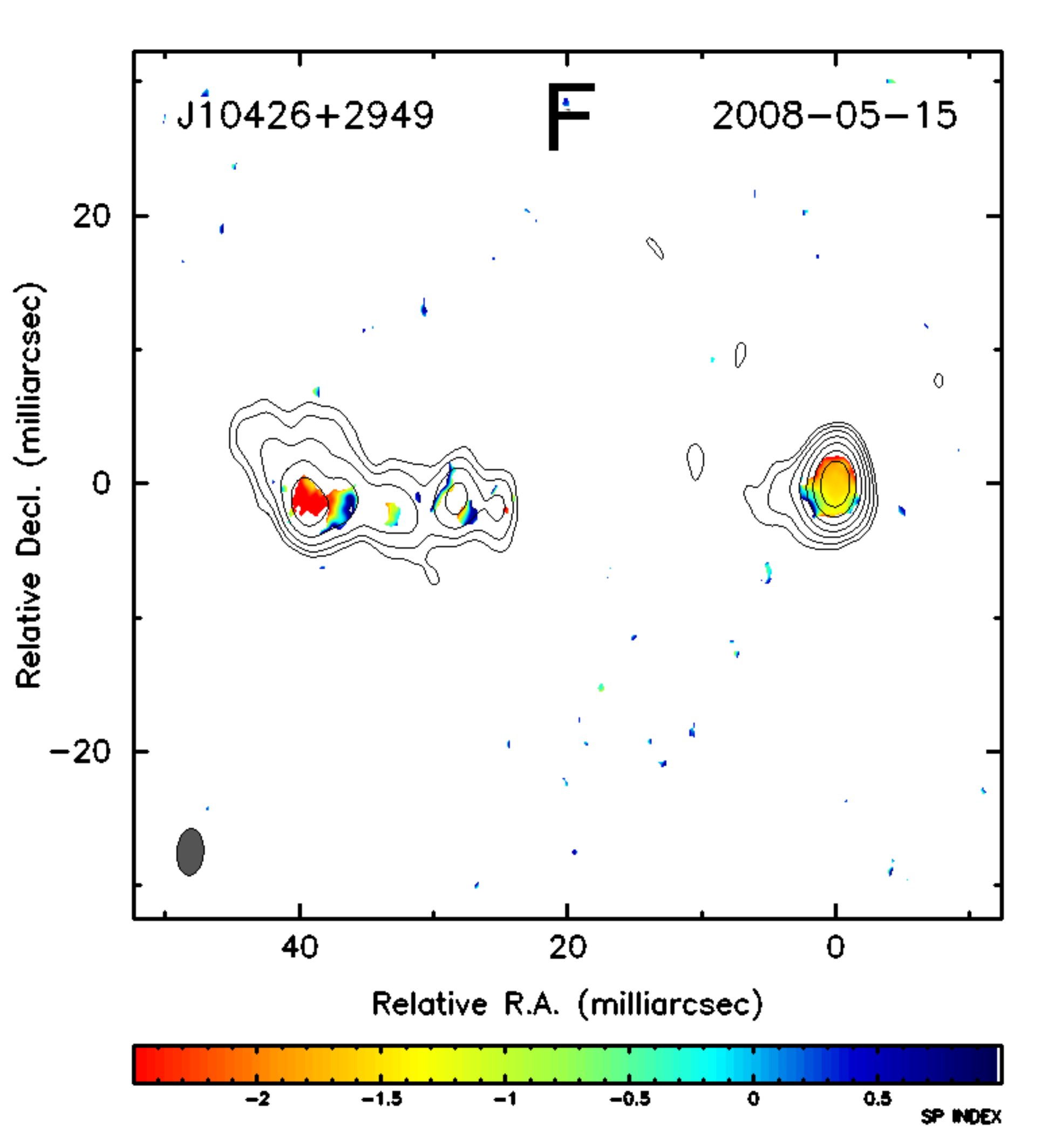}
	}	
~\hfill
	\centering
	{%
	\includegraphics[width=0.32273\textwidth]{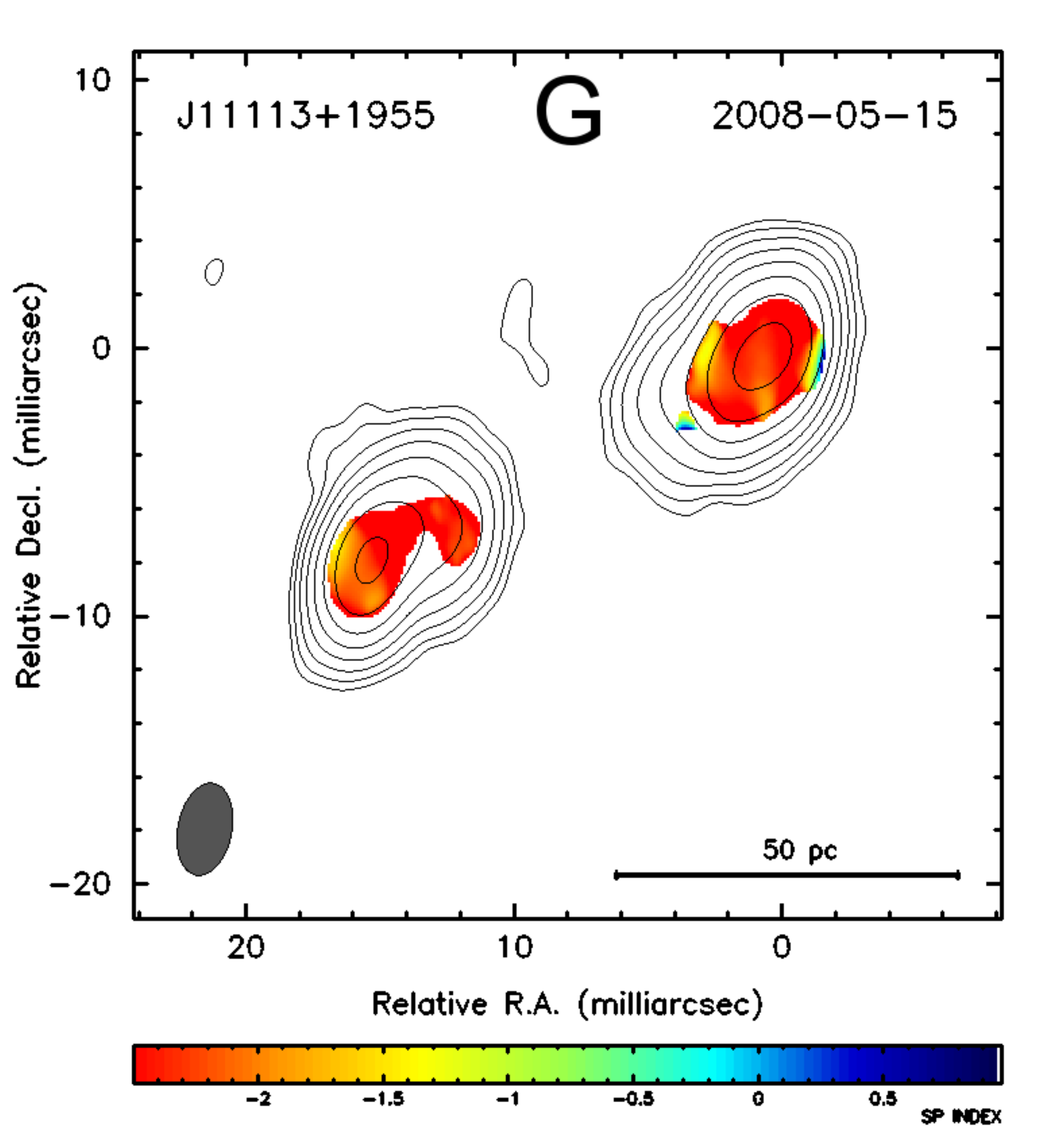}
	}	
~\hfill
	\centering
	{%
	\includegraphics[width=0.32273\textwidth]{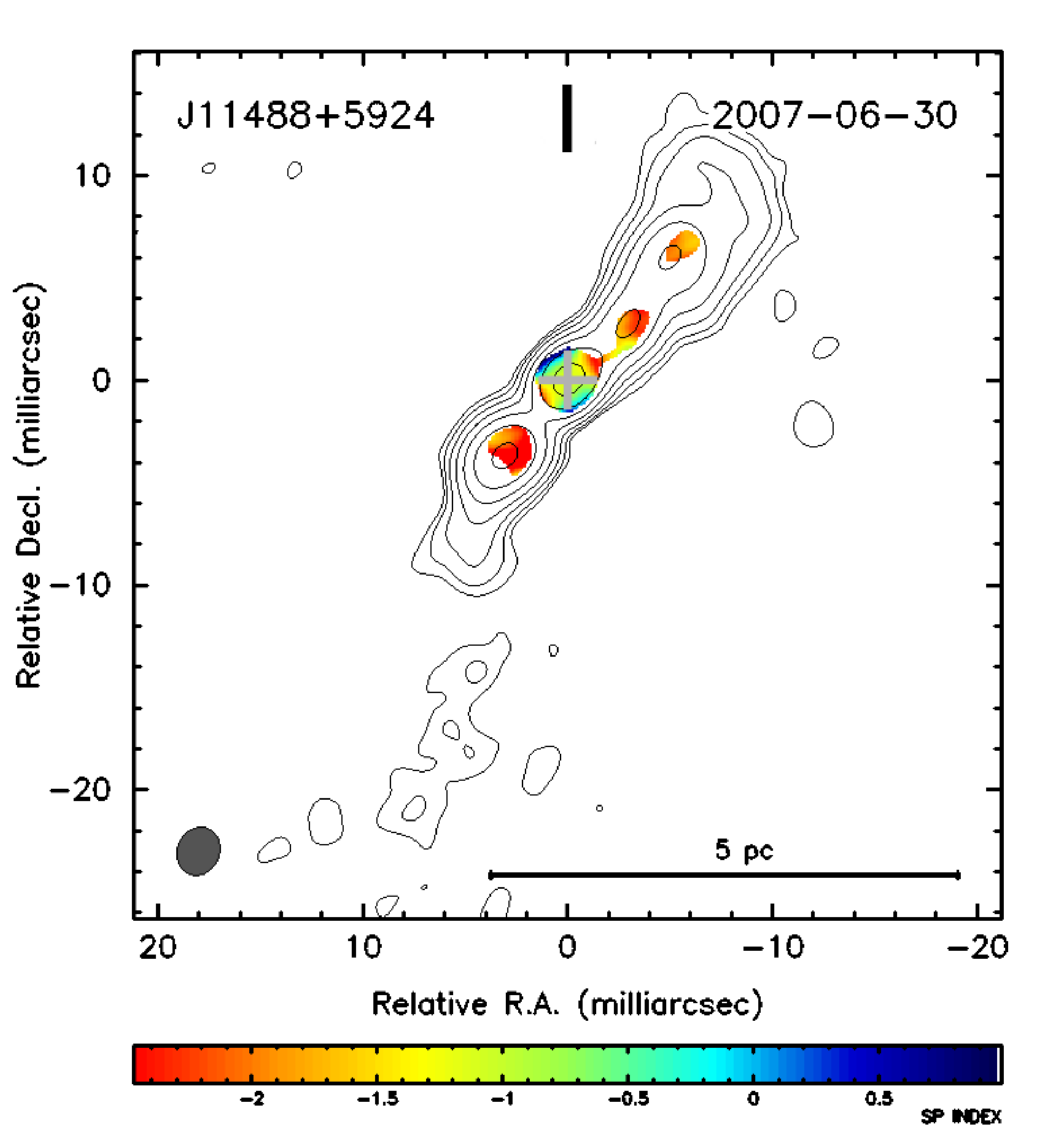}
	}
~\hfill
	\centering
	{%
	\includegraphics[width=0.32273\textwidth]{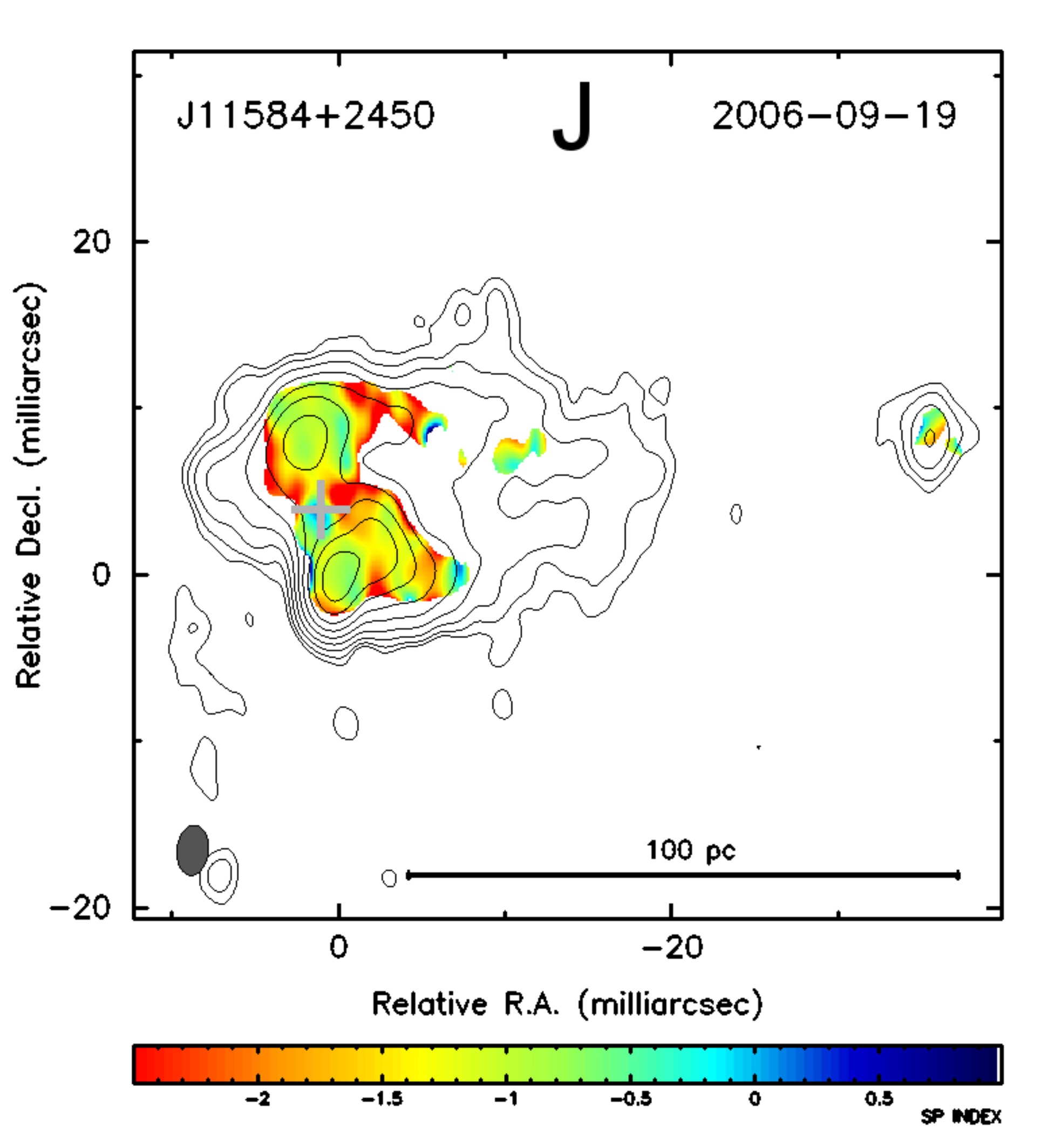}
	}		
~\hfill
	\centering
	{%
	\includegraphics[width=0.32273\textwidth]{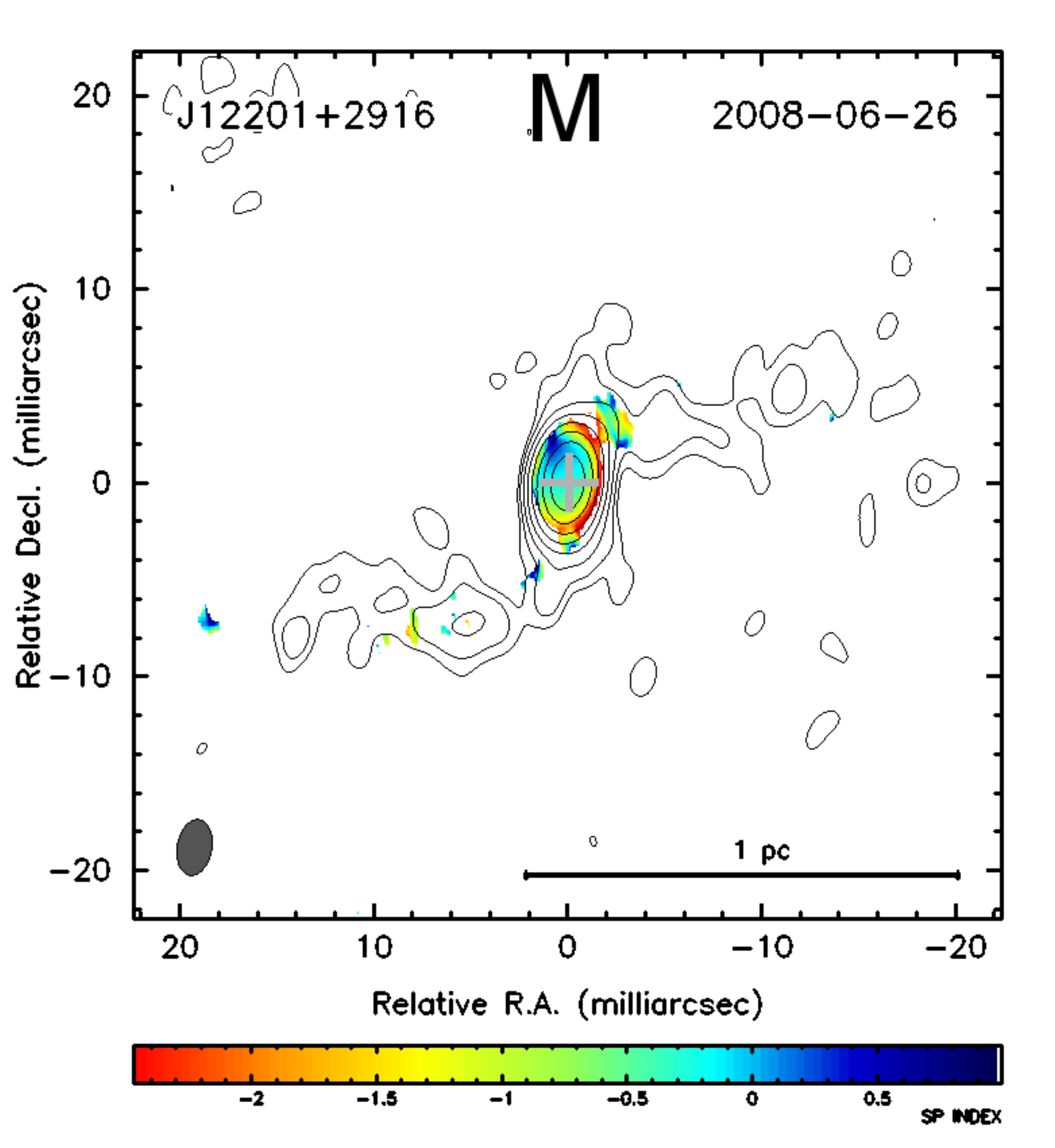}
	}	
	\caption{5 GHz contour maps of VIPS CSOs with 8-15 GHz spectral index map overlays. The contour levels begin at thrice the theoretical noise (typically $\sim$ 0.4 mJy for BT088 and 1.0 mJy for BT094) and increases by powers of 2. The colour scale is fixed from -2.5 to 1 to facilitate comparison. When detected, a grey cross is placed over the core to guide the reader's eye. Sources with confirmed spectral redshifts have associated distance bars for linear scale.}
\label{fig:cso_xuspix}
\end{figure*}

\begin{figure*}
\ContinuedFloat
\captionsetup{list=off}
	\centering
	 {%
	 \includegraphics[width=0.32273\textwidth]{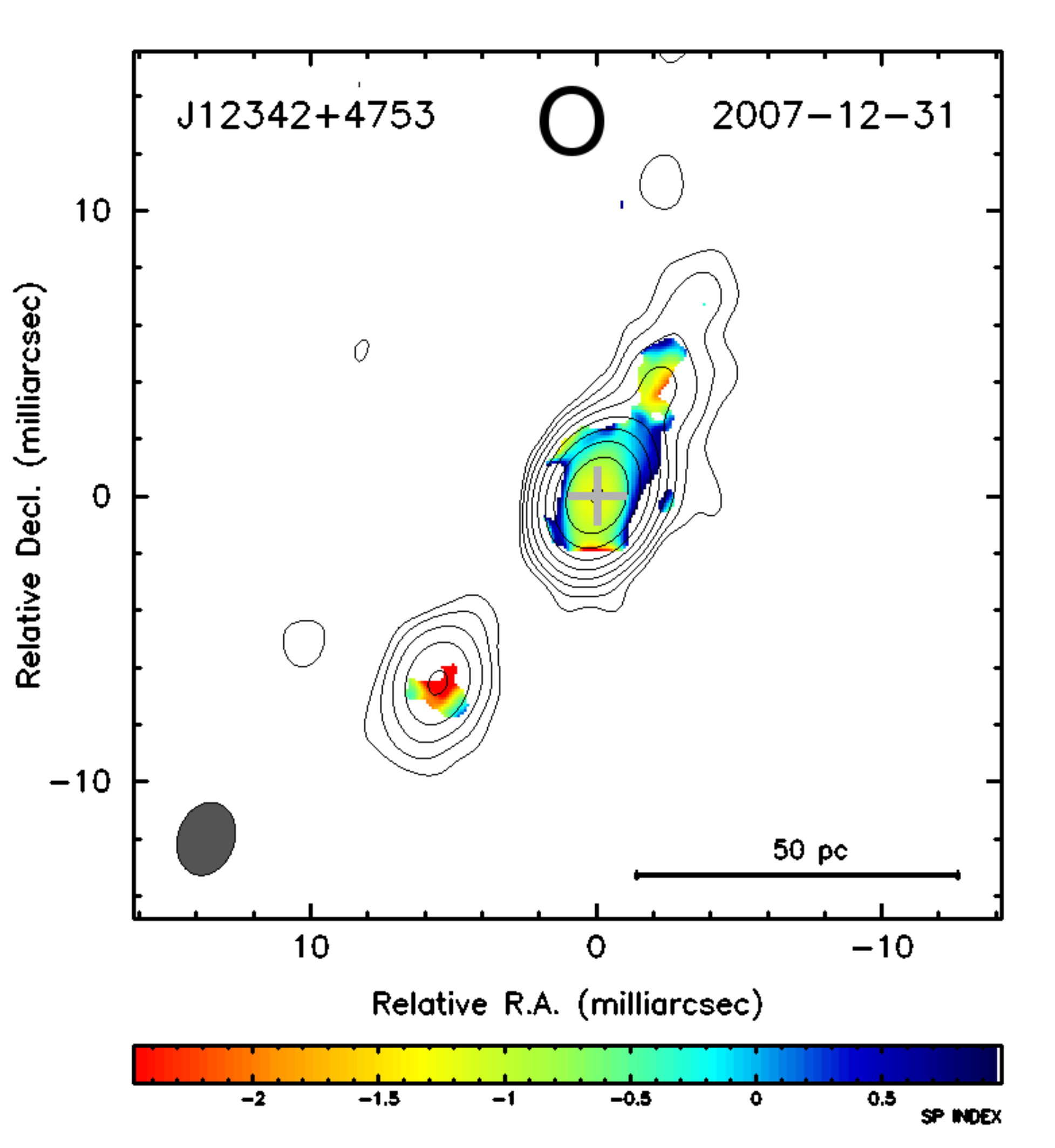}
	}
~\hfill
	\centering
	{%
	\includegraphics[width=0.32273\textwidth]{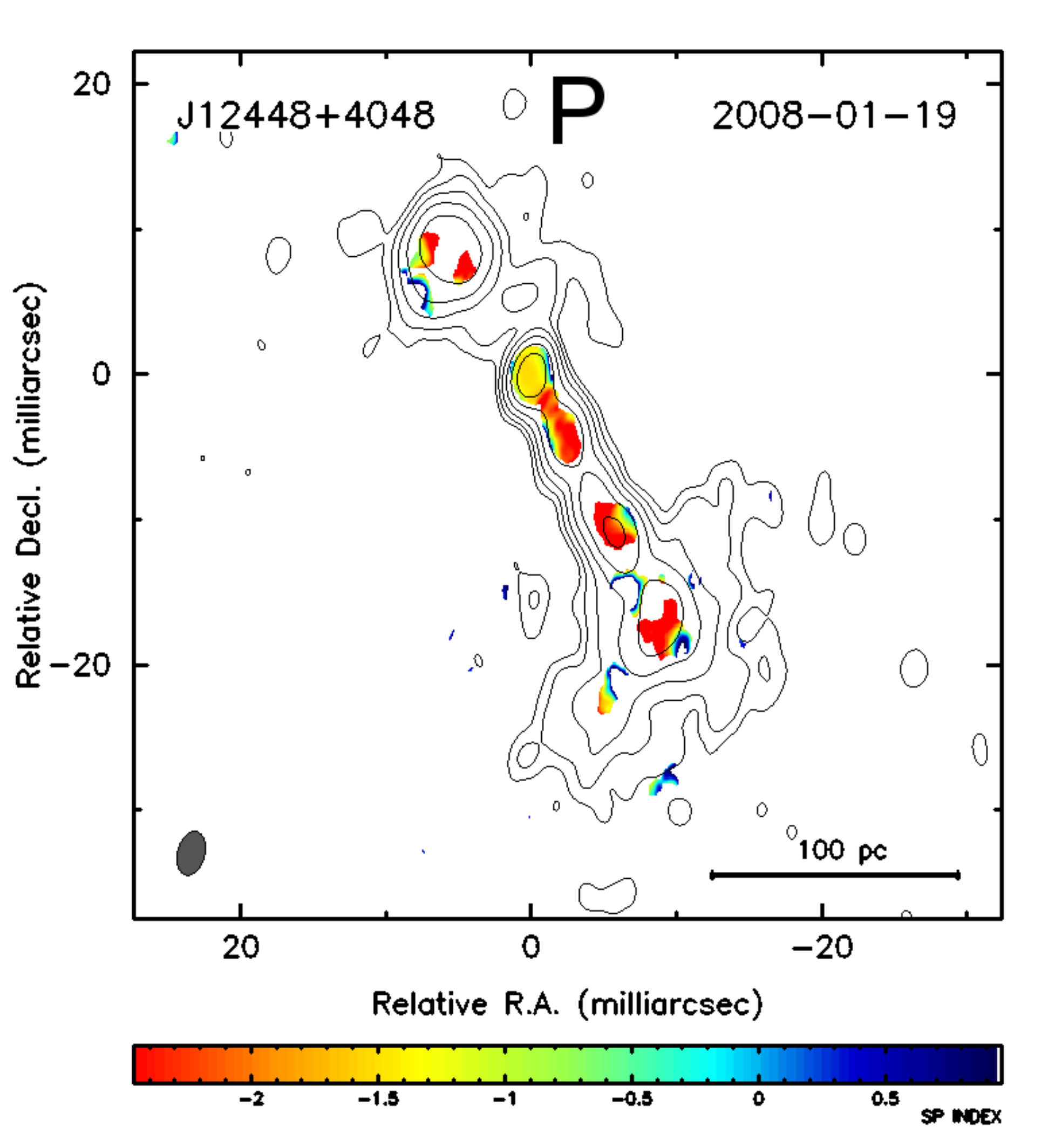}
	}
~\hfill
	\centering
	{%
	\includegraphics[width=0.32273\textwidth]{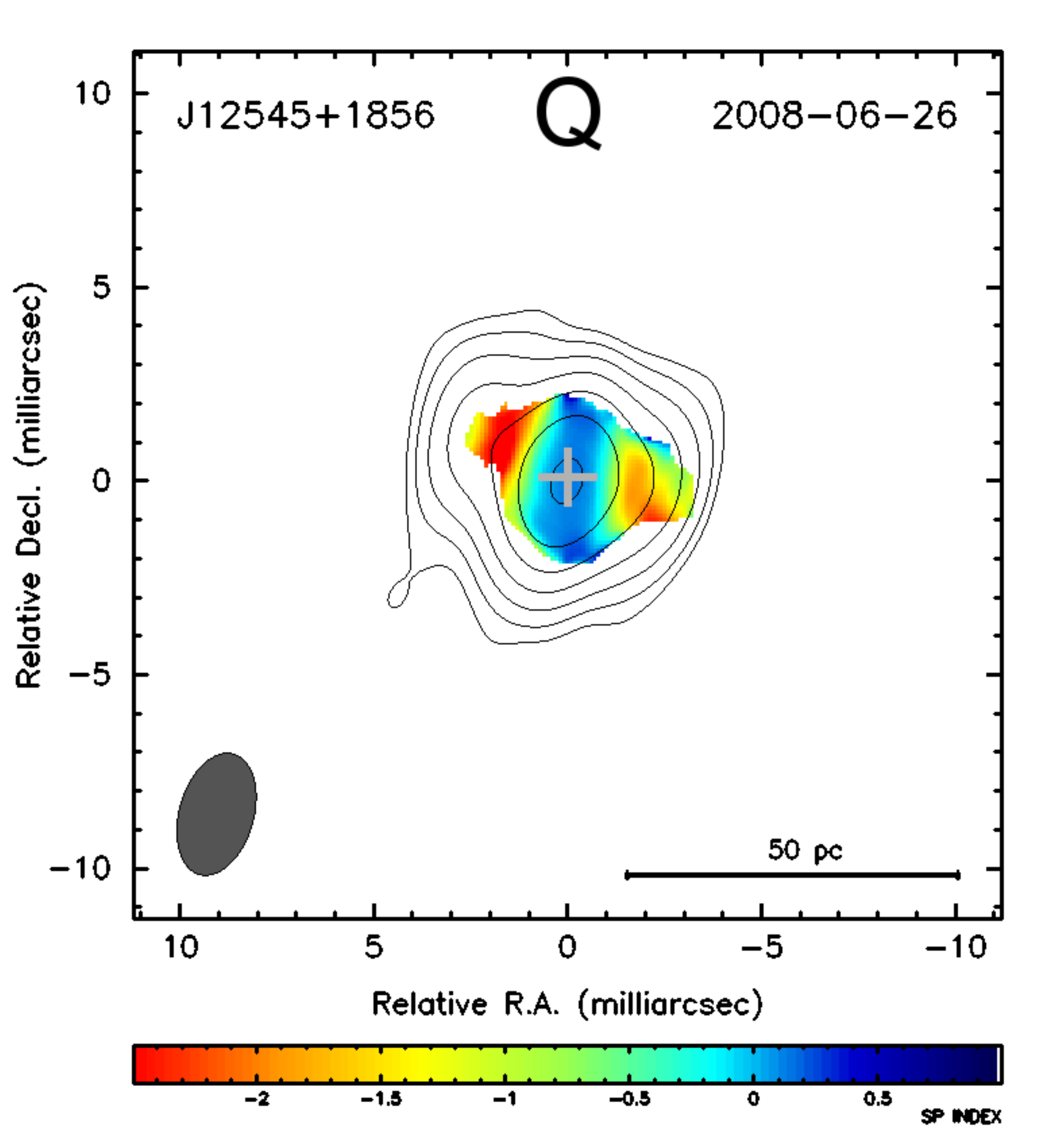}
	}	
~\hfill
	\centering
	{%
	\includegraphics[width=0.32273\textwidth]{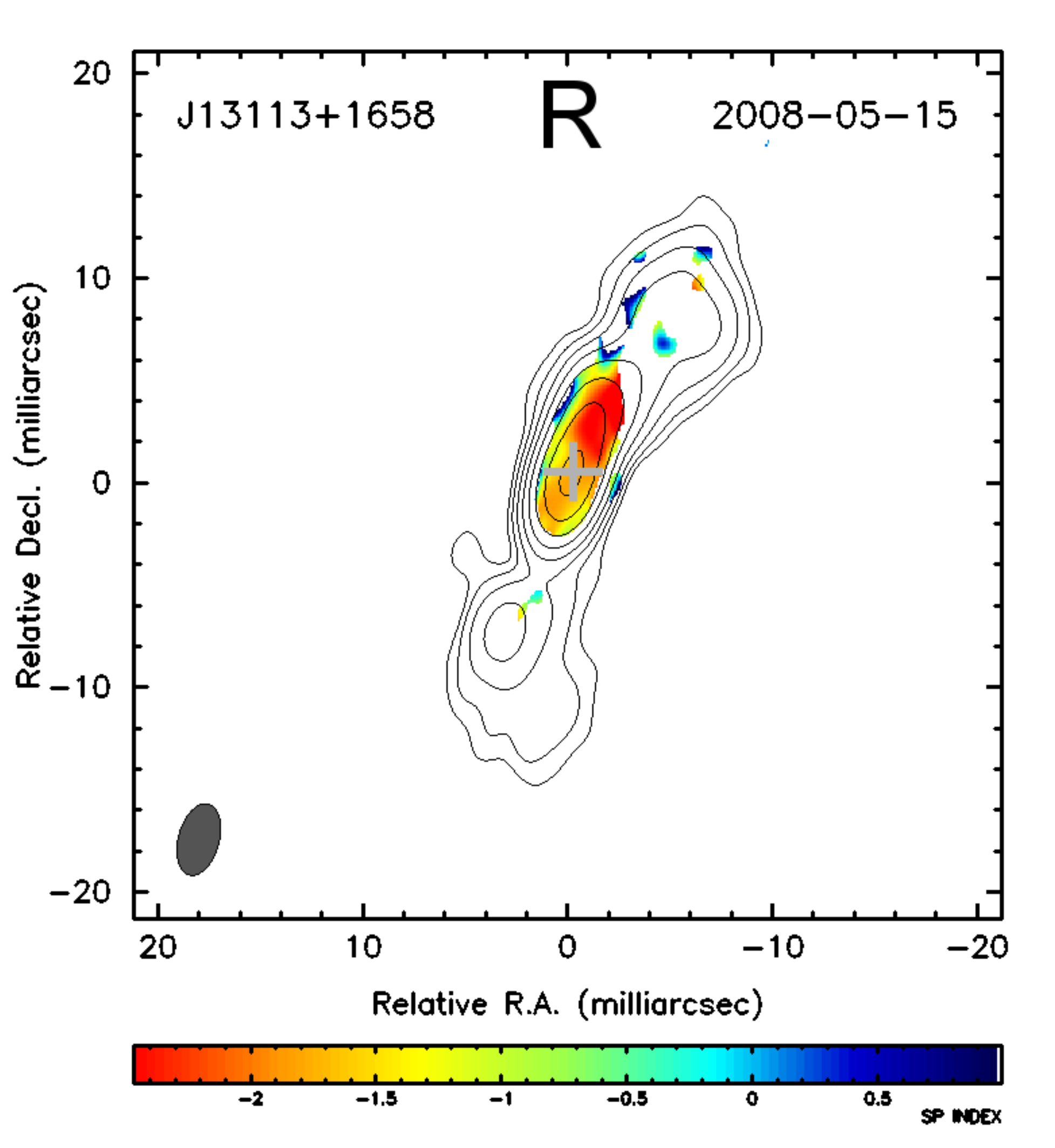}
	}	
~\hfill
	\centering
	{%
	\includegraphics[width=0.32273\textwidth]{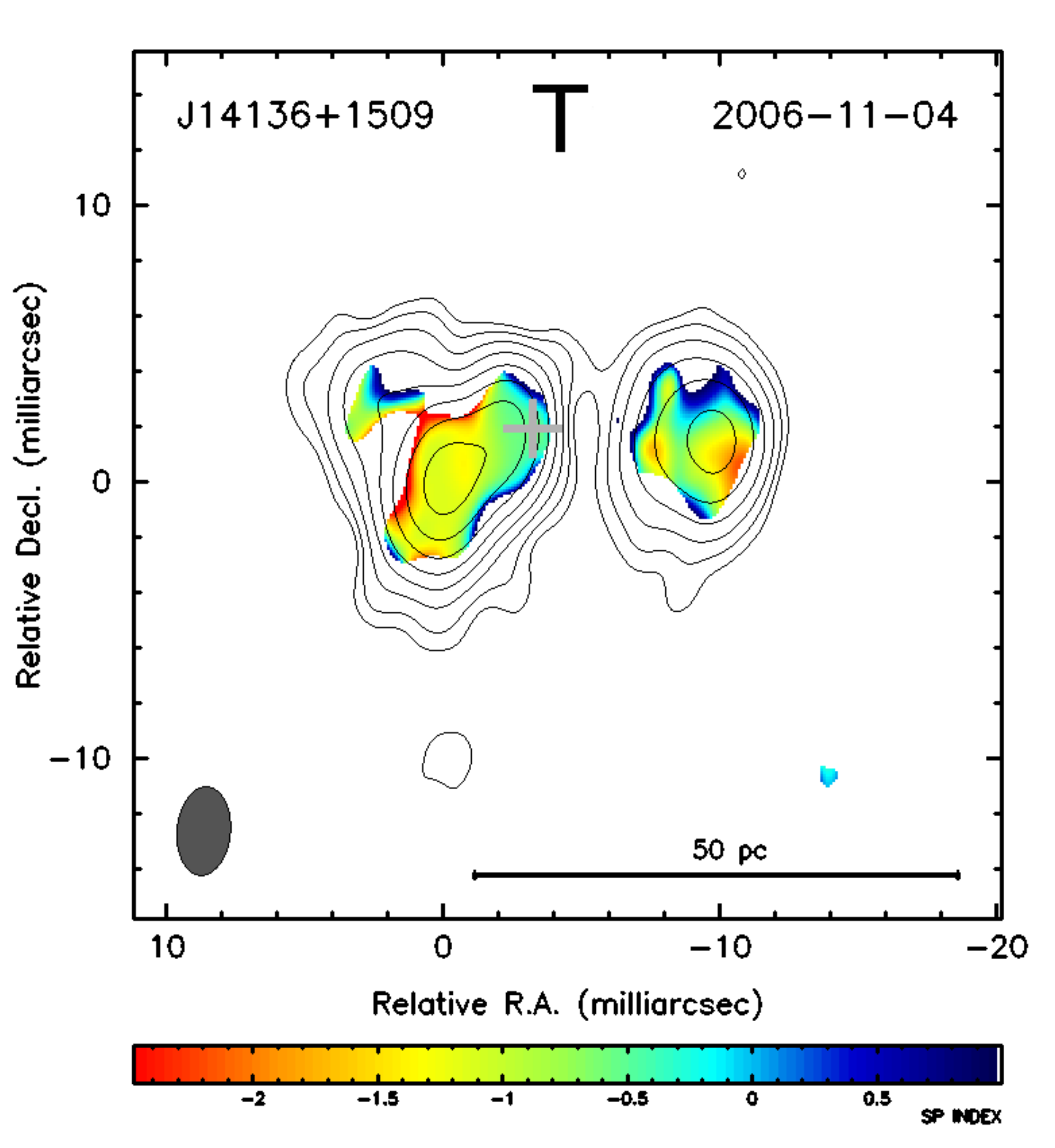}
	}	
~\hfill
	\centering
	{%
	\includegraphics[width=0.32273\textwidth]{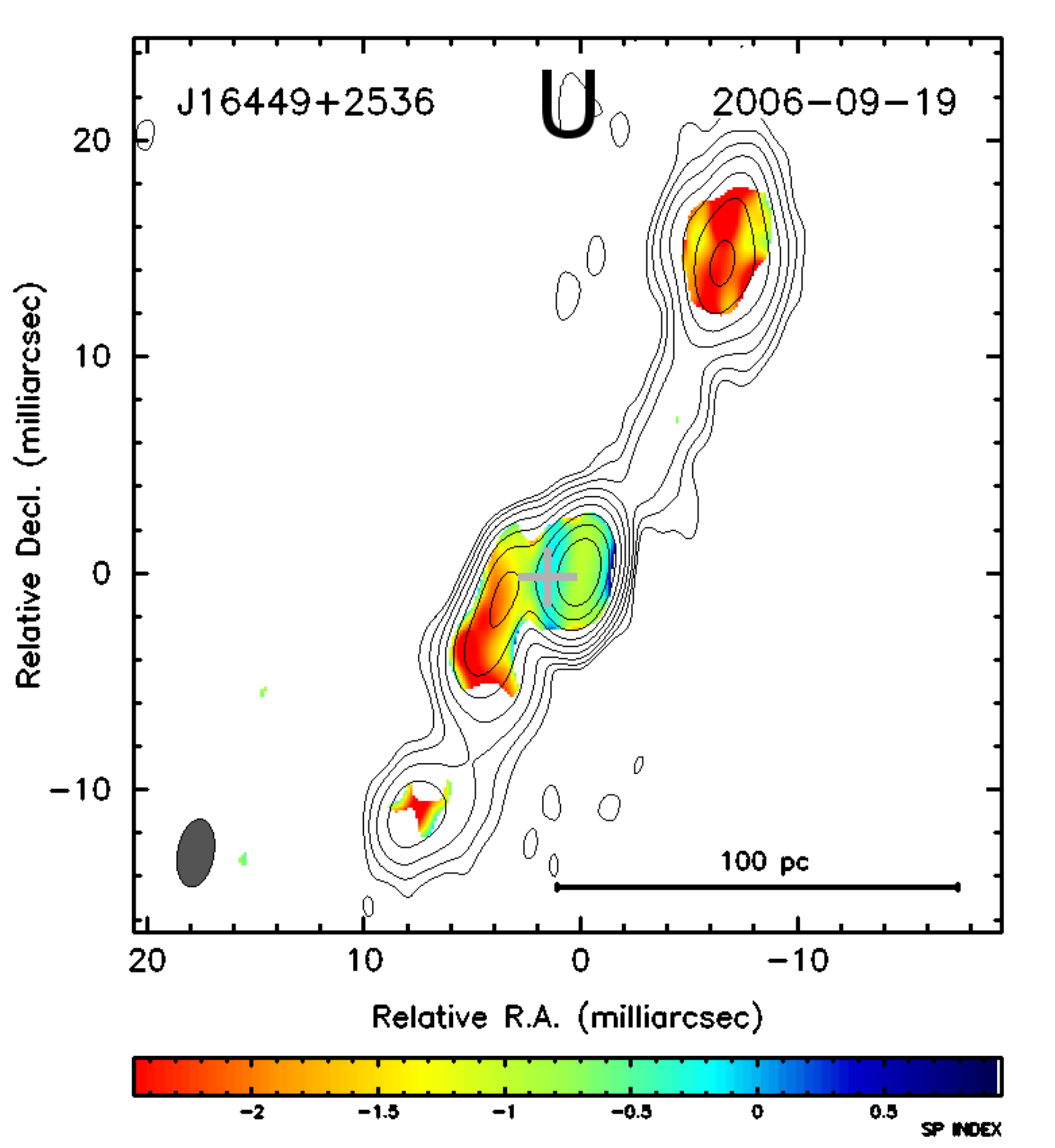}
	}	
	
	\caption{\textit{\textbf{- Continued}} 5 GHz contour maps of VIPS CSOs with 8-15 GHz spectral index map overlays. The contour levels begin at thrice the theoretical noise (typically $\sim$ 0.4 mJy for BT088 and 1.0 mJy for BT094) and increases by powers of 2. The colour scale is fixed from -2.5 to 1 to facilitate comparison. When detected, a grey cross is placed over the core to guide the reader's eye. Sources with confirmed spectral redshifts have associated distance bars for linear scale.}
\label{fig:cso_xuspix}
\end{figure*}

 \begin{figure*}

	\centering
	 {%
	 \includegraphics[width=0.32273\textwidth]{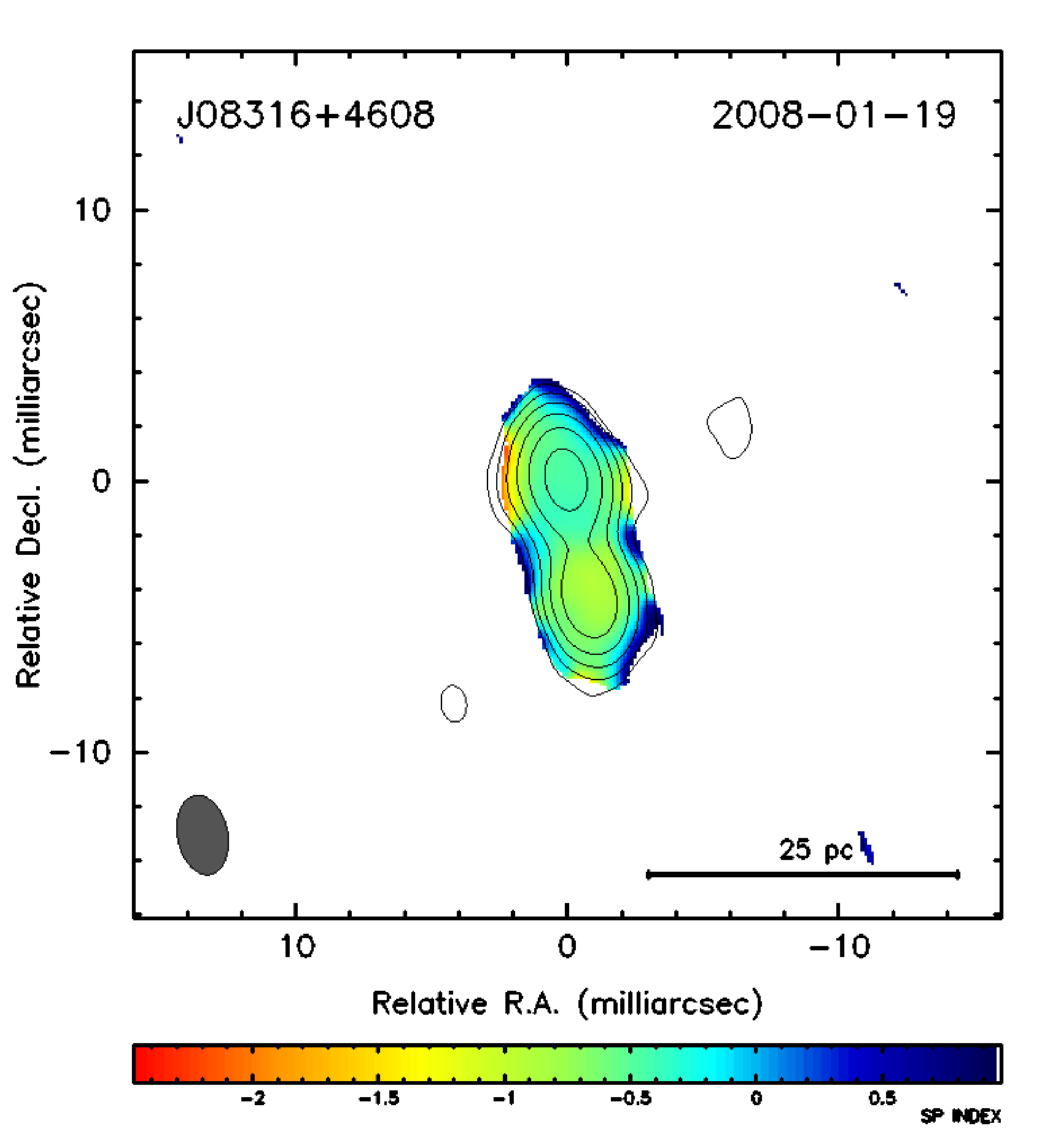}
	}
~\hfill
	\centering
	{%
	\includegraphics[width=0.32273\textwidth]{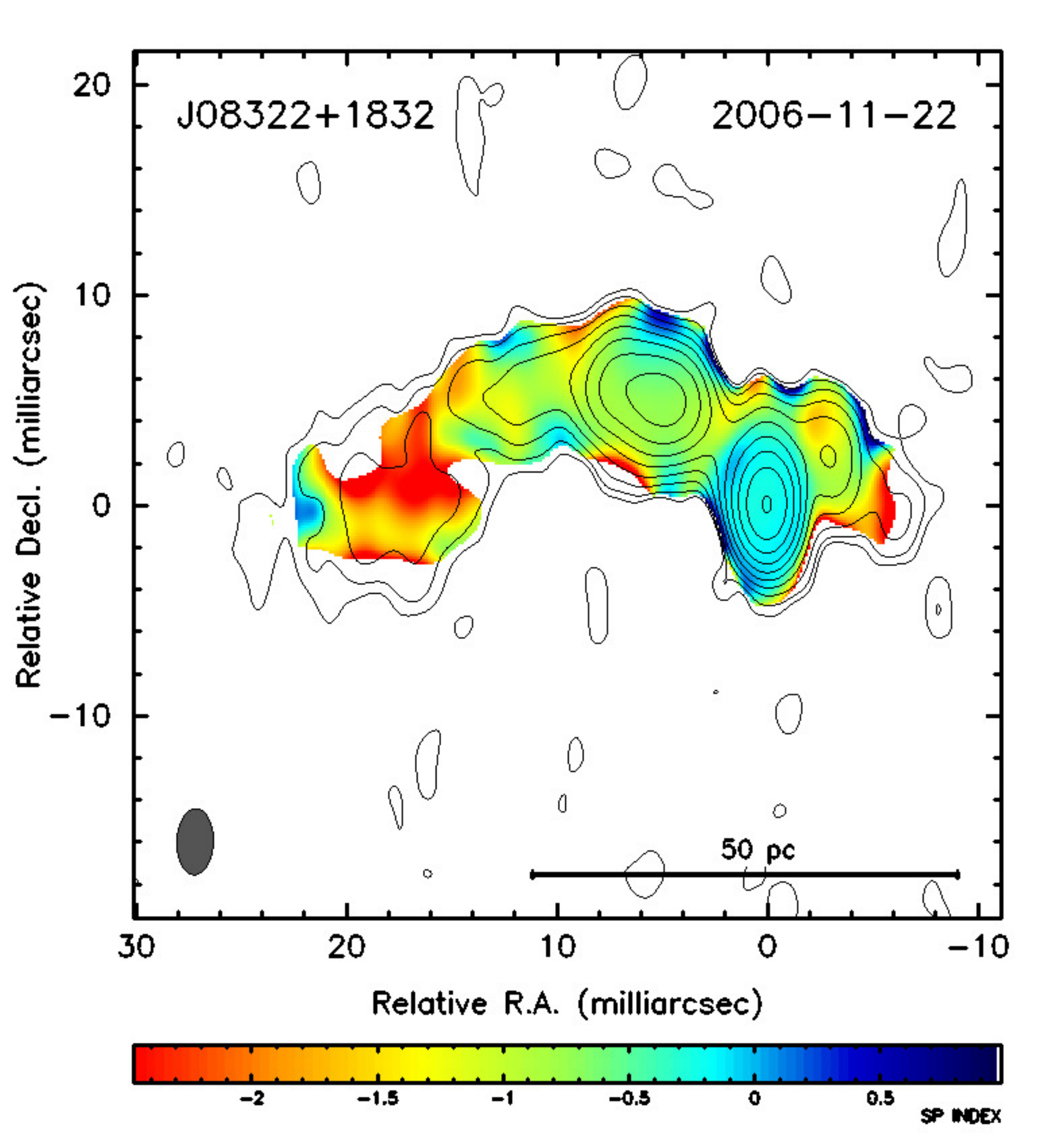}
	}
~\hfill
	\centering
	{%
	\includegraphics[width=0.32273\textwidth]{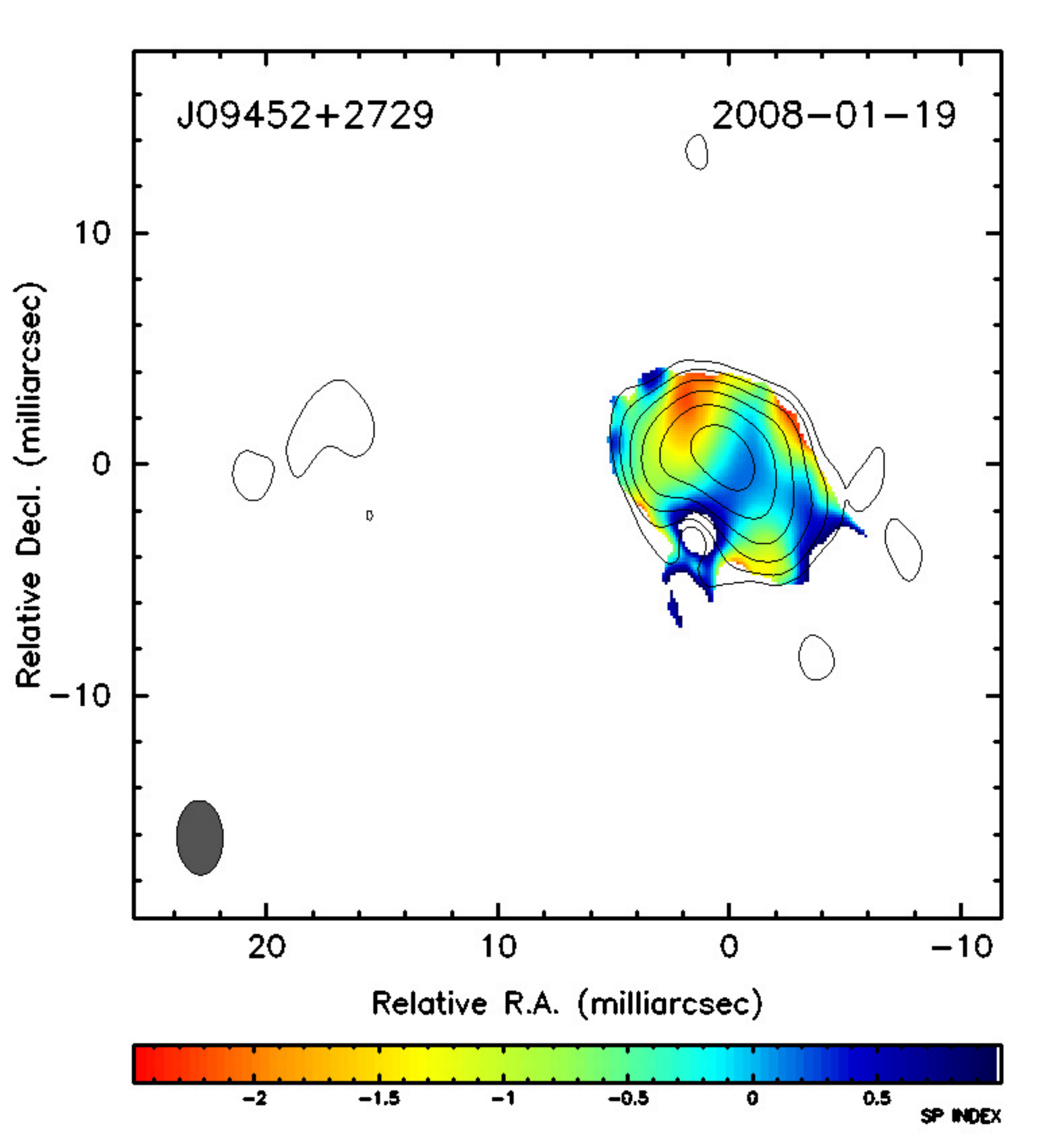}
	}	
~\hfill
	\centering
	{%
	\includegraphics[width=0.32273\textwidth]{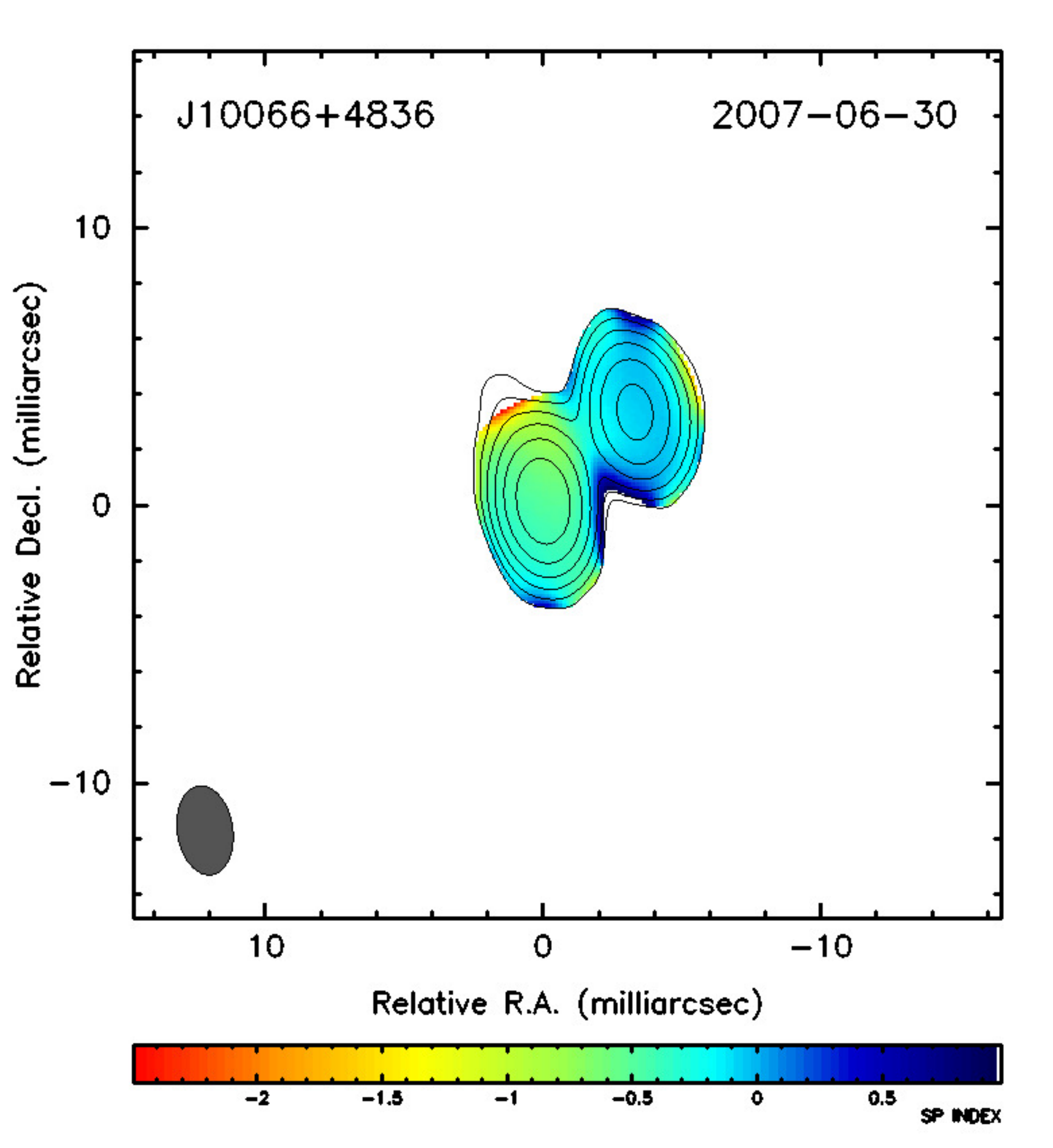}
	}	
~\hfill
	\centering
	{%
	\includegraphics[width=0.32273\textwidth]{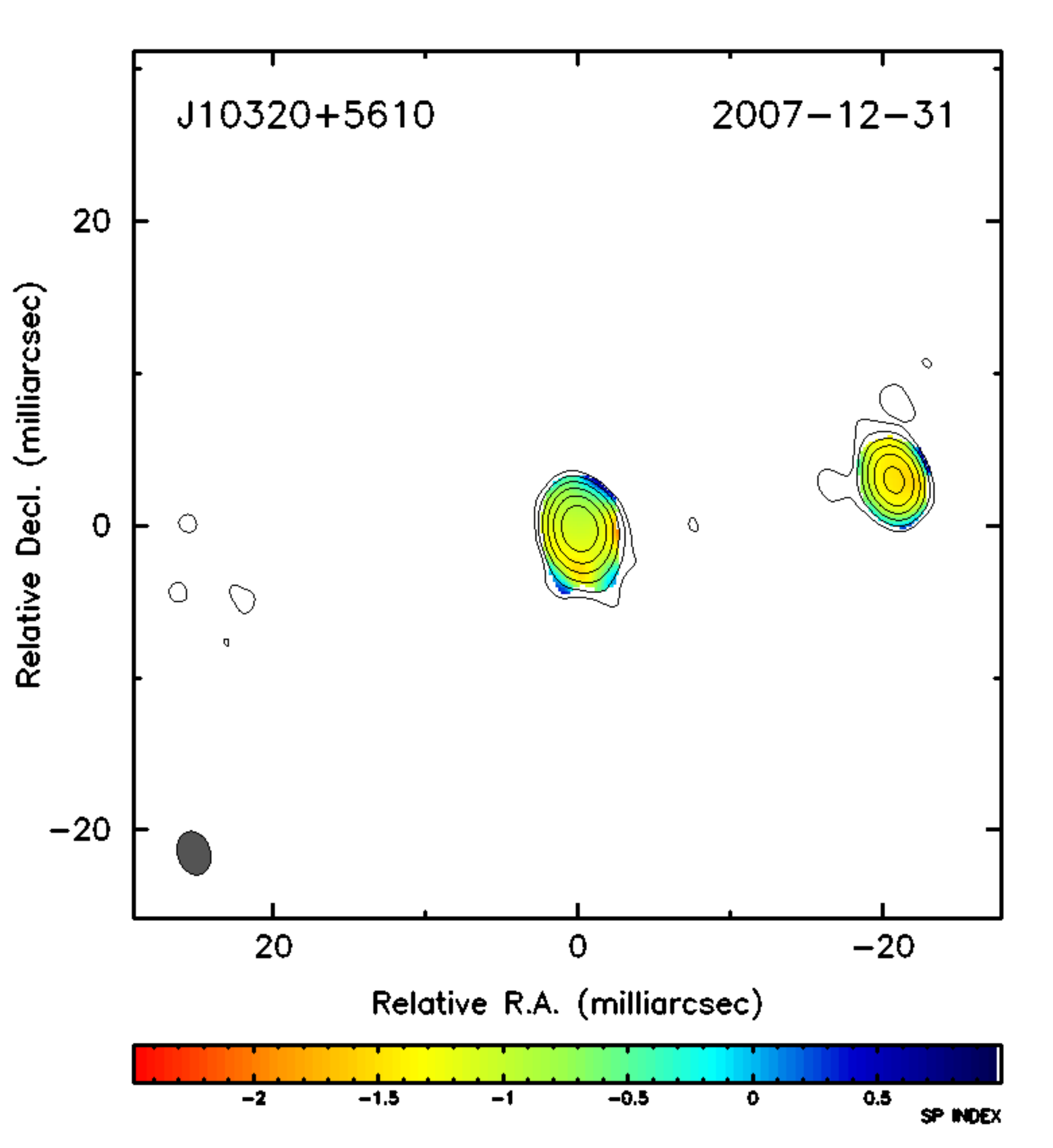}
	}	
~\hfill
	\centering
	{%
	\includegraphics[width=0.32273\textwidth]{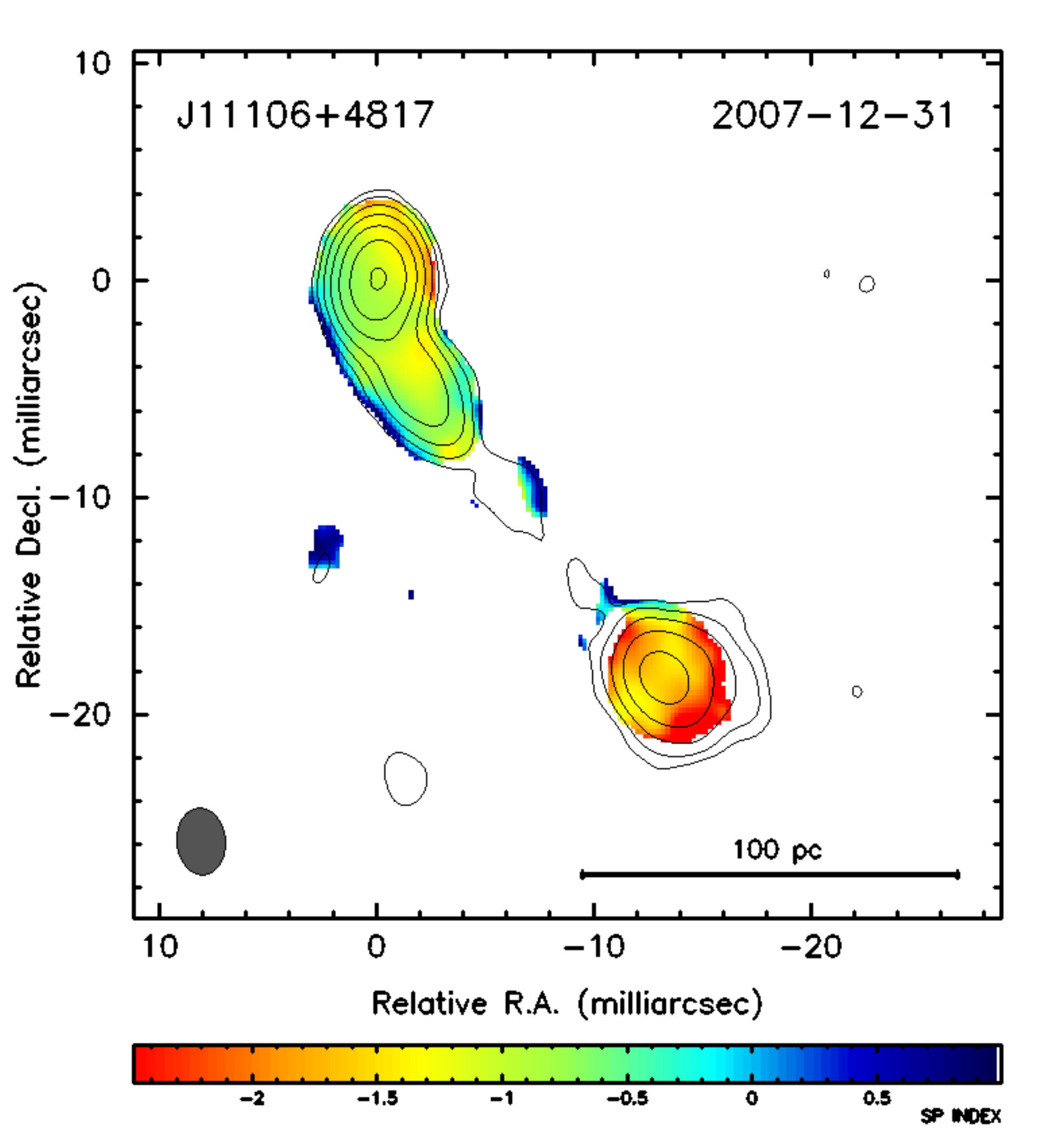}
	}	
~\hfill
	\centering
	{%
	\includegraphics[width=0.32273\textwidth]{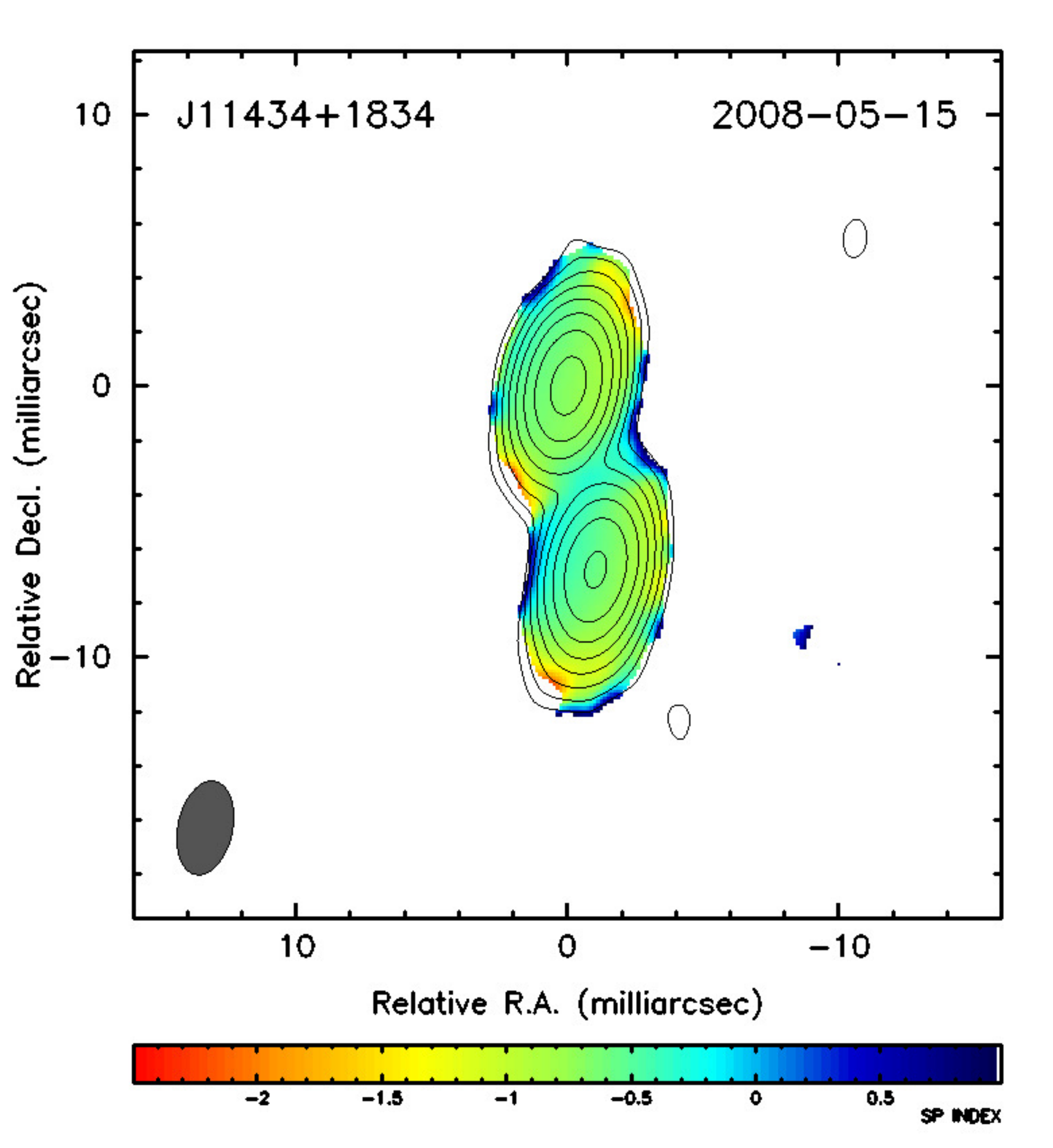}
	}
~\hfill
	\centering
	{%
	\includegraphics[width=0.32273\textwidth]{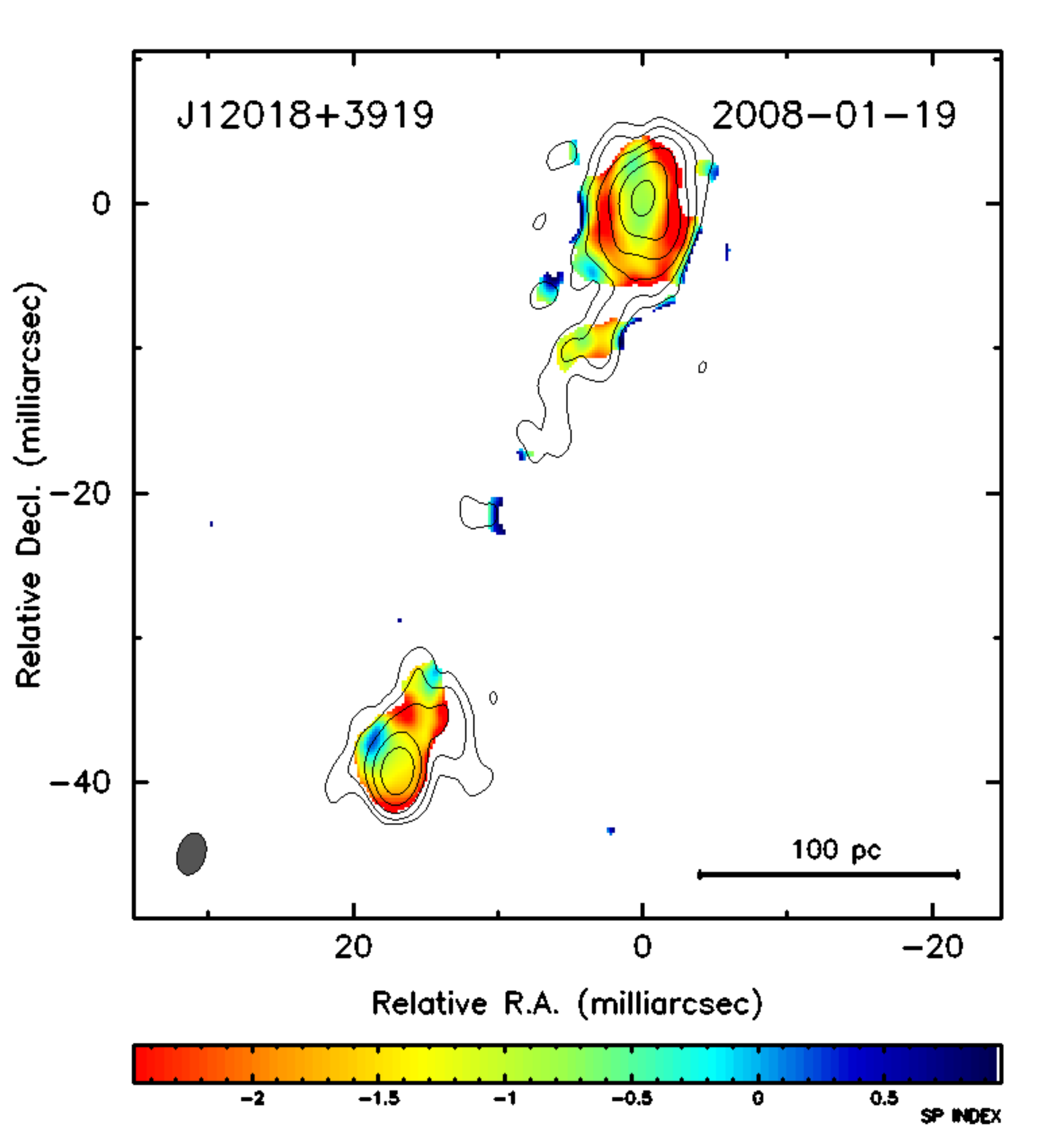}
	}		
~\hfill
	\centering
	{%
	\includegraphics[width=0.32273\textwidth]{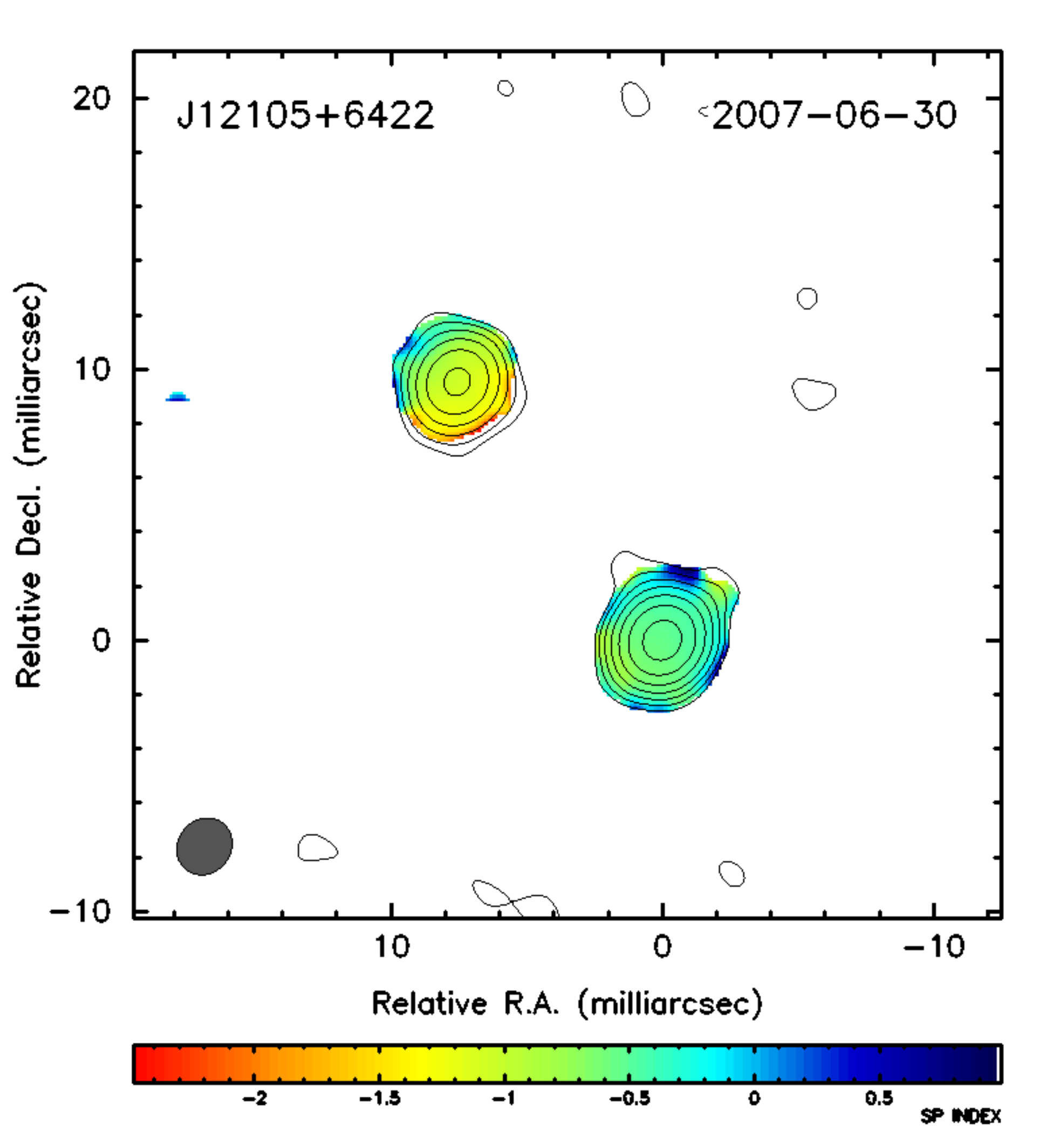}
	}	
	\caption{5 GHz contour maps of the remaining CSO candidates with 5-8 GHz spectral index map overlays. The contour levels begin at thrice the theoretical noise (typically $\sim$ 0.4 mJy for BT088 and 1.0 mJy for BT094) and increases by powers of 2. The colour scale is fixed from -2.5 to 1 to facilitate comparison.  Sources with confirmed spectral redshifts have associated distance bars for linear scale. (A colour version and the complete figure set is available in the online journal.)}
\label{fig:cand_cxspix}
\end{figure*}

\begin{figure*}

	\centering
	 {%
	 \includegraphics[width=0.32273\textwidth]{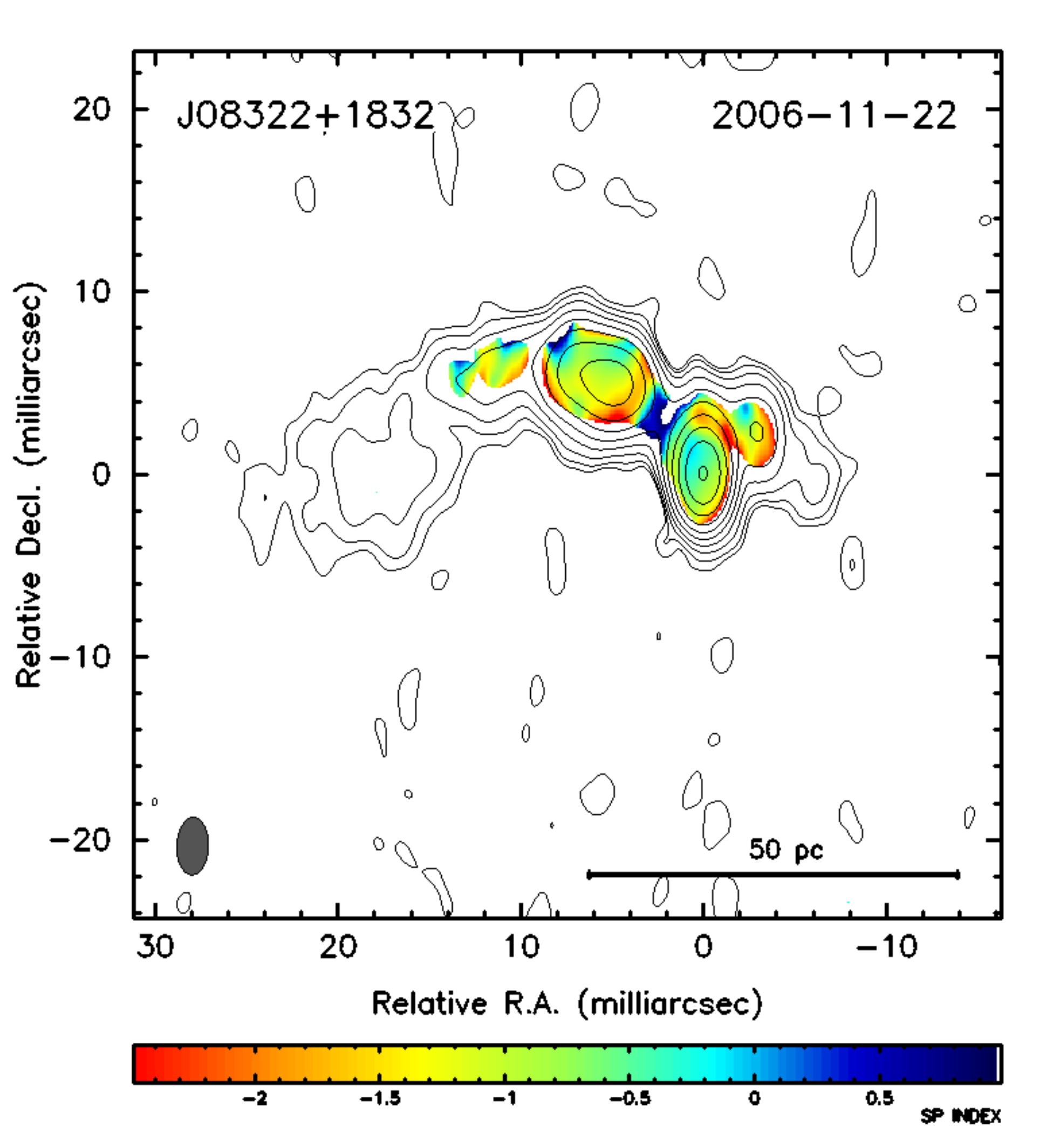}
	}
~\hfill
	\centering
	{%
	\includegraphics[width=0.32273\textwidth]{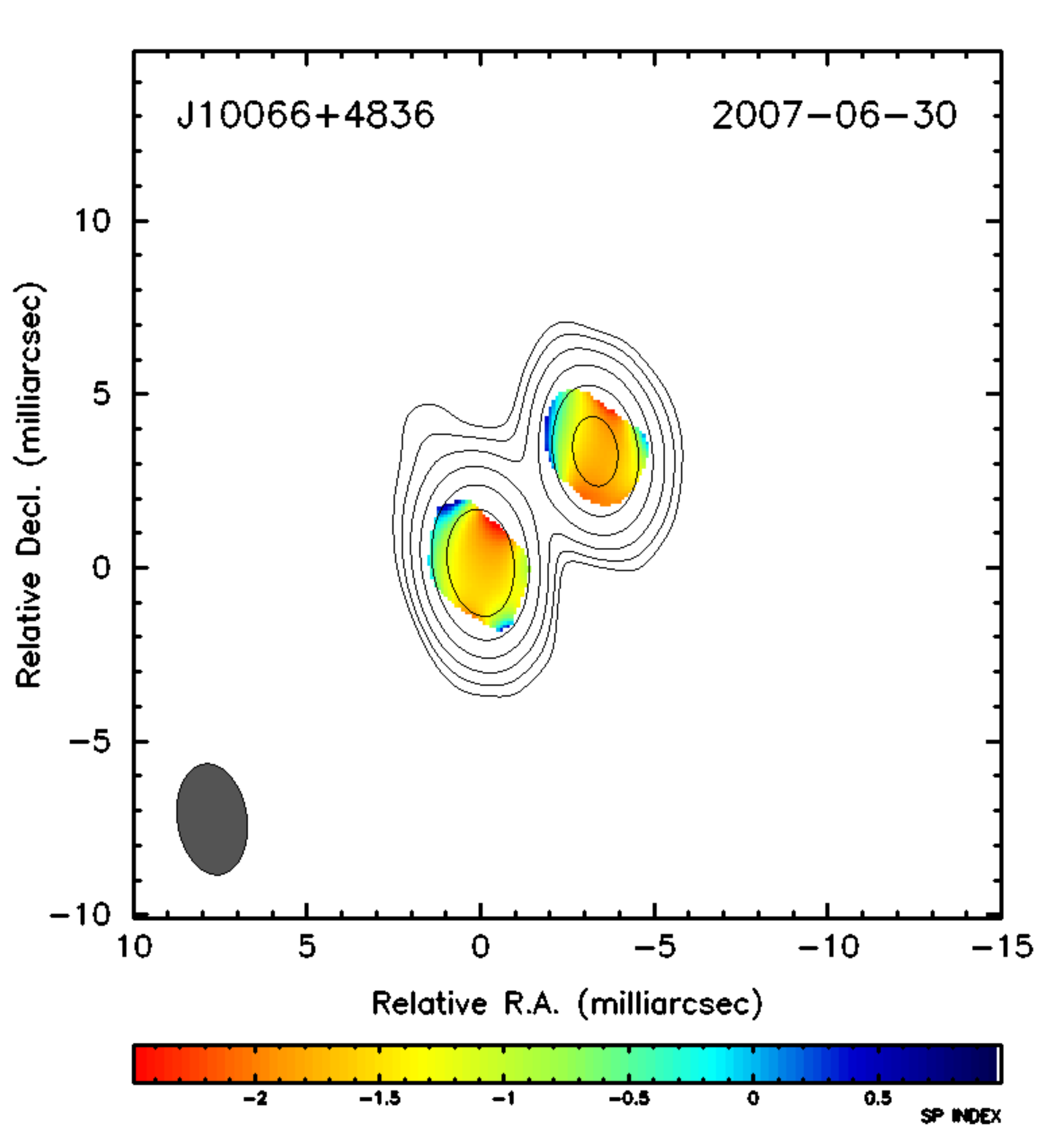}
	}
~\hfill
	\centering
	{%
	\includegraphics[width=0.32273\textwidth]{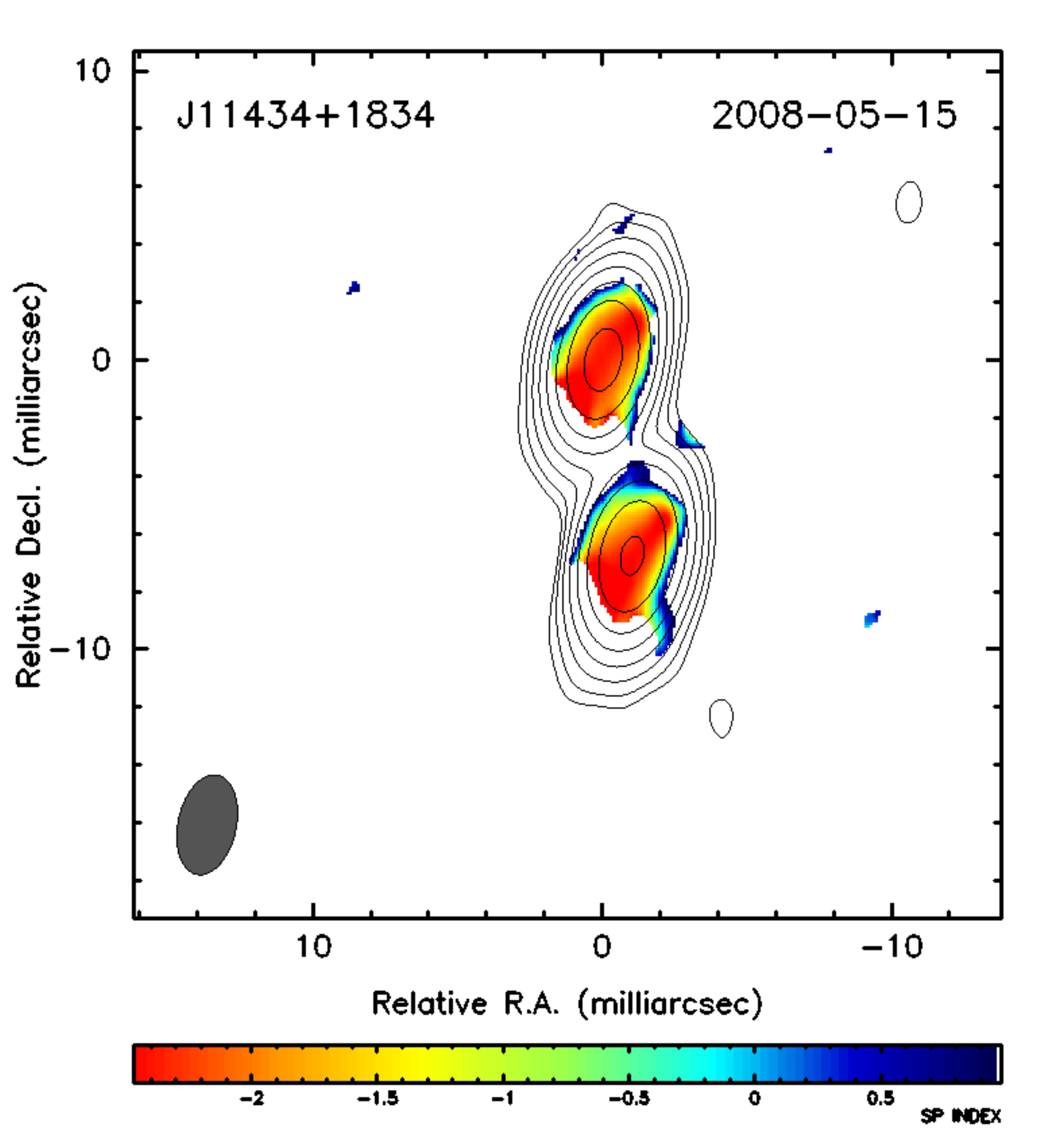}
	}	
~\hfill
	\centering
	{%
	\includegraphics[width=0.32273\textwidth]{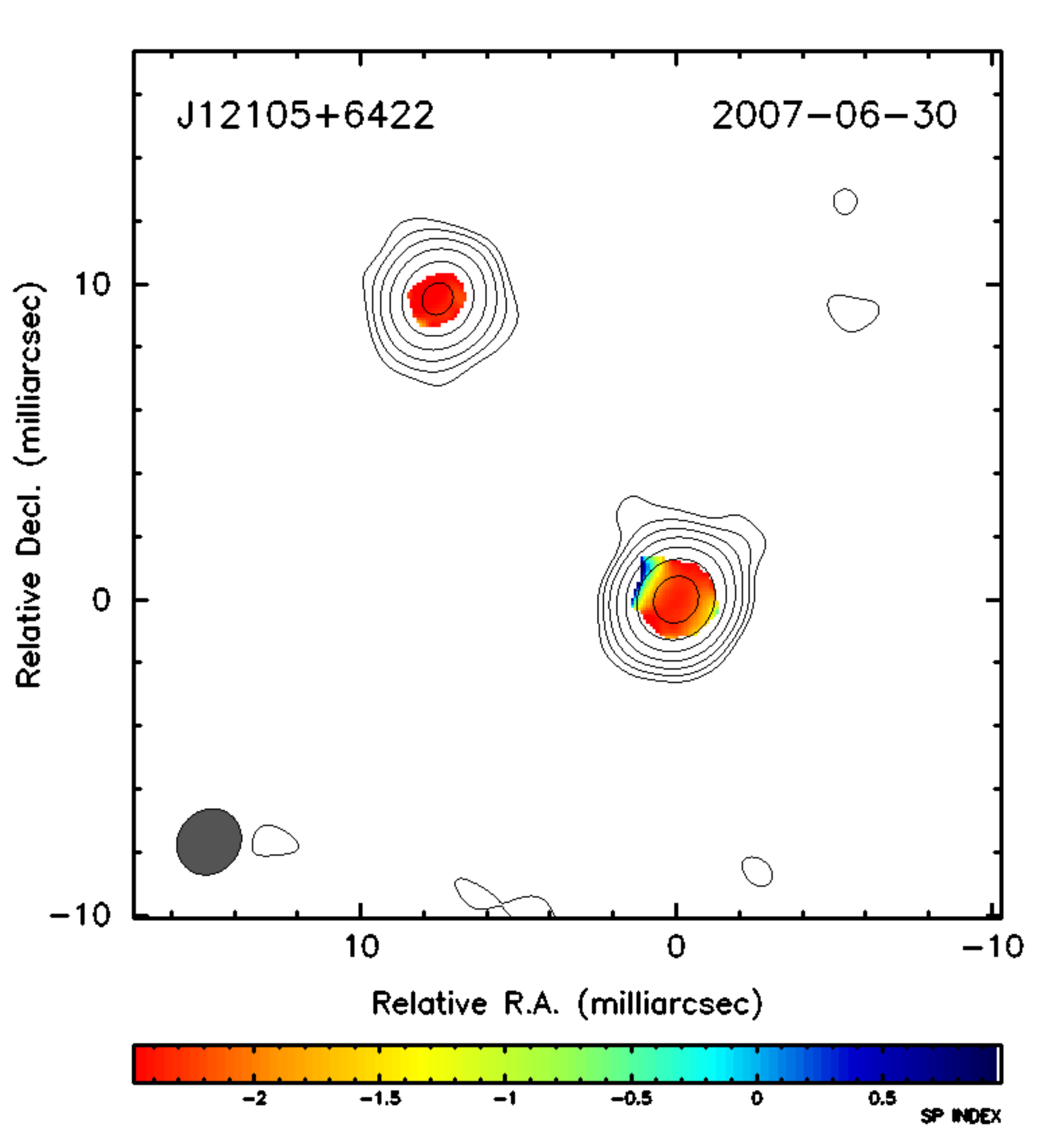}
	}	
~\hfill
	\centering
	{%
	\includegraphics[width=0.32273\textwidth]{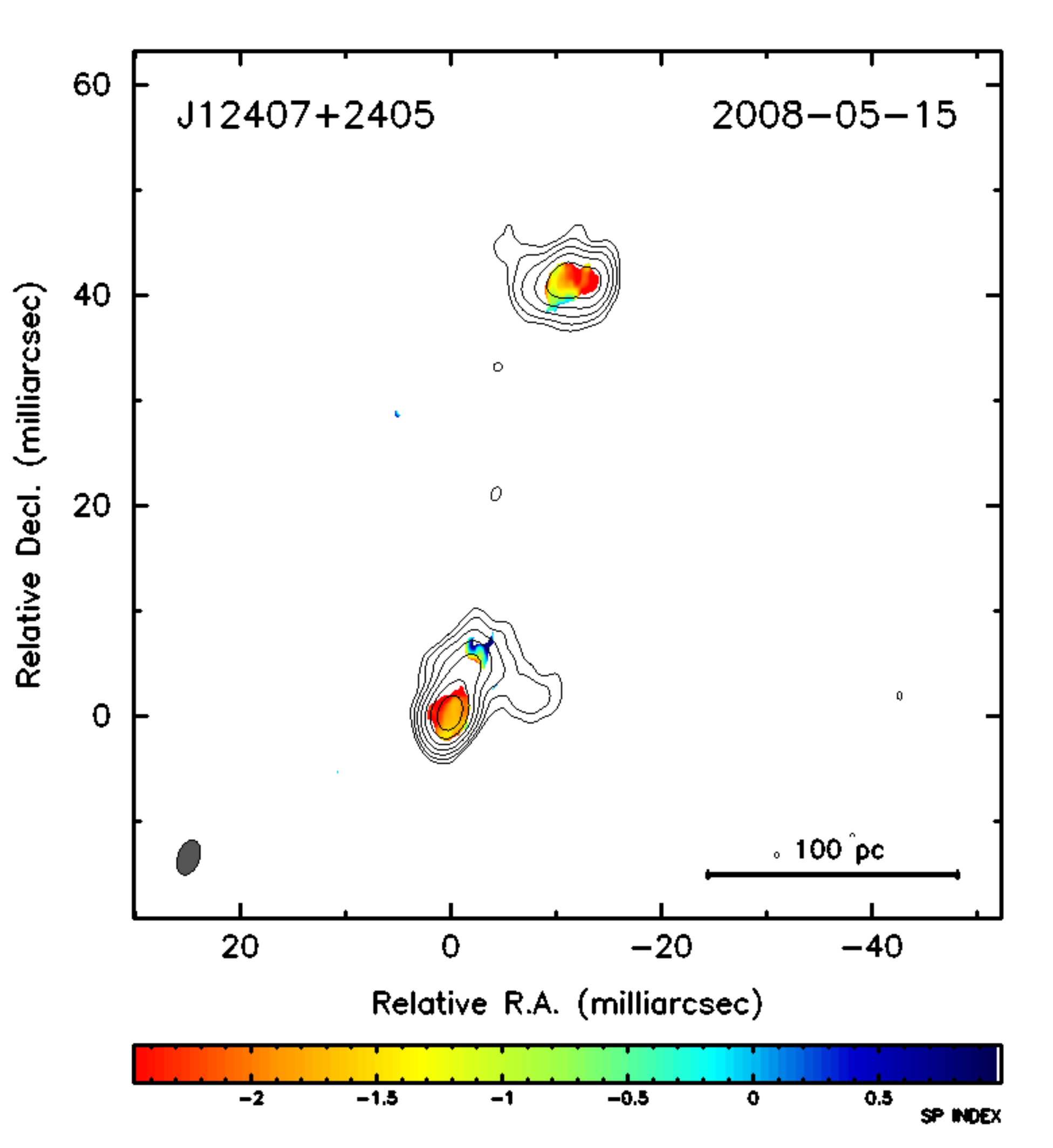}
	}	
~\hfill
	\centering
	{%
	\includegraphics[width=0.32273\textwidth]{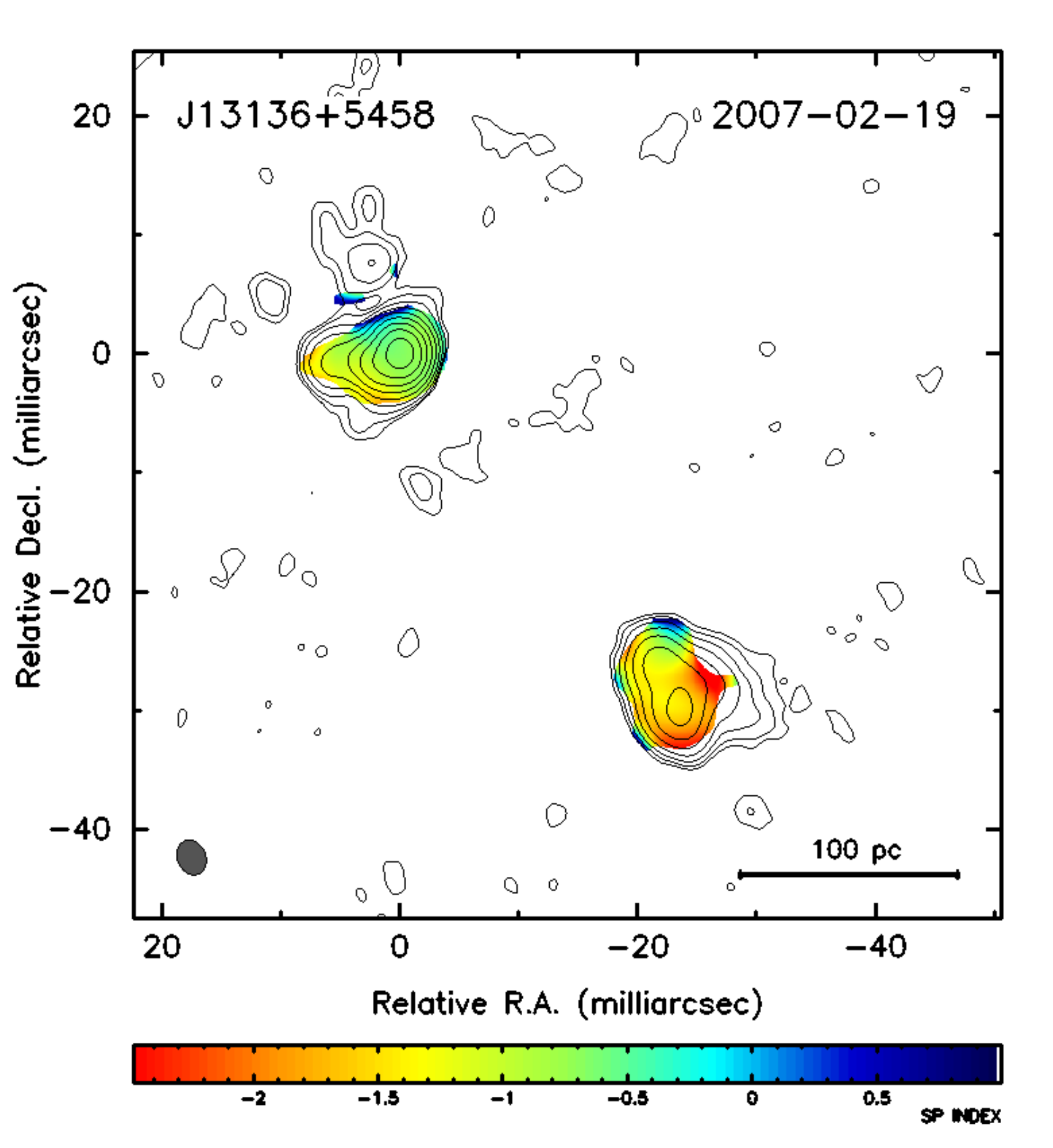}
	}	
~\hfill
	\centering
	{%
	\includegraphics[width=0.32273\textwidth]{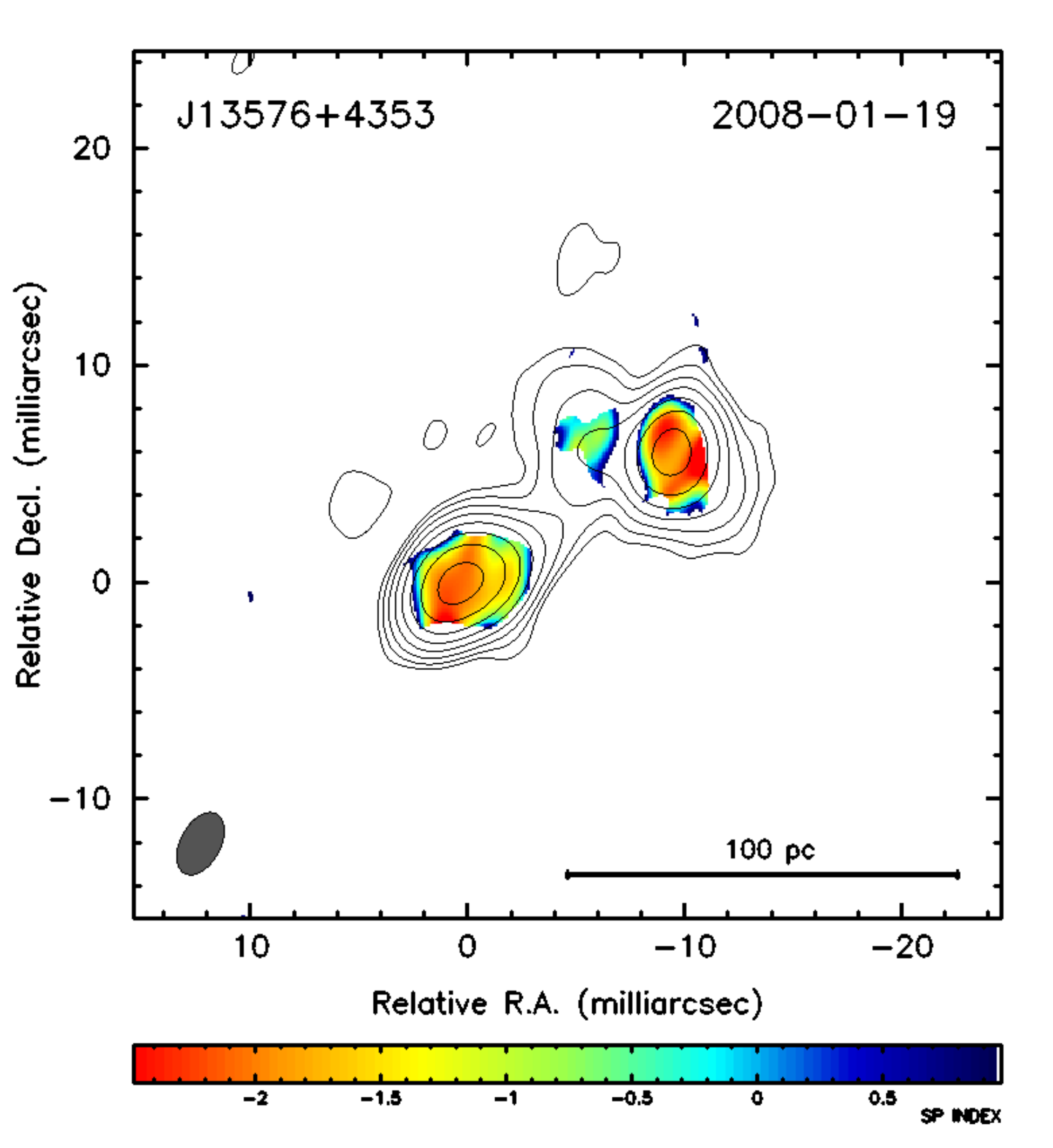}
	}
~\hfill
	\centering
	{%
	\includegraphics[width=0.32273\textwidth]{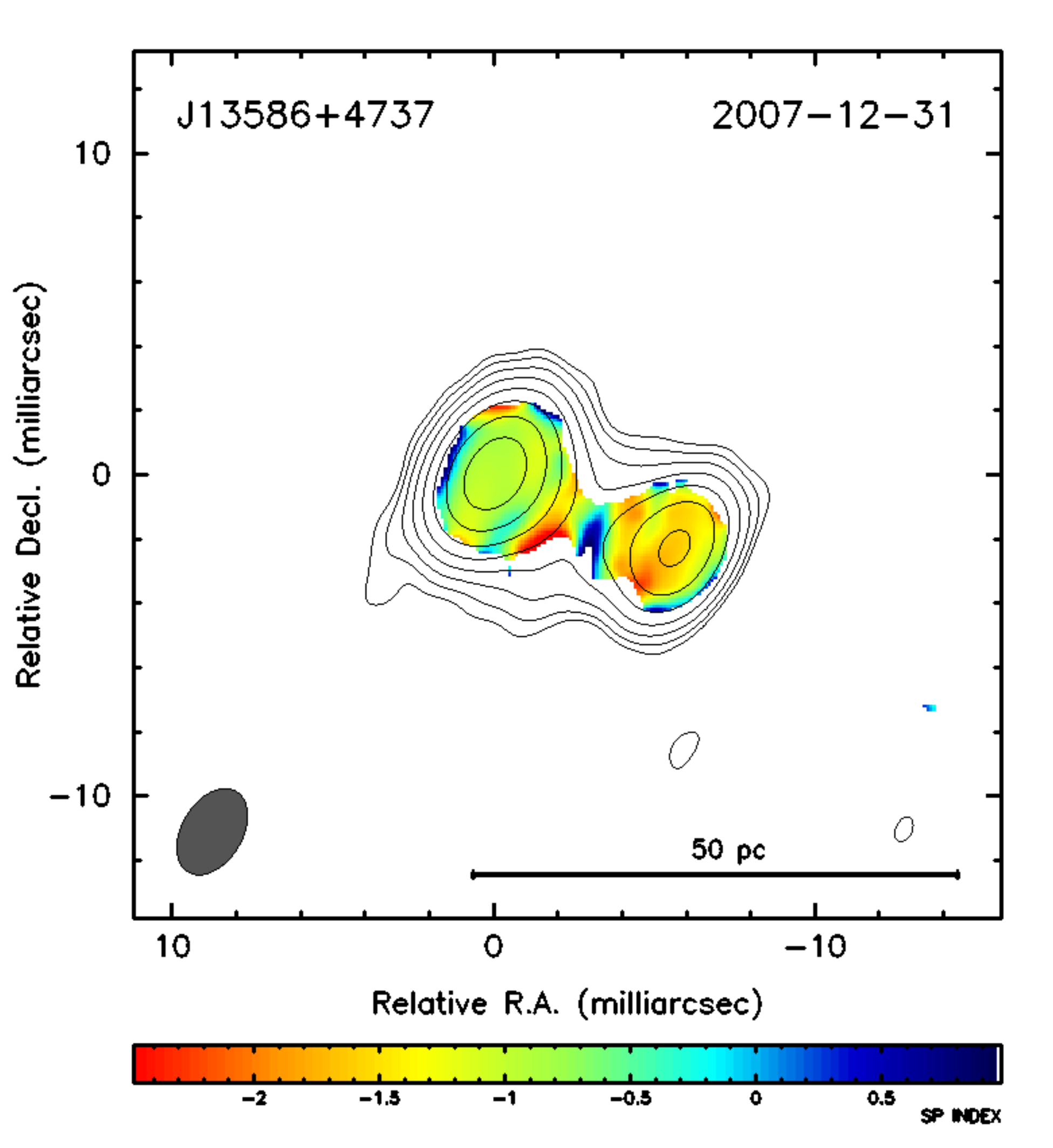}
	}		
~\hfill
	\centering
	{%
	\includegraphics[width=0.32273\textwidth]{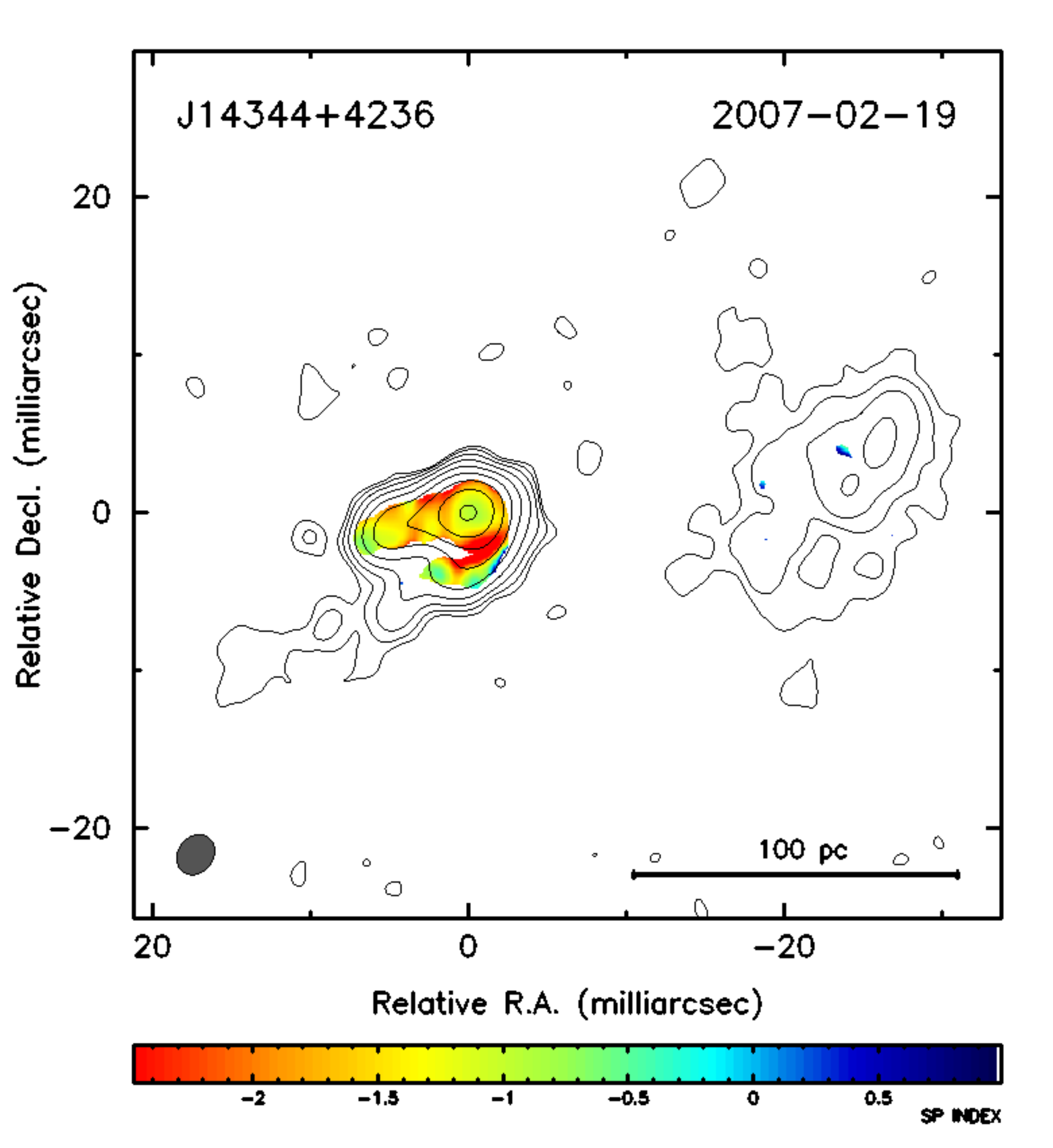}
	}	
	\caption{5 GHz contour maps of the remaining CSO candidates with 8-15 GHz spectral index map overlays. The contour levels begin at thrice the theoretical noise (typically $\sim$ 0.4 mJy for BT088 and 1.0 mJy for BT094) and increases by powers of 2. The colour scale is fixed from -2.5 to 1 to facilitate comparison.  Sources with confirmed spectral redshifts have associated distance bars for linear scale. (A colour version and the complete figure set is available in the online journal.)}
\label{fig:cand_xuspix}
\end{figure*}

\begin{figure*}

	\centering
	 {%
	 \includegraphics[width=0.32273\textwidth]{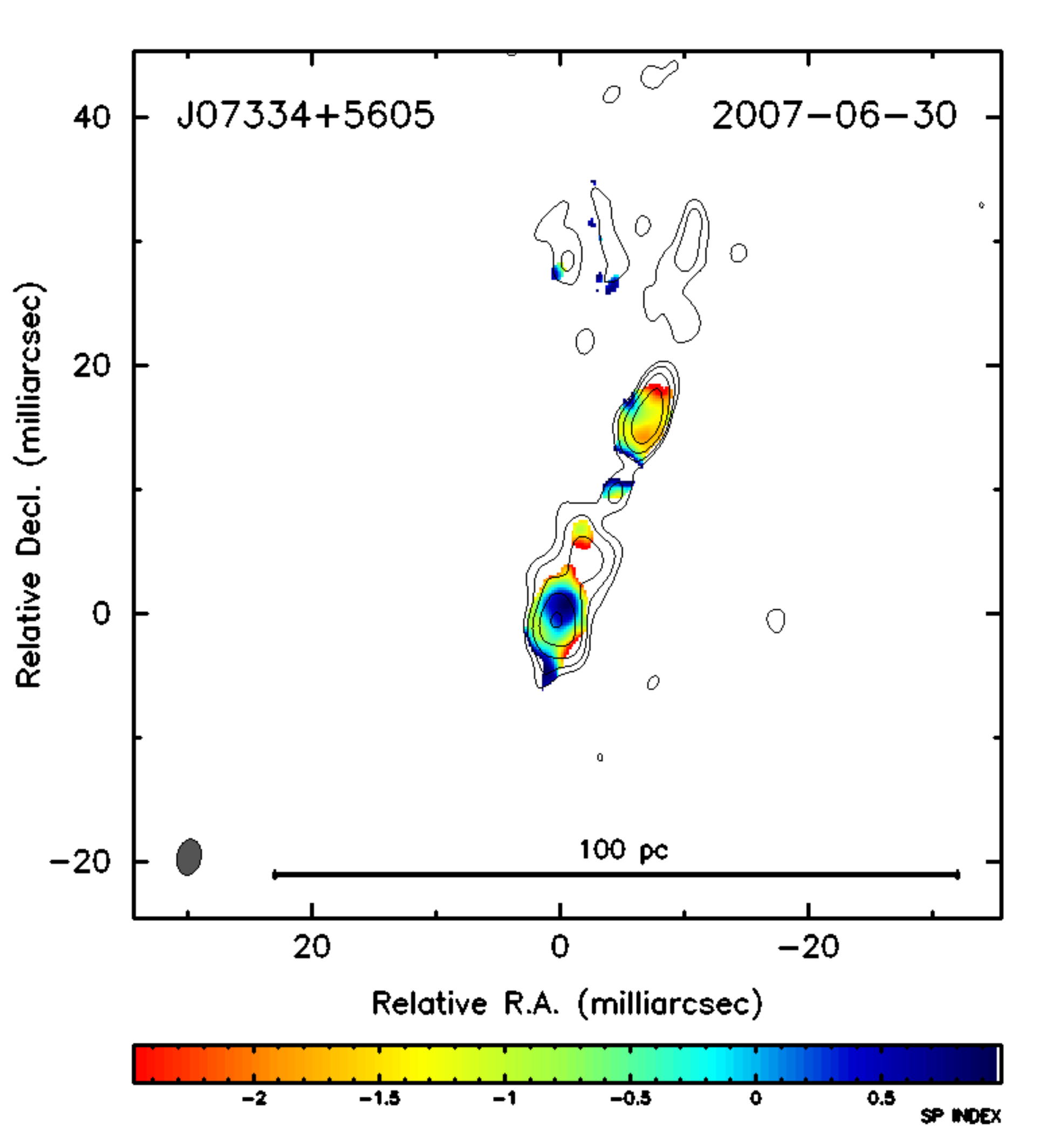}
	}
~\hfill
	\centering
	{%
	\includegraphics[width=0.32273\textwidth]{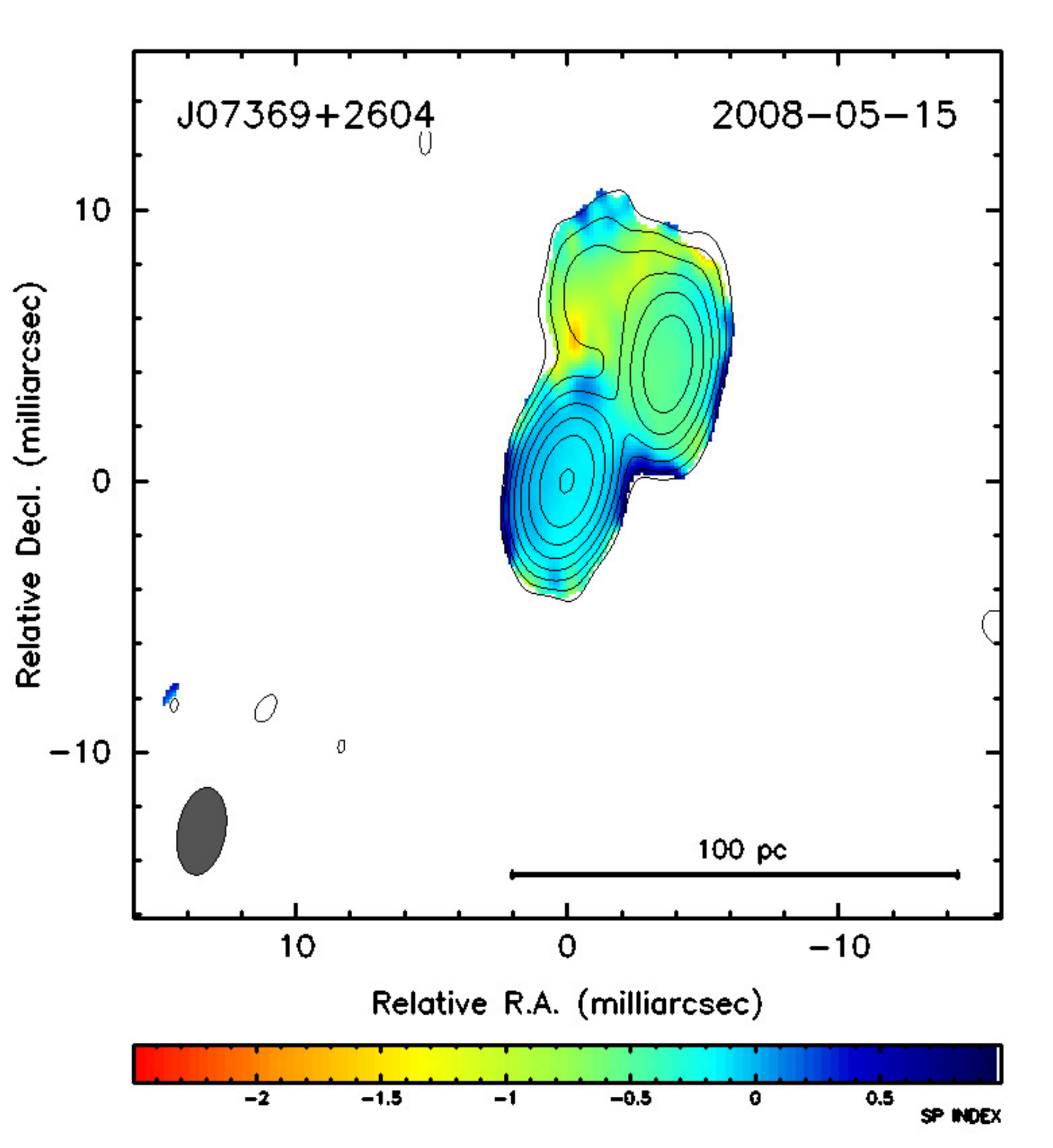}
	}
~\hfill
	\centering
	{%
	\includegraphics[width=0.32273\textwidth]{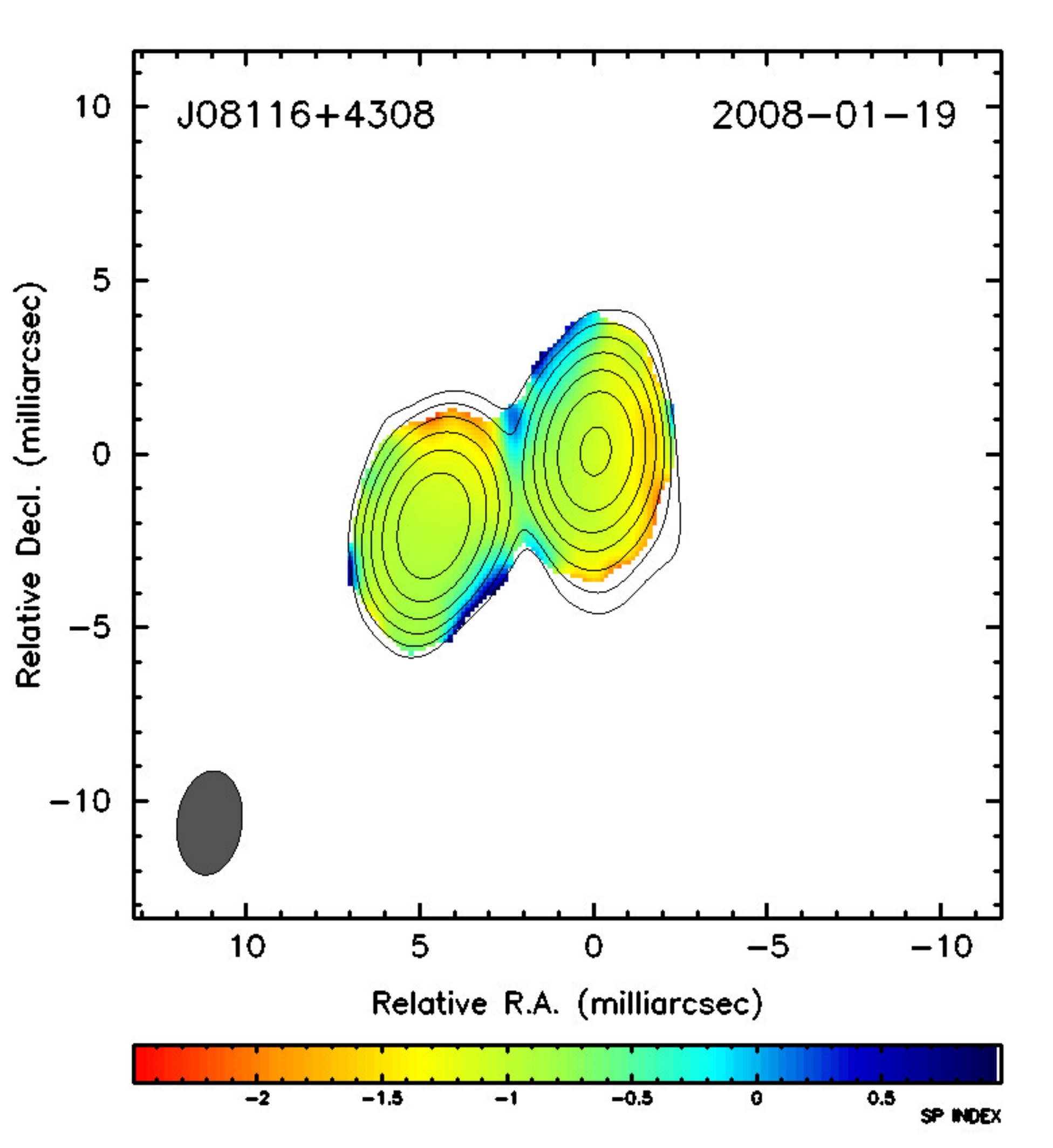}
	}	
~\hfill
	\centering
	{%
	\includegraphics[width=0.32273\textwidth]{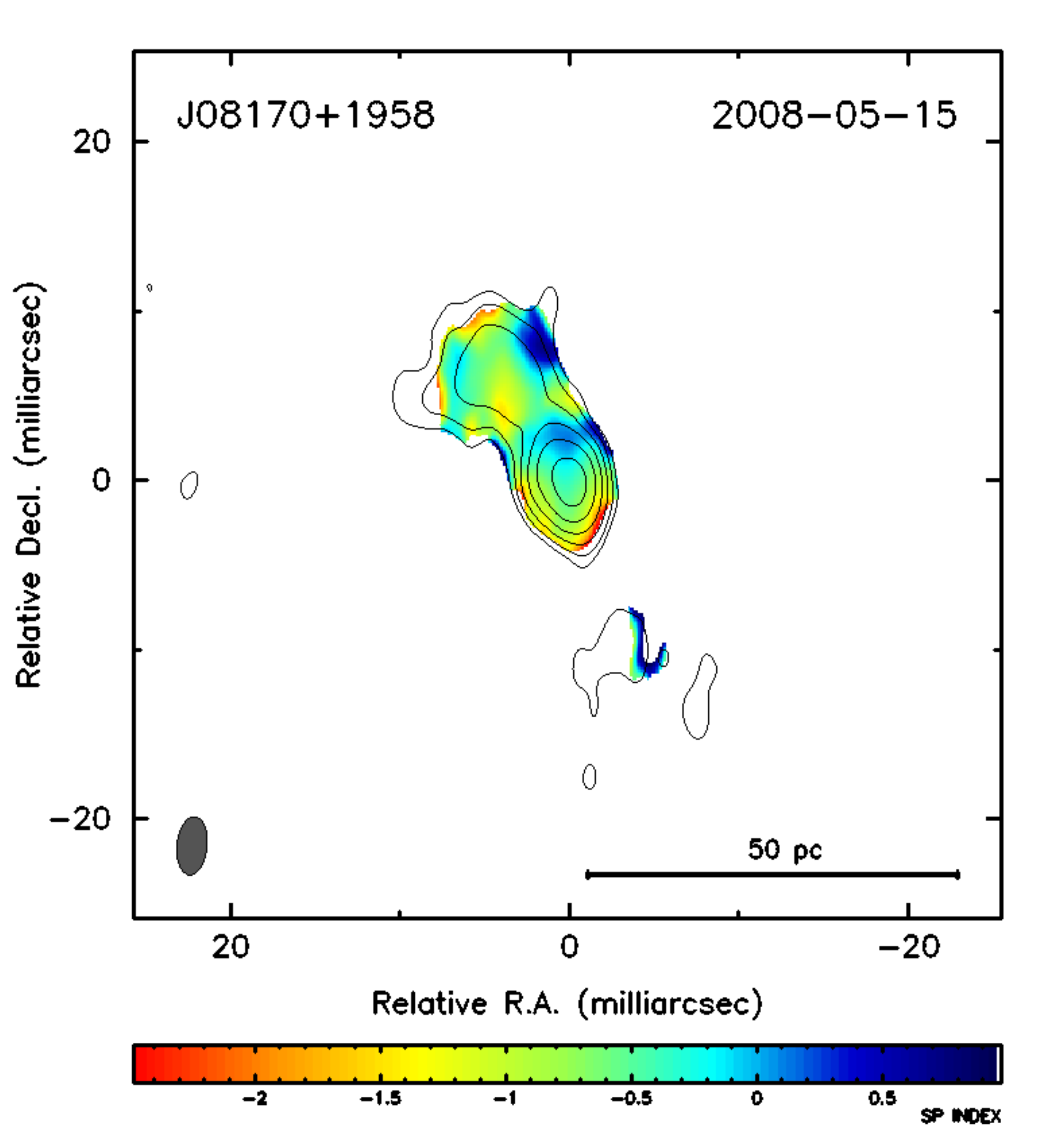}
	}	
~\hfill
	\centering
	{%
	\includegraphics[width=0.32273\textwidth]{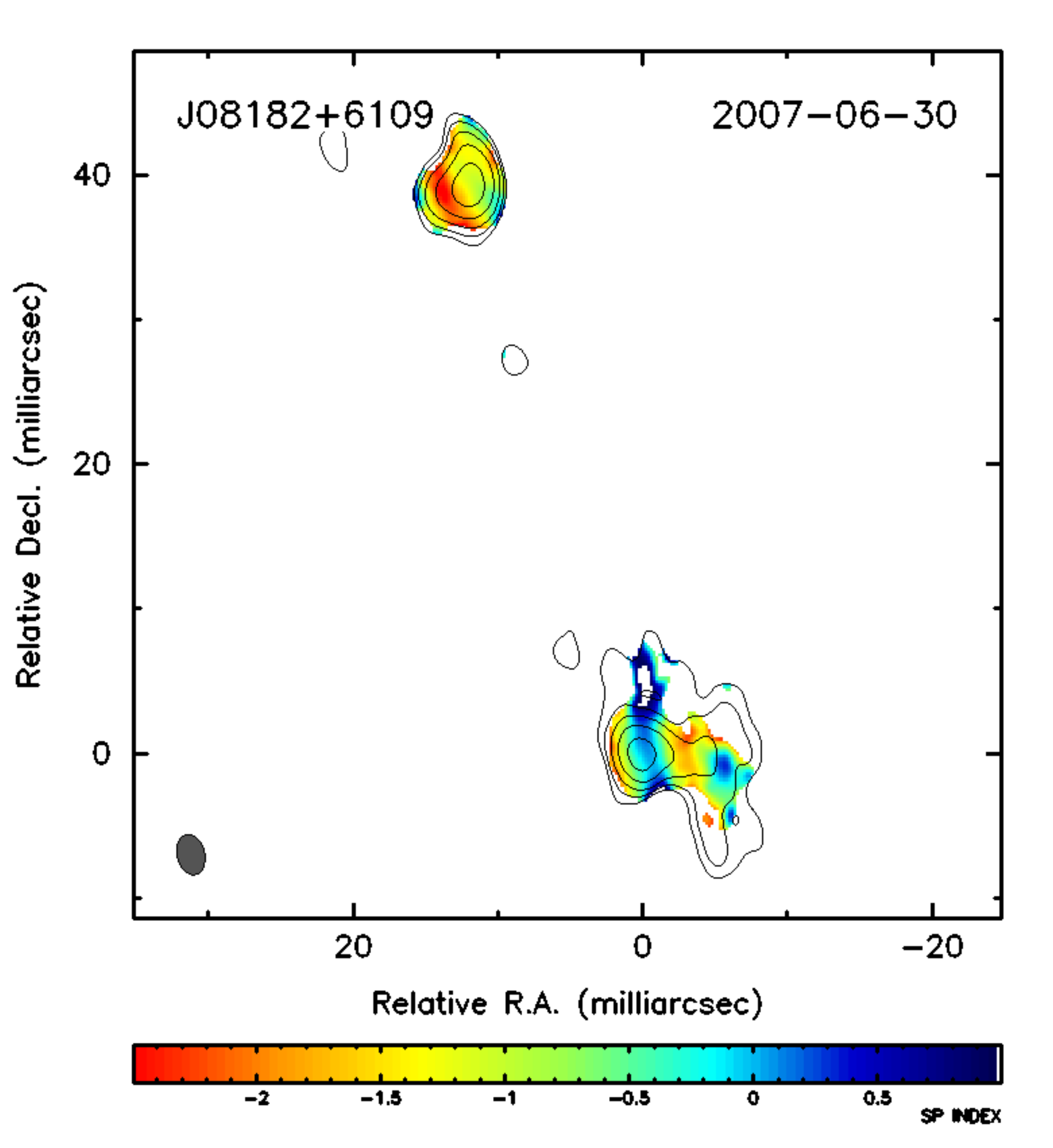}
	}	
~\hfill
	\centering
	{%
	\includegraphics[width=0.32273\textwidth]{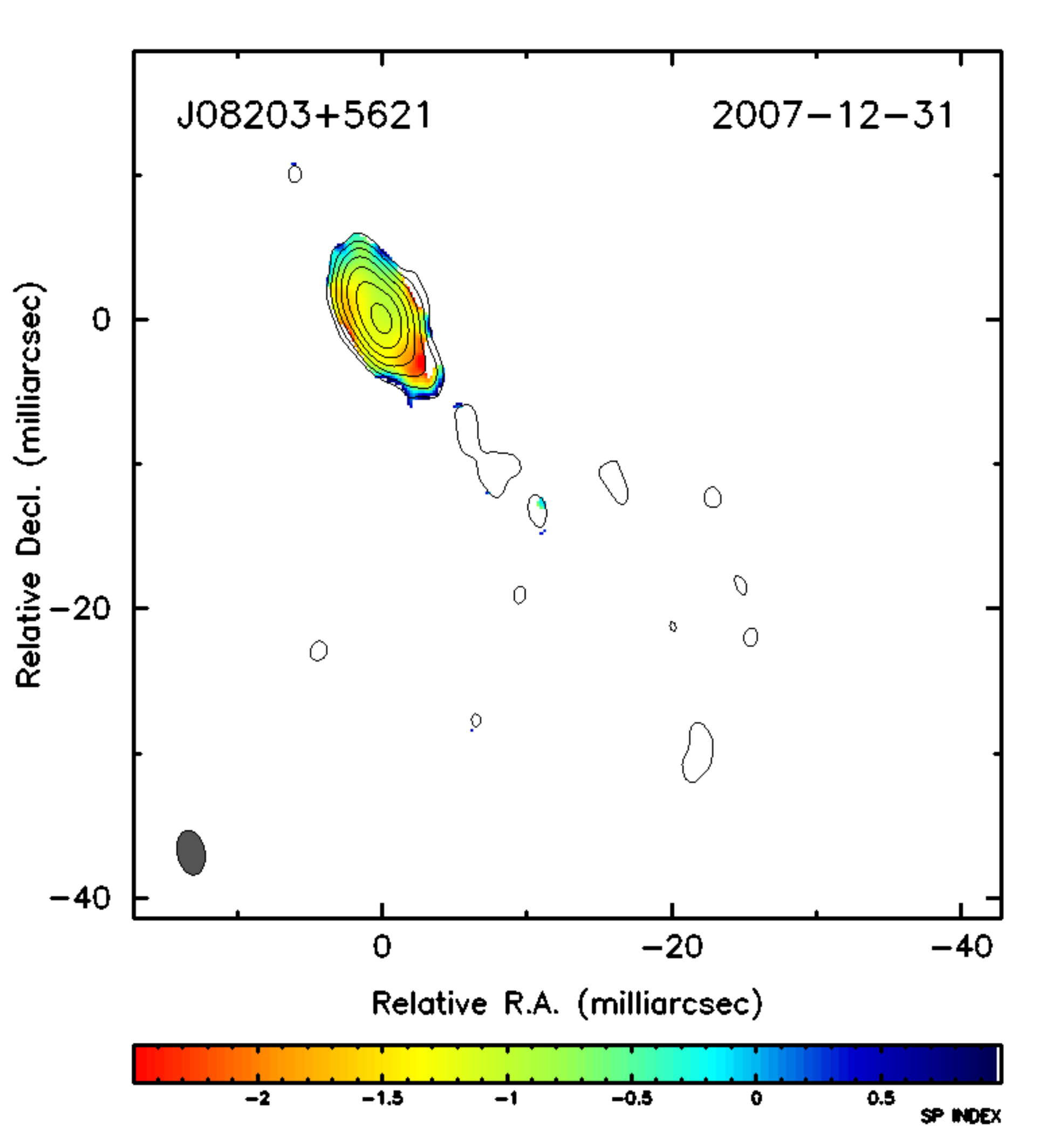}
	}	
~\hfill
	\centering
	{%
	\includegraphics[width=0.32273\textwidth]{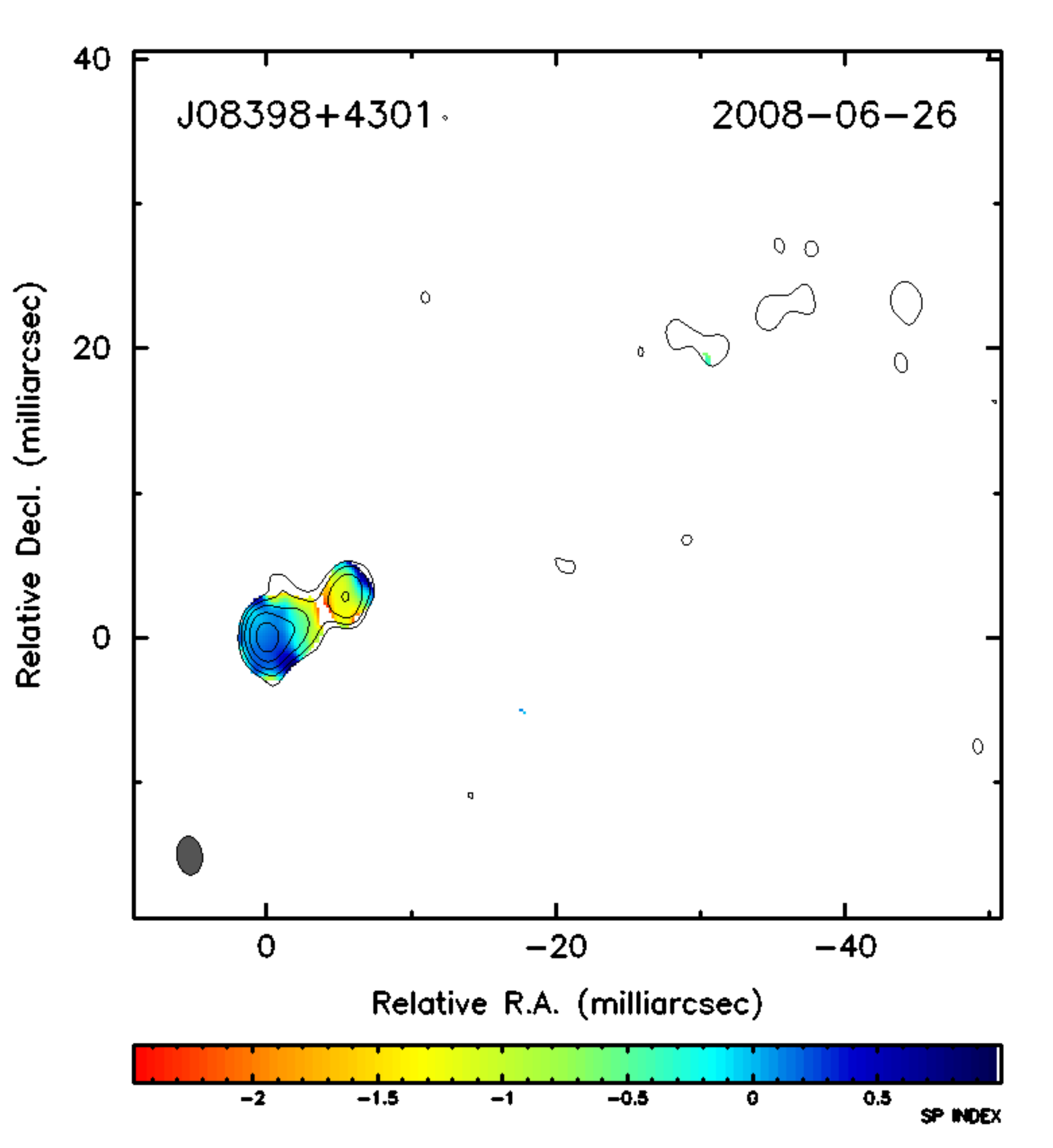}
	}
~\hfill
	\centering
	{%
	\includegraphics[width=0.32273\textwidth]{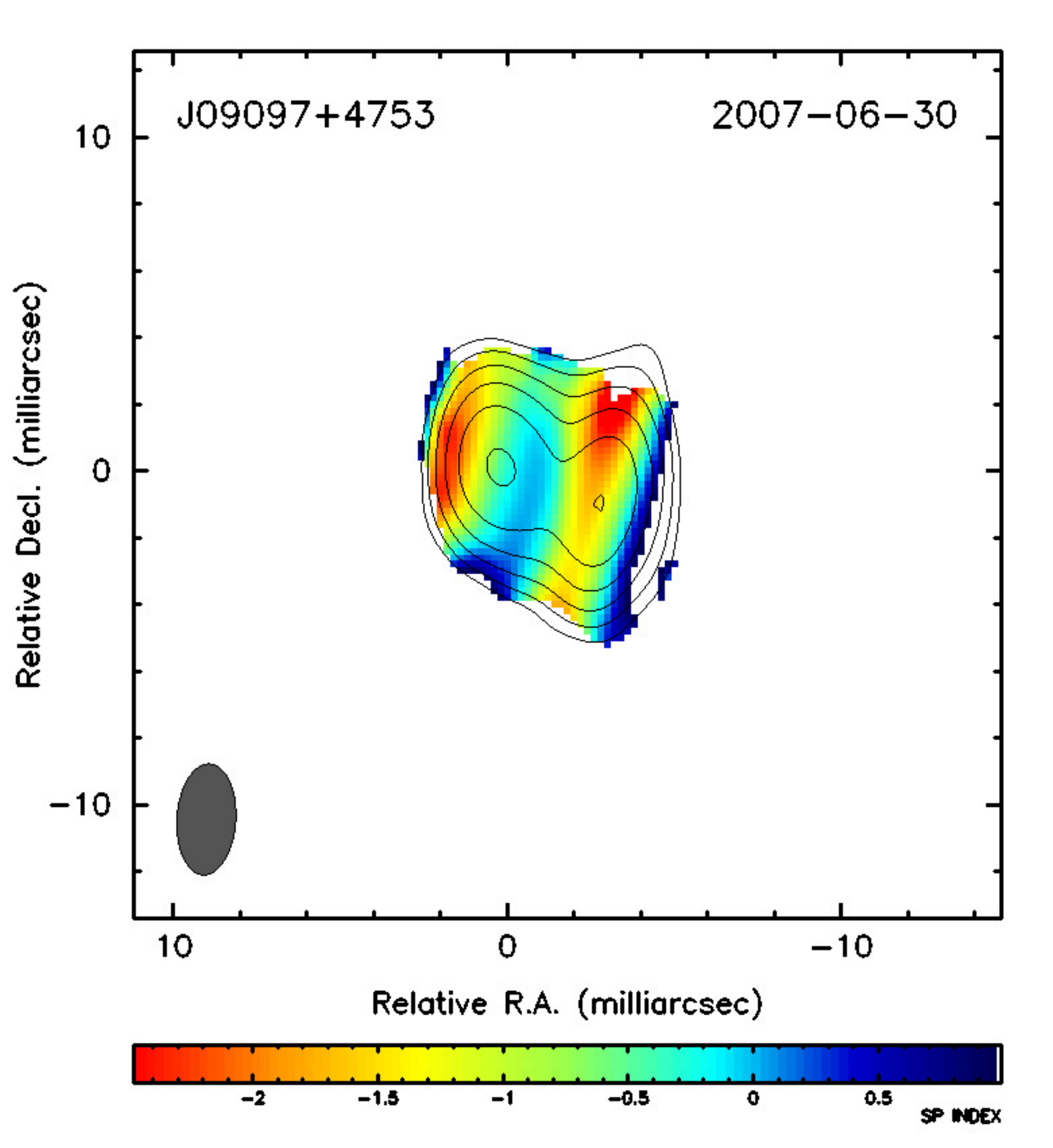}
	}		
~\hfill
	\centering
	{%
	\includegraphics[width=0.32273\textwidth]{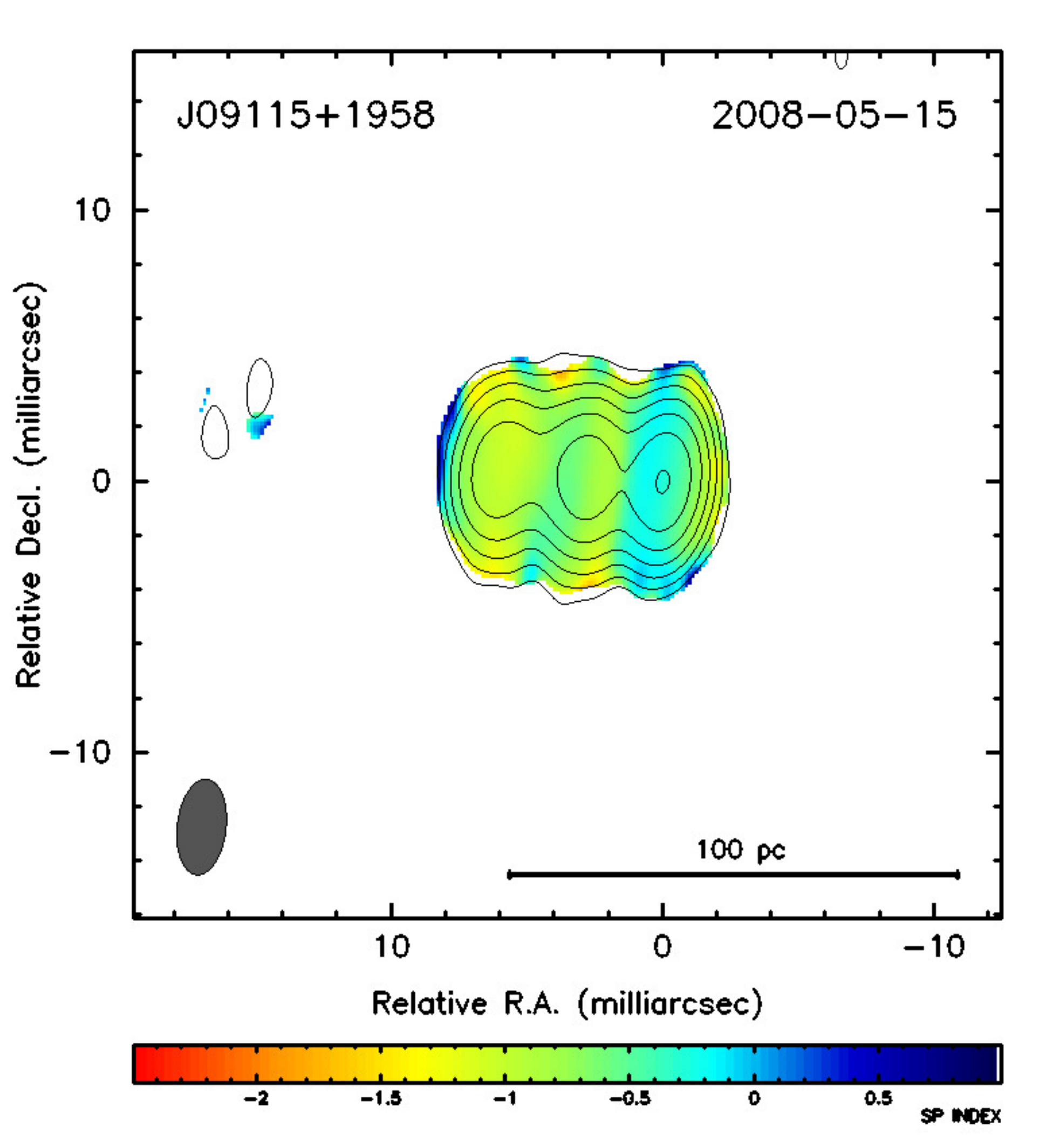}
	}	
	\caption{5 GHz contour maps of refuted CSOs with 5-8 GHz spectral index map overlays. The contour levels begin at thrice the theoretical noise (typically $\sim$ 0.4 mJy for BT088 and 1.0 mJy for BT094) and increases by powers of 2. The colour scale is fixed from -2.5 to 1 to facilitate comparison.  Sources with confirmed spectral redshifts have associated distance bars for linear scale. (A colour version and the complete figure set is available in the online journal.)}
\label{fig:refuted_cxspix}
\end{figure*}

 \begin{figure*}

	\centering
	 {%
	 \includegraphics[width=0.32273\textwidth]{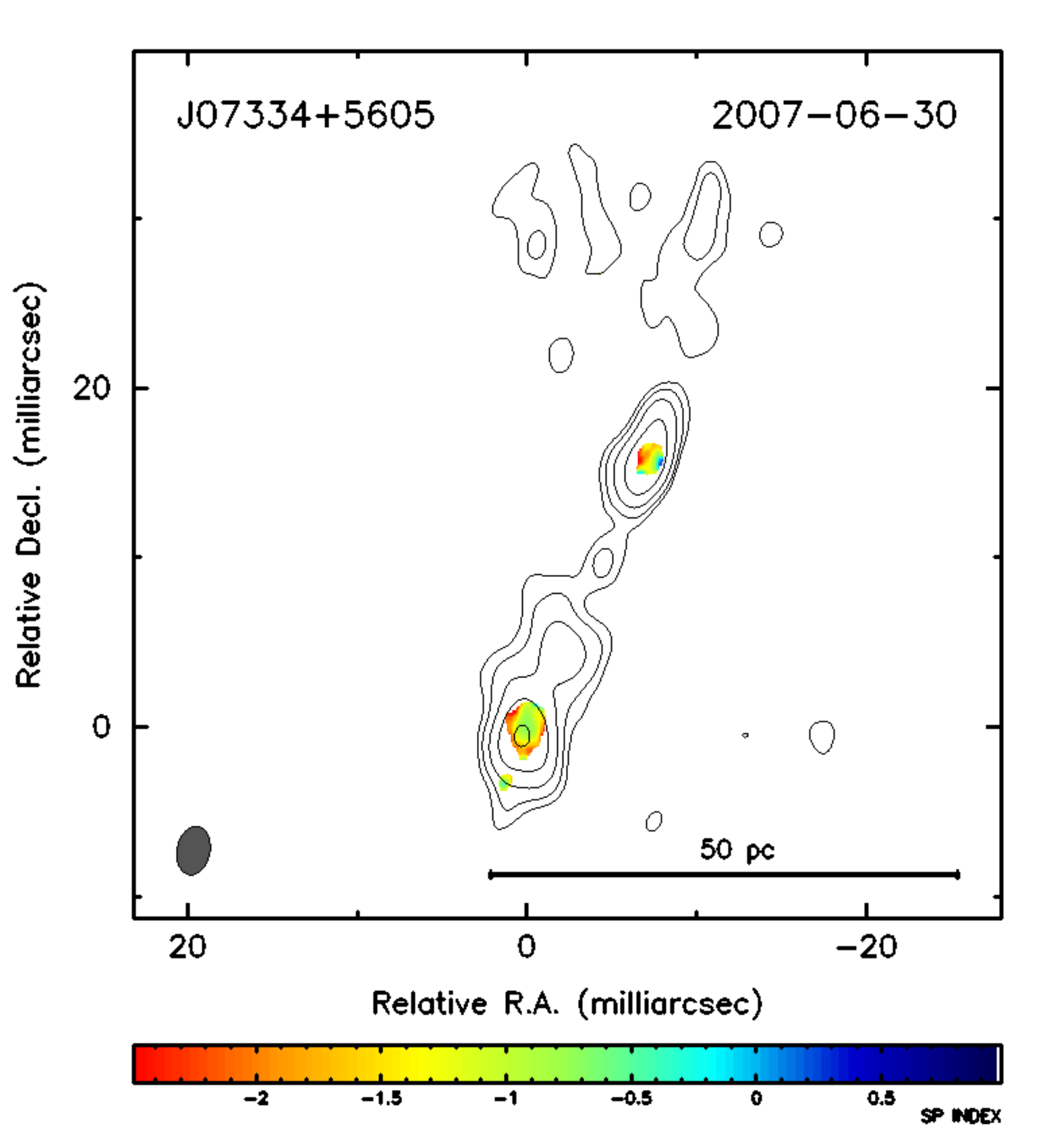}
	}
~\hfill
	\centering
	{%
	\includegraphics[width=0.32273\textwidth]{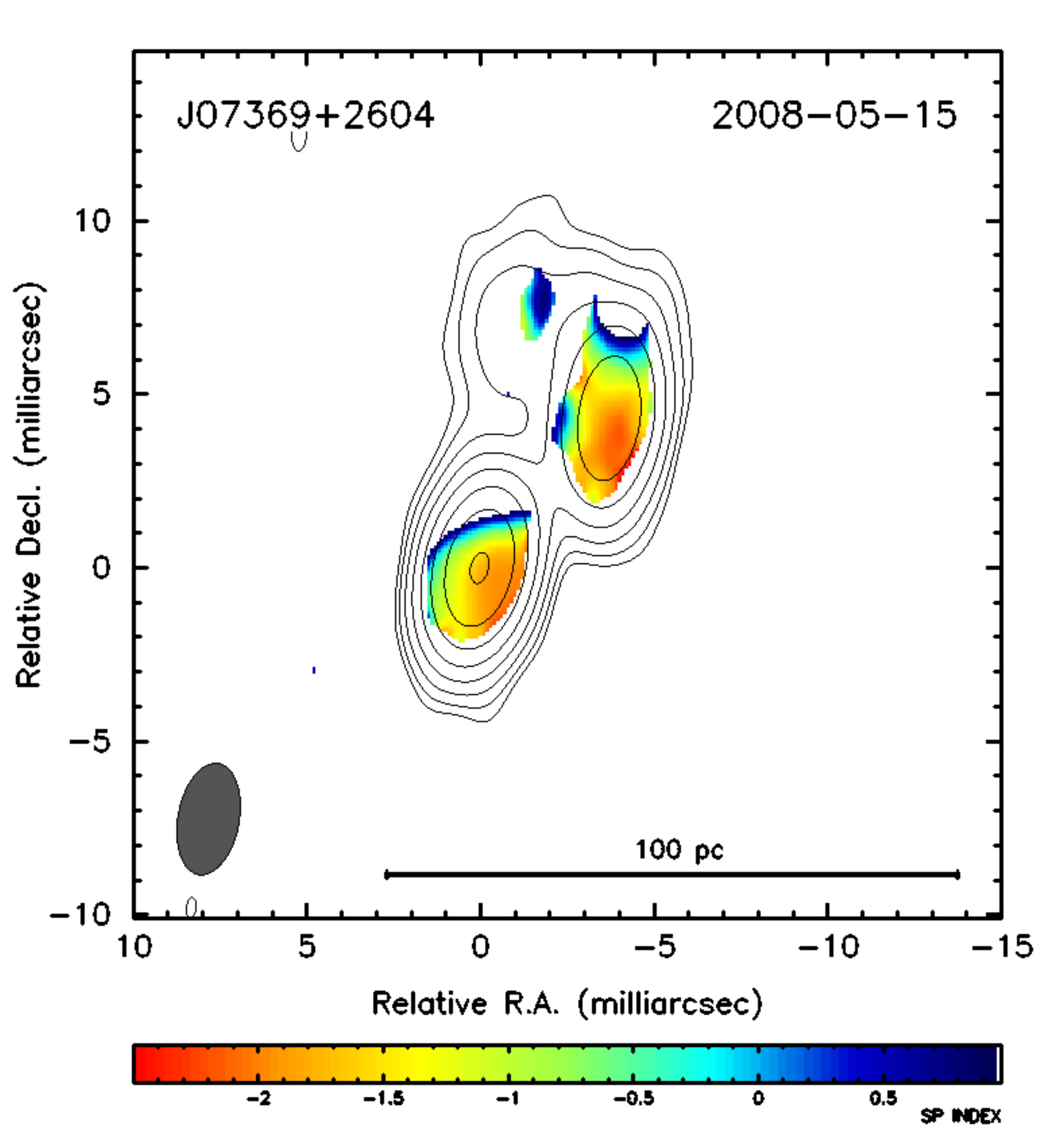}
	}
~\hfill
	\centering
	{%
	\includegraphics[width=0.32273\textwidth]{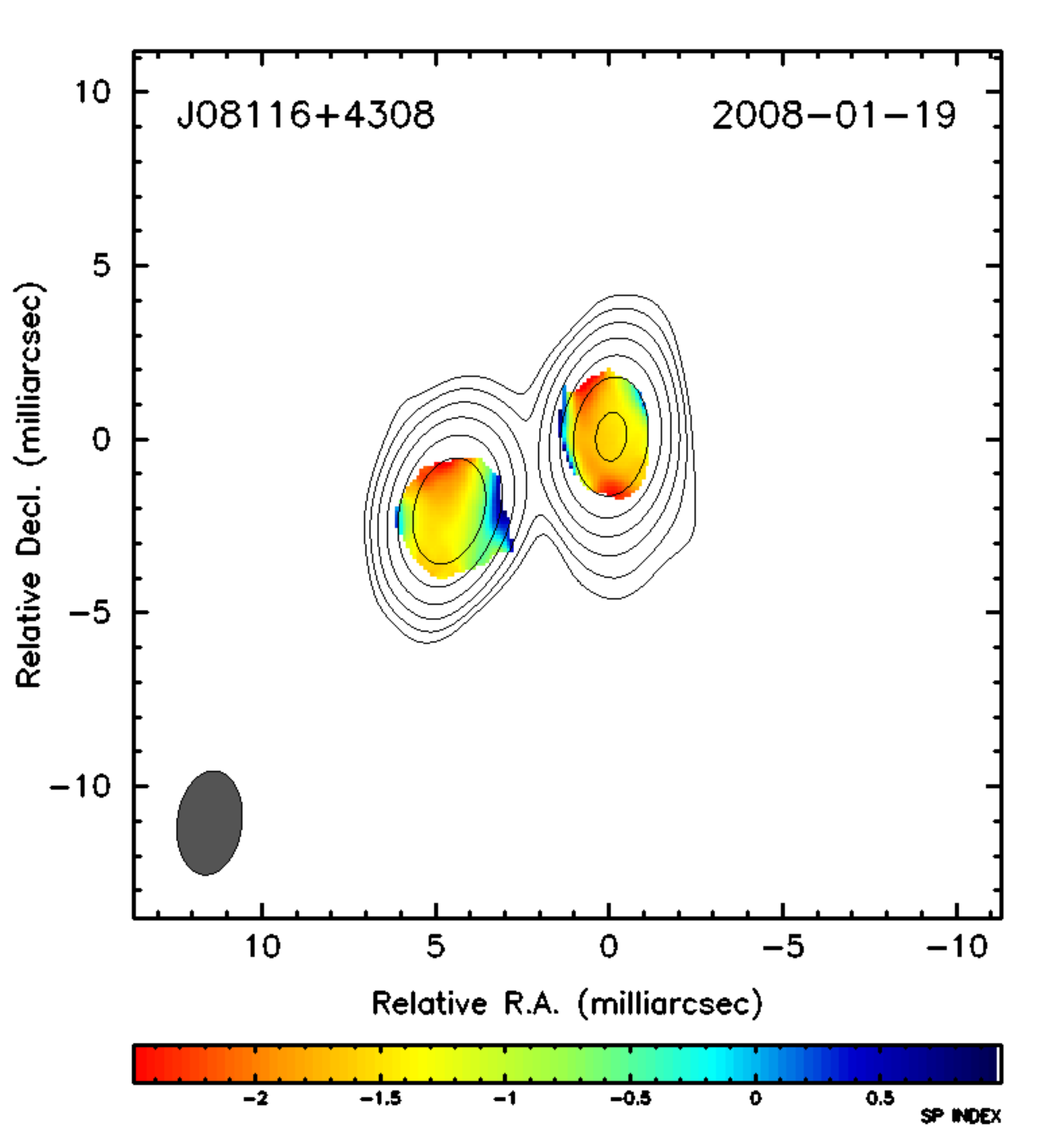}
	}	
~\hfill
	\centering
	{%
	\includegraphics[width=0.32273\textwidth]{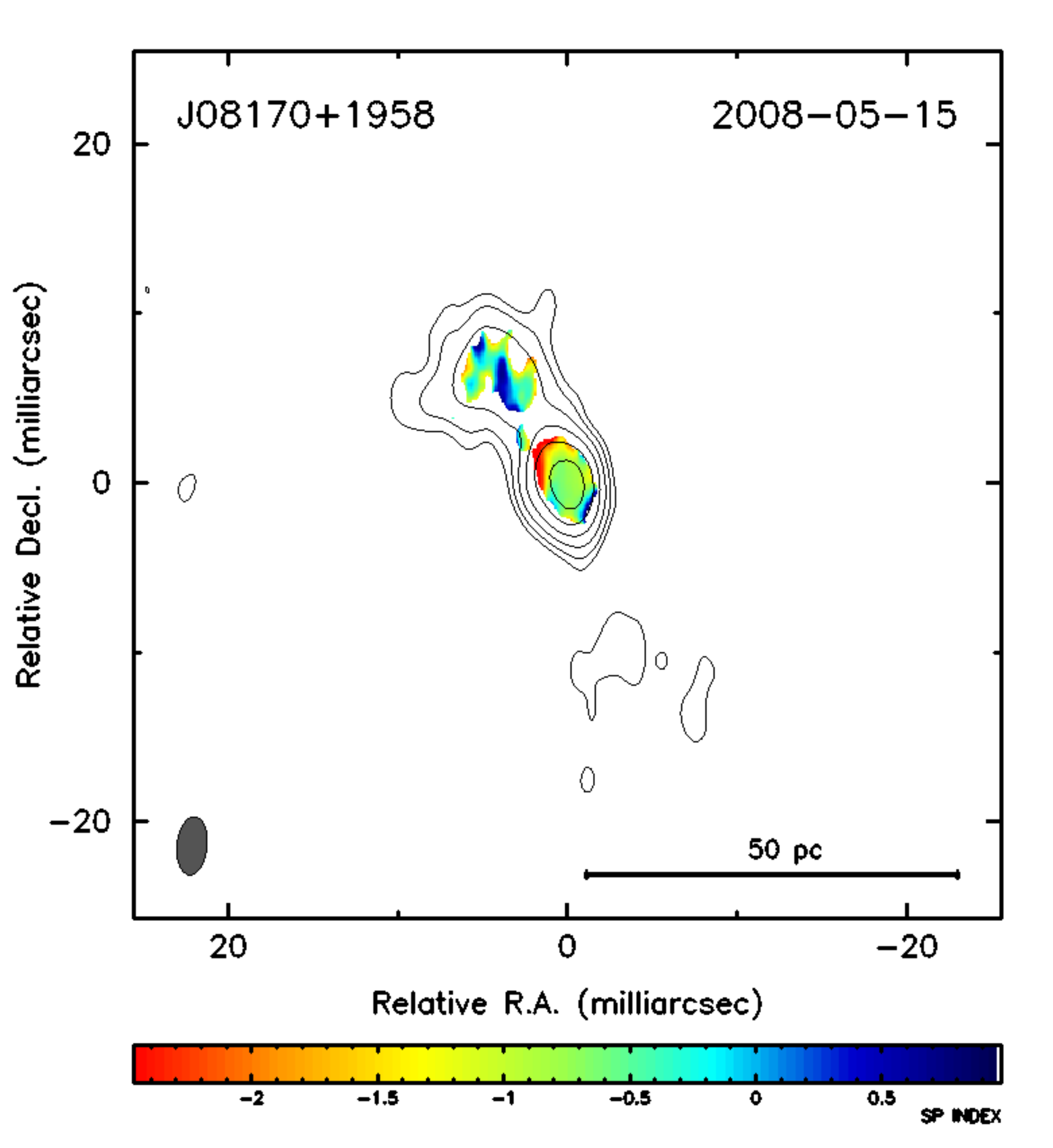}
	}	
~\hfill
	\centering
	{%
	\includegraphics[width=0.32273\textwidth]{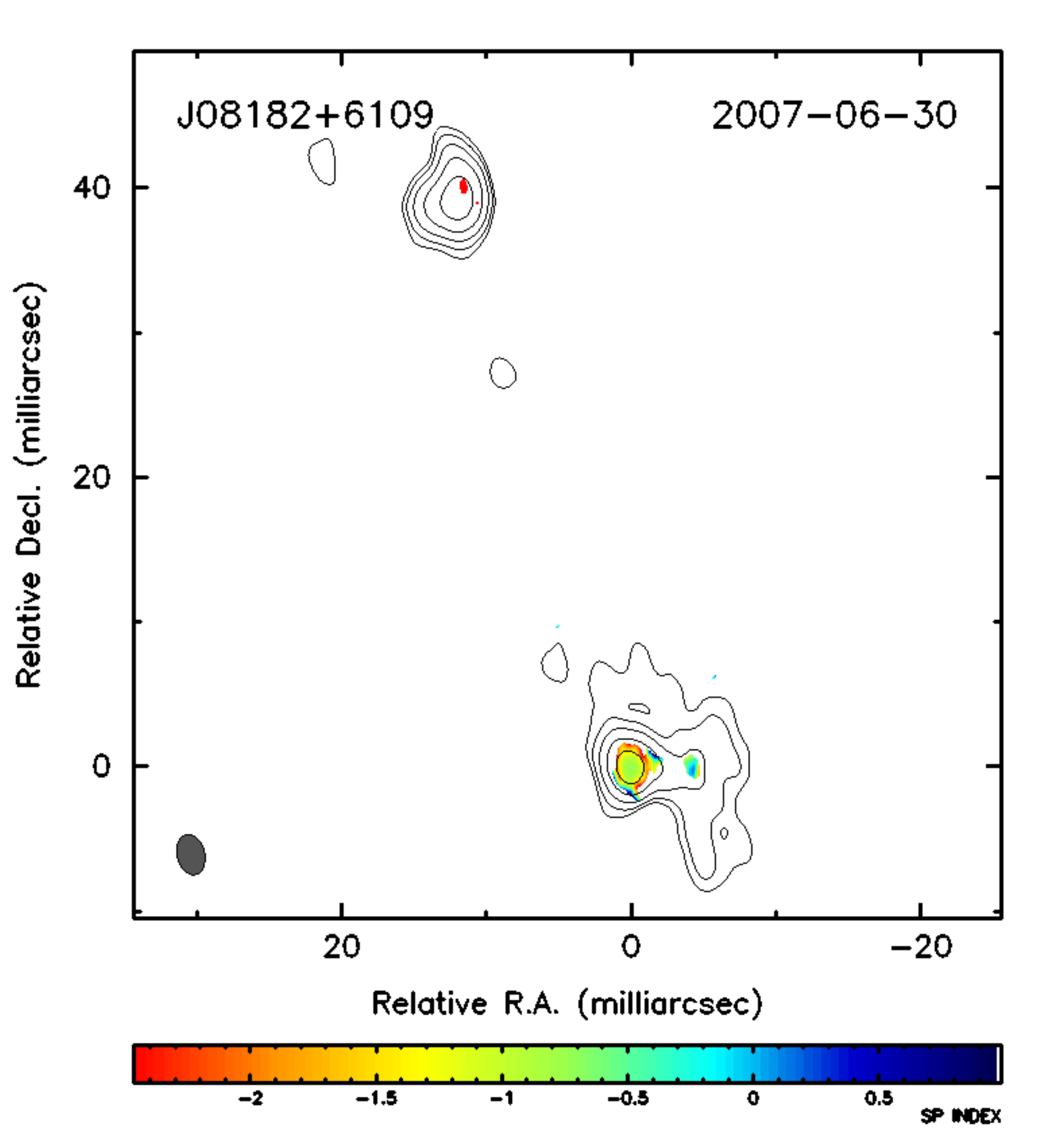}
	}	
~\hfill
	\centering
	{%
	\includegraphics[width=0.32273\textwidth]{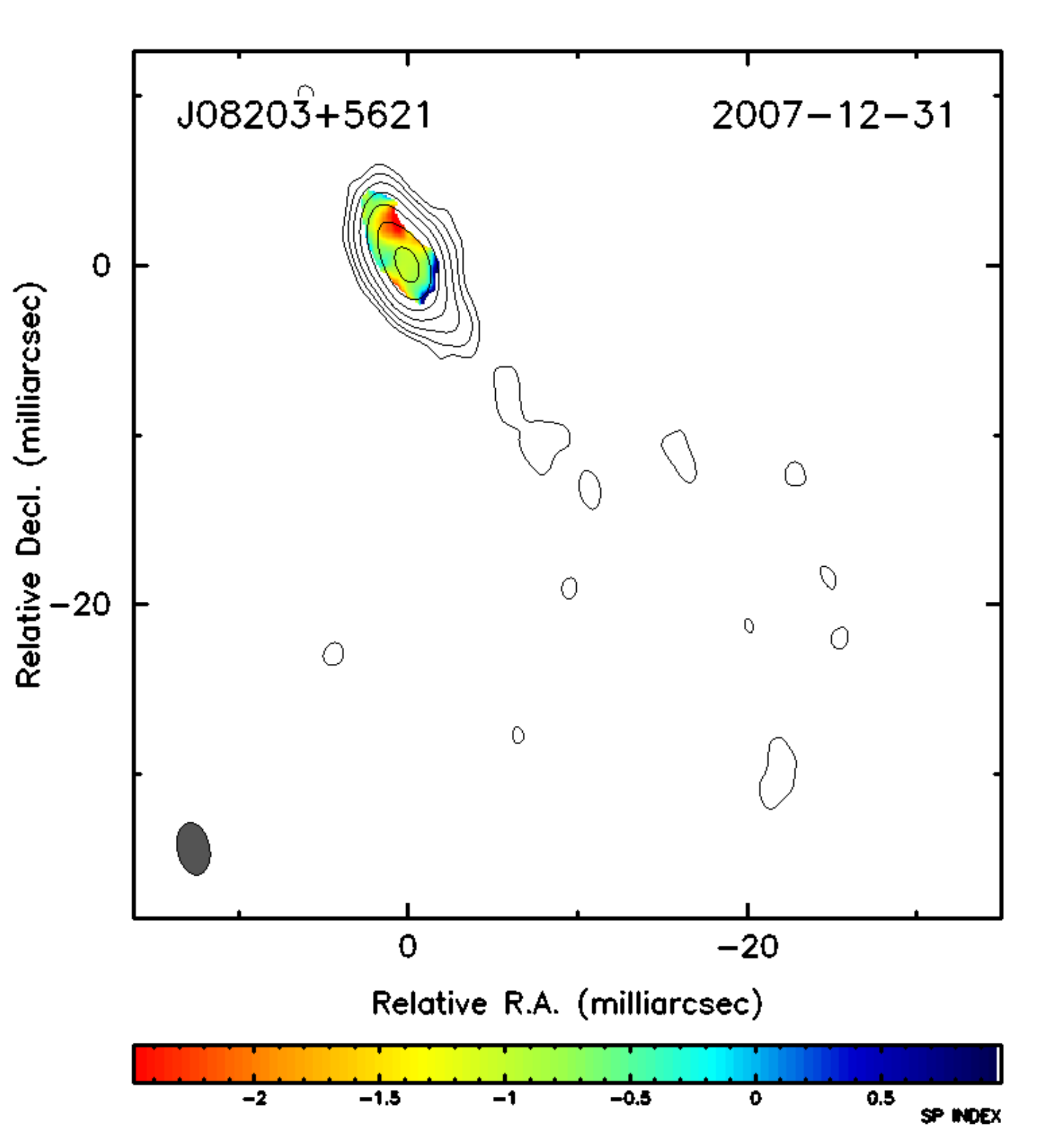}
	}	
~\hfill
	\centering
	{%
	\includegraphics[width=0.32273\textwidth]{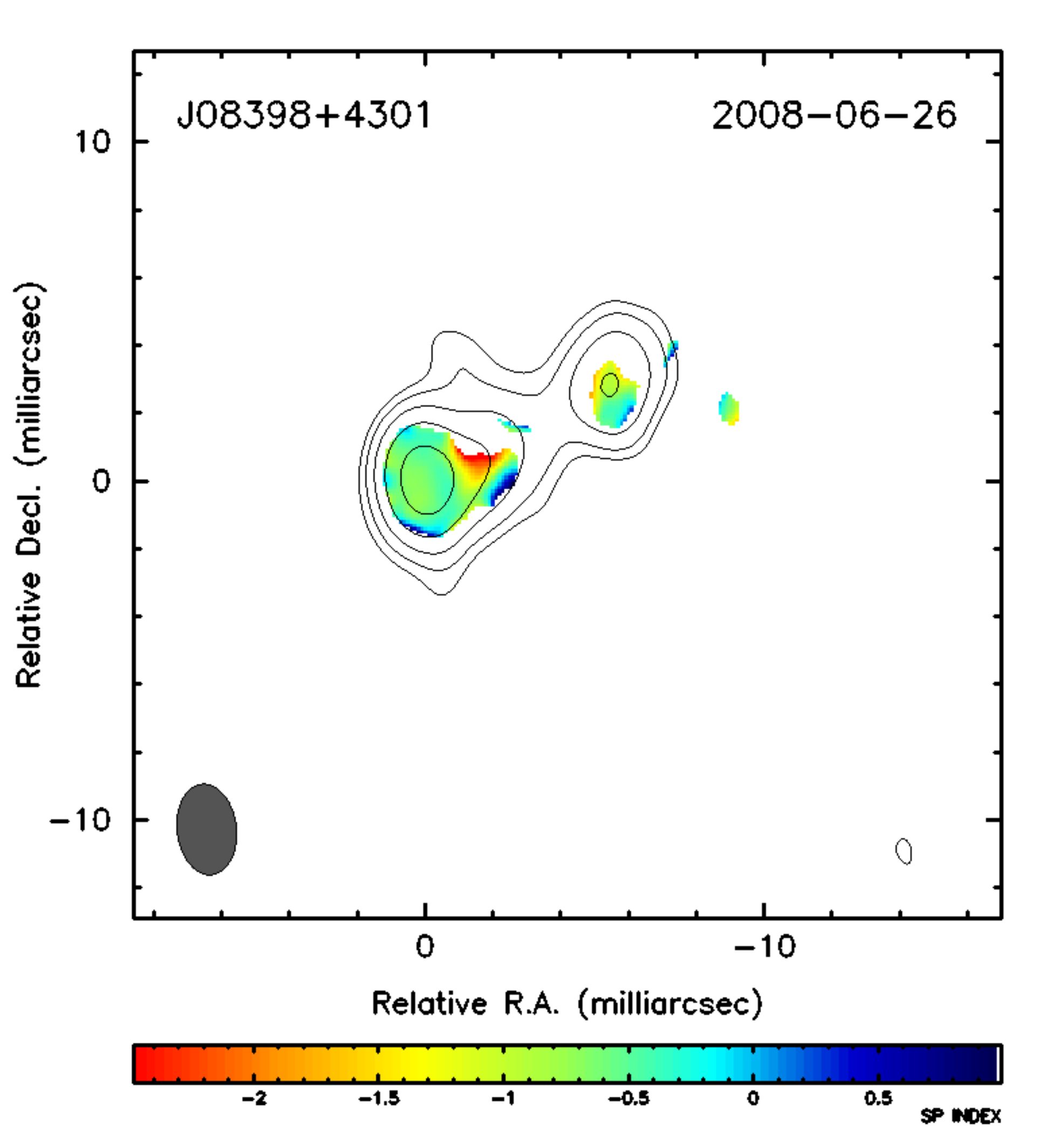}
	}
~\hfill
	\centering
	{%
	\includegraphics[width=0.32273\textwidth]{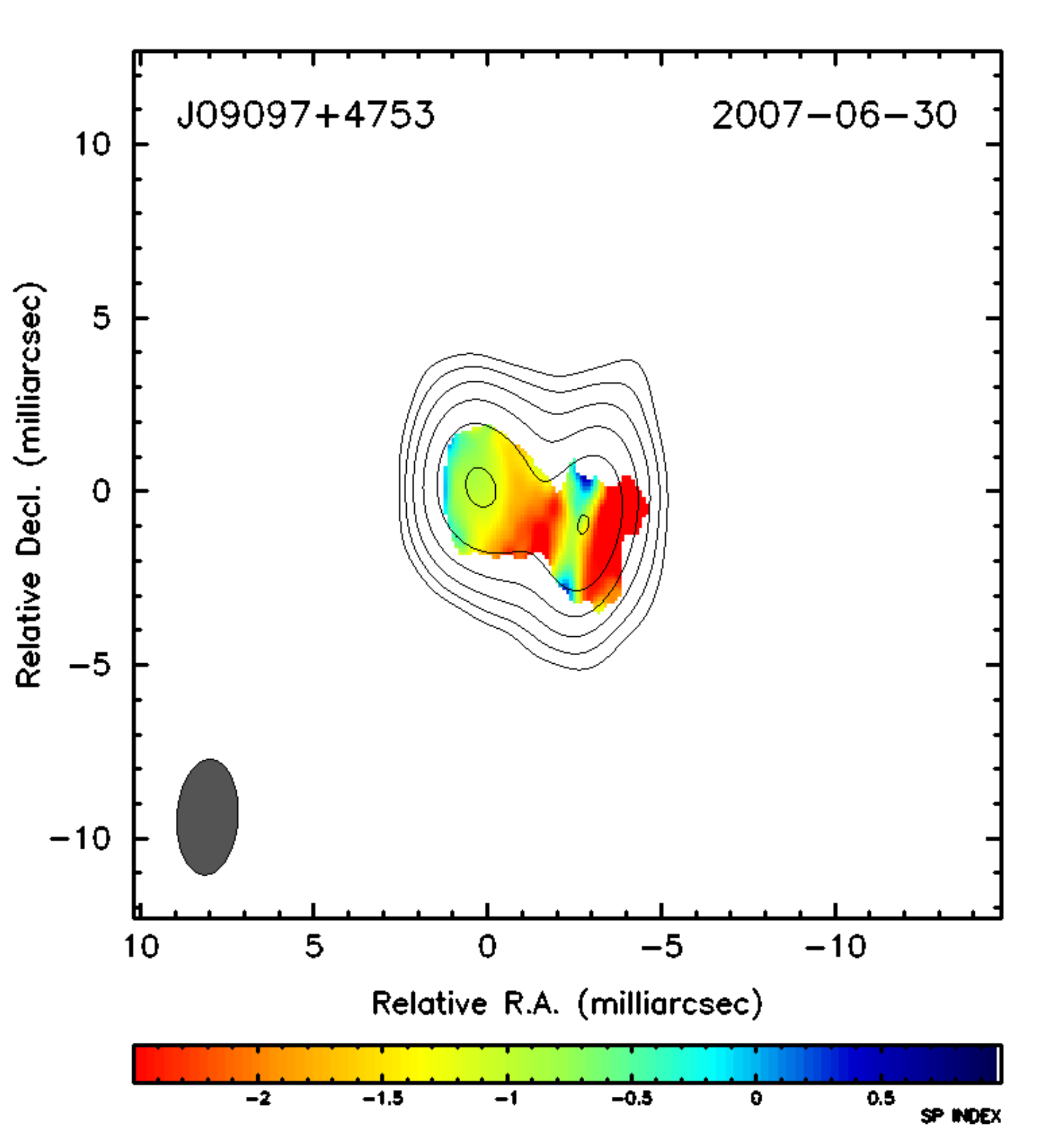}
	}		
~\hfill
	\centering
	{%
	\includegraphics[width=0.32273\textwidth]{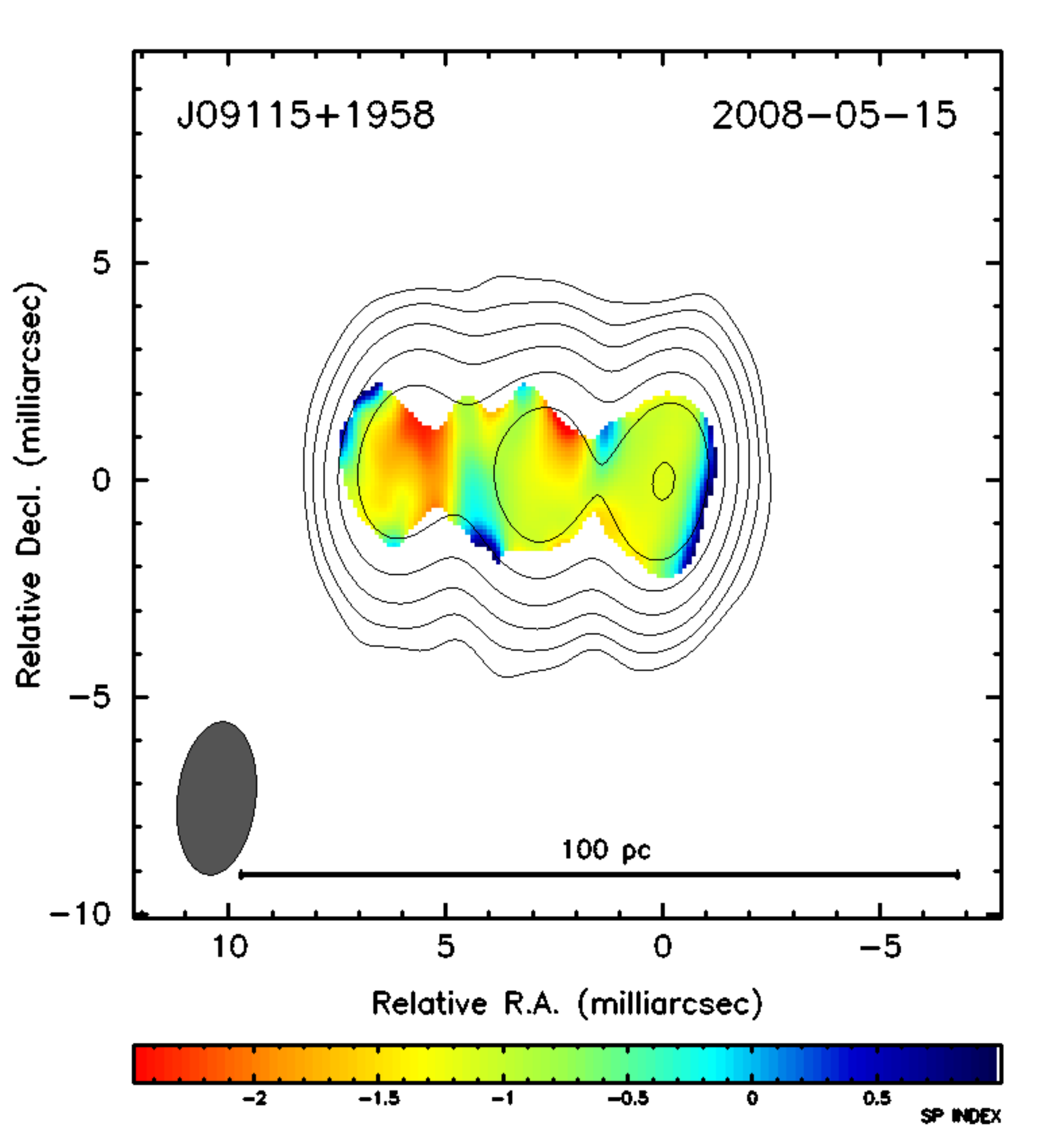}
	}	
	\caption{5 GHz contour maps of refuted CSOs with 8-15 GHz spectral index map overlays. The contour levels begin at thrice the theoretical noise (typically $\sim$ 0.4 mJy for BT088 and 1.0 mJy for BT094) and increases by powers of 2. The colour scale is fixed from -2.5 to 1 to facilitate comparison.  Sources with confirmed spectral redshifts have associated distance bars for linear scale. (A colour version and the complete figure set is available in the online journal.)}
\label{fig:refuted_xuspix}
\end{figure*}

The 5 and 8 GHz, as well as the 8 and 15 GHz images were matched in resolution in order to obtain a spectral index distribution across the source, where we take $F_\nu\propto \nu^\alpha$. Each spectral index map was overlaid onto 5 GHz contours to understand the distribution of spectral index within each source structure (Figures \ref{fig:cso_cxspix}, \ref{fig:cso_xuspix}, \ref{fig:cand_cxspix}, \ref{fig:cand_xuspix}, \ref{fig:refuted_cxspix} and \ref{fig:refuted_xuspix}). The spectral index maps for CSOs have been `lettered' to facilitate discussion and referencing of the images, for the remaining maps for candidates and refuted CSOs the source name will be sufficient. For consistency, this analysis was automated using scripts which did the following:
\begin{itemize}
\item The resolution of the images was matched by first adding a $(u,v)$ taper to the higher resolution maps until the beam was close (within roughly 0.03 mas for 5-8 GHz spectral index and 0.25 mas for 8-15 GHz spectral index) to the size of the lower frequency. 
\item The beam size was then forced to match exactly using the Difmap command RESTORE, suppressing any residual large $(u,v)$ values from the higher frequency.  The map dimensions were also matched, then spectral index maps were made using the AIPS task COMB. 
\item The Caltech package program MAPLOT was then used to overlay the spectral index maps onto the 5 GHz contours to identify where the spectral characteristics were located within the overall structure of each source. 

\end{itemize}
Error maps associated with each spectral index map were produced using the root mean square error of the two input maps. In all these maps, the pixels on the edge of each feature have a higher error than the centre of the features since the border features are not sampled by the  full synthesised beam. The average error in $\alpha$ for the maps, calculated by measuring the root mean square of the central features and the boundary of the components separately, are $\pm0.03, \pm0.15$ and $\pm0.03, \pm0.14$  for the 5 to 8 and 8 to 15 spectral index maps from BT088 and $\pm0.03, \pm0.15$ and $\pm0.04, \pm0.13$ for the 5 to 8 and 8 to 15 spectral index maps from BT094 respectively. This means that variation of that amplitude, particularly on size scales smaller than the beam width, should not be interpreted as realistic features on these maps. 

\subsection{VLBI Recovery of Flux Density}
\label{vlbi_recovery}
\begin{figure}
        \includegraphics[width=0.51\textwidth,center]{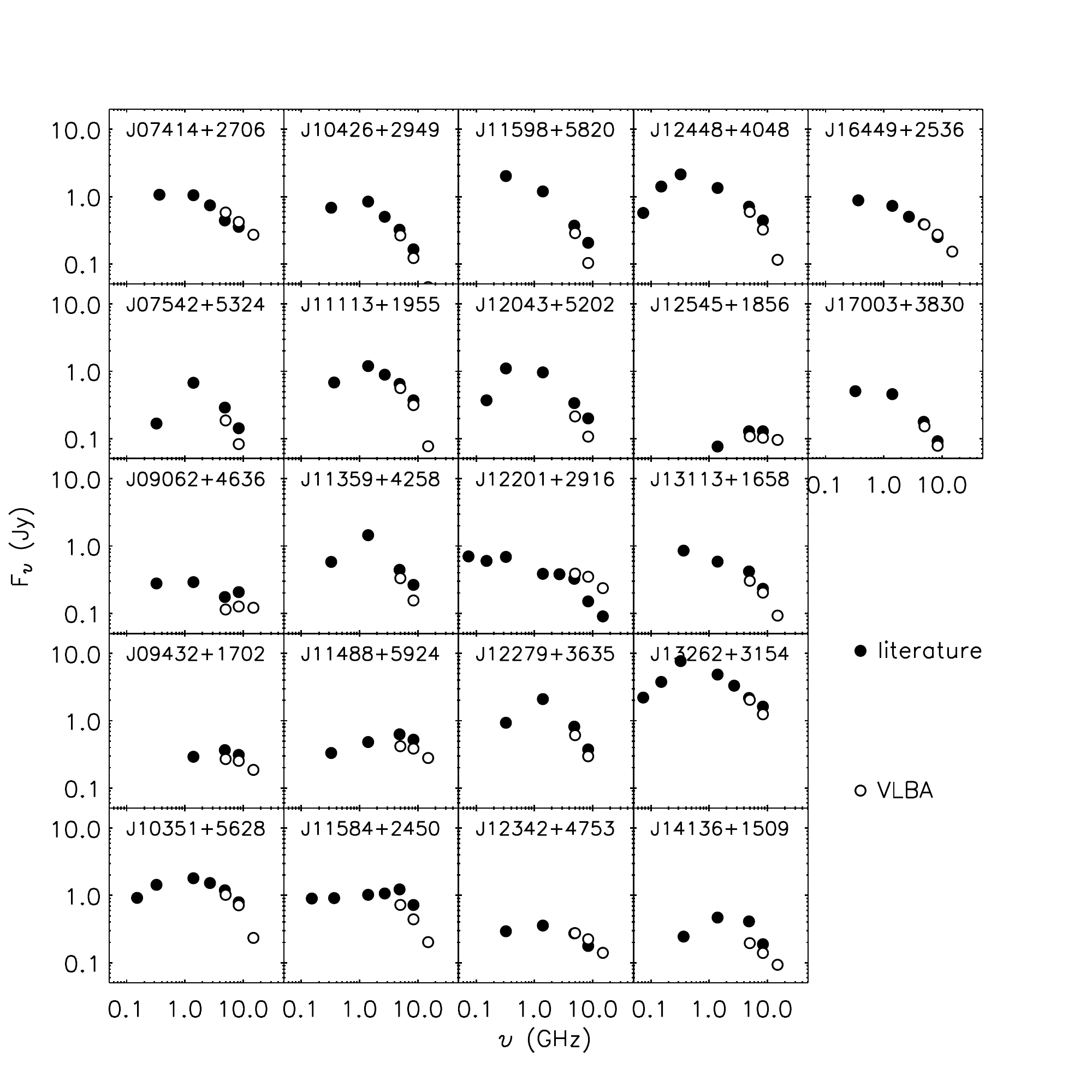}
\caption{Plotting VLBA total flux for each CSO along with non-simultaneous VLA and single dish archival observations. This shows the spectral `turnover' of each source, and also how much of the expected flux density (filled dots) we are recovering with our VLBA observations, which are marked by open dots.The archival fluxes in MHz are: 74 (VLSS; Cohen et al. 2007), 151 (6C; Hales et al. 1990), 325 (WENSS; Rengelink et al. 1997), 365 only used when no WENSS (TXS; Douglas et al. 1996), 1400 (NVSS; Condon et al. 1998), 2695 used only when no PKS (Condon et al. 1978), 2700 (PKS; Wall 1975), 4850 (GB6; Gregory \& Condon 1991), 8400 (CLASS; Myers et al. 2003). The only exception is for J1221+2916 which used Cohen et al. (2004) and  Nagar et al. (2005) for 74 and 15000 MHz respectively. Errors are smaller than the symbols in the plot.}
\label{fig:SED}
\end{figure}

Comparing the summed flux densities of the CSOs at each frequency (Table  \ref{tab:cso}) with the previously observed flux densities from surveys at single dish and shorter  baseline interferometers (i.e. VLSS, WENSS, NVSS, GB6) we can test whether or not any emission is being missed by the VLBA observations.  Sources could resolve out in these high resolution observations if emission is on larger scales that is visible to the single dish, but not to the VLBA. In general the agreement is excellent (Fig. \ref{fig:SED}).  When compared to the 5 GHz single-dish observations of \citet{1991ApJS...75....1B}, half of the sources recover the total flux within the formal errors and 21 sources agree within an additional 15\%.  The few exceptions (J09062+4636, J12043+5202, and J12201+2916) could be the result of variability since there is a 15 year separation between the observations. 

Most of the CSOs exhibit a peak in these spectral energy distributions at $\sim$1 GHz, as expected for lobe dominated emission on size scales of $\sim$10 mas   \citep{1998PASP..110..493O}.  A second turnover, such as that hinted at in the SED for J12201+2916 below 300 MHz (Fig. \ref{fig:SED}), is indicative of larger-scale (older) emission not associated with the younger radio source shown in Figures 2 \& 3.  Finally, it is worth pointing out that a few sources, notably J09062+4636, J12545+1856, J13113+1658, and J16449+2536, lack a clear peak.  Since Fig.\ref{fig:SED} contains only confirmed CSOs, this illustrates the point that not all CSOs necessarily have a gigahertz peaked spectrum (GPS) though most (82\% or 18 of 22) do.  The GPS peak tells us about the dominant component size, but does not directly confirm a CSO nature or a young age for a given source.

%

\subsection{Polarisation}
\label{polarisation}
\begin{figure*}
	\centering
	 {%
	 \includegraphics[width=0.41\textwidth]{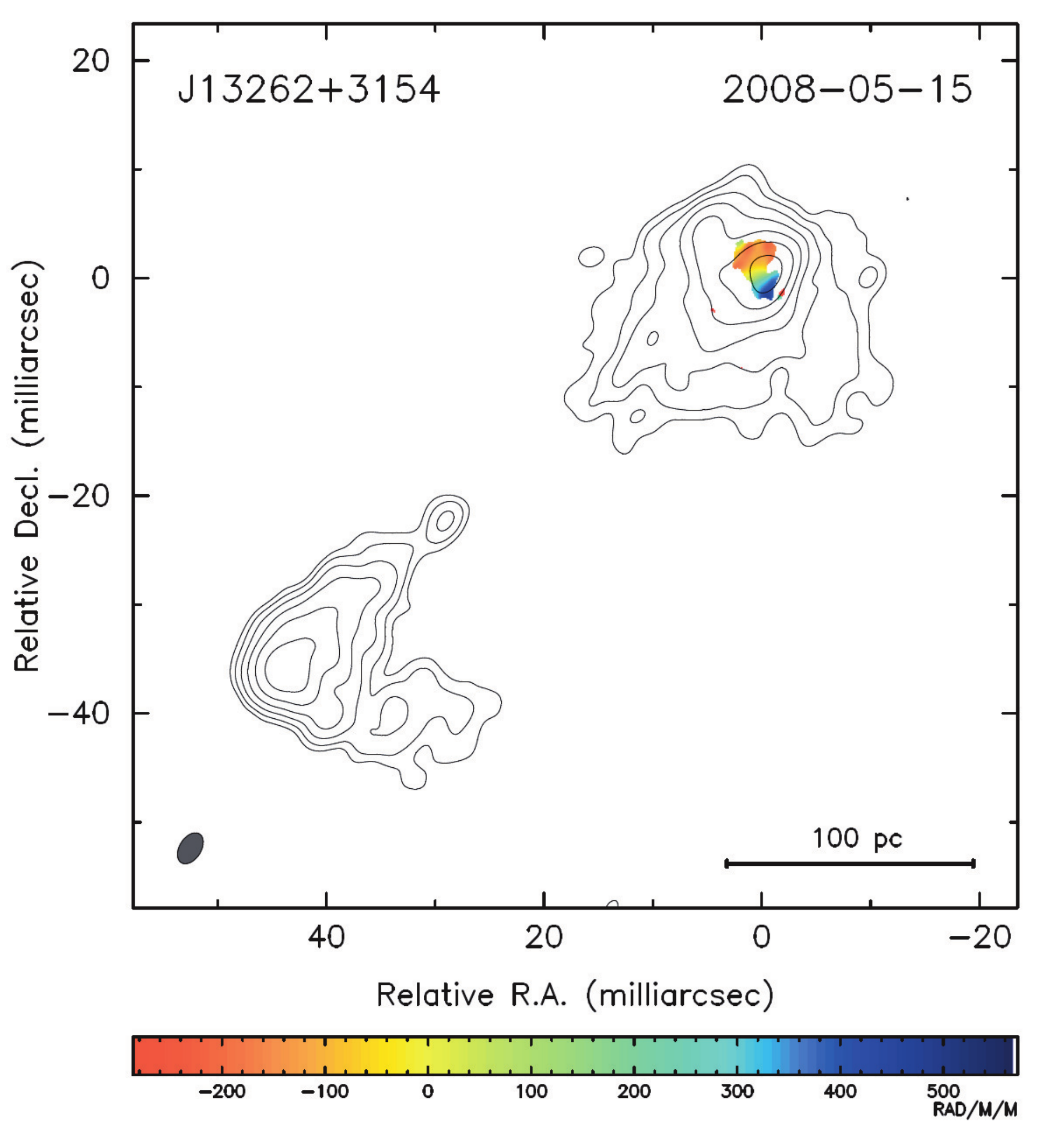}
	}
~\hfill
	\centering
	{%
	\includegraphics[width=0.51\textwidth]{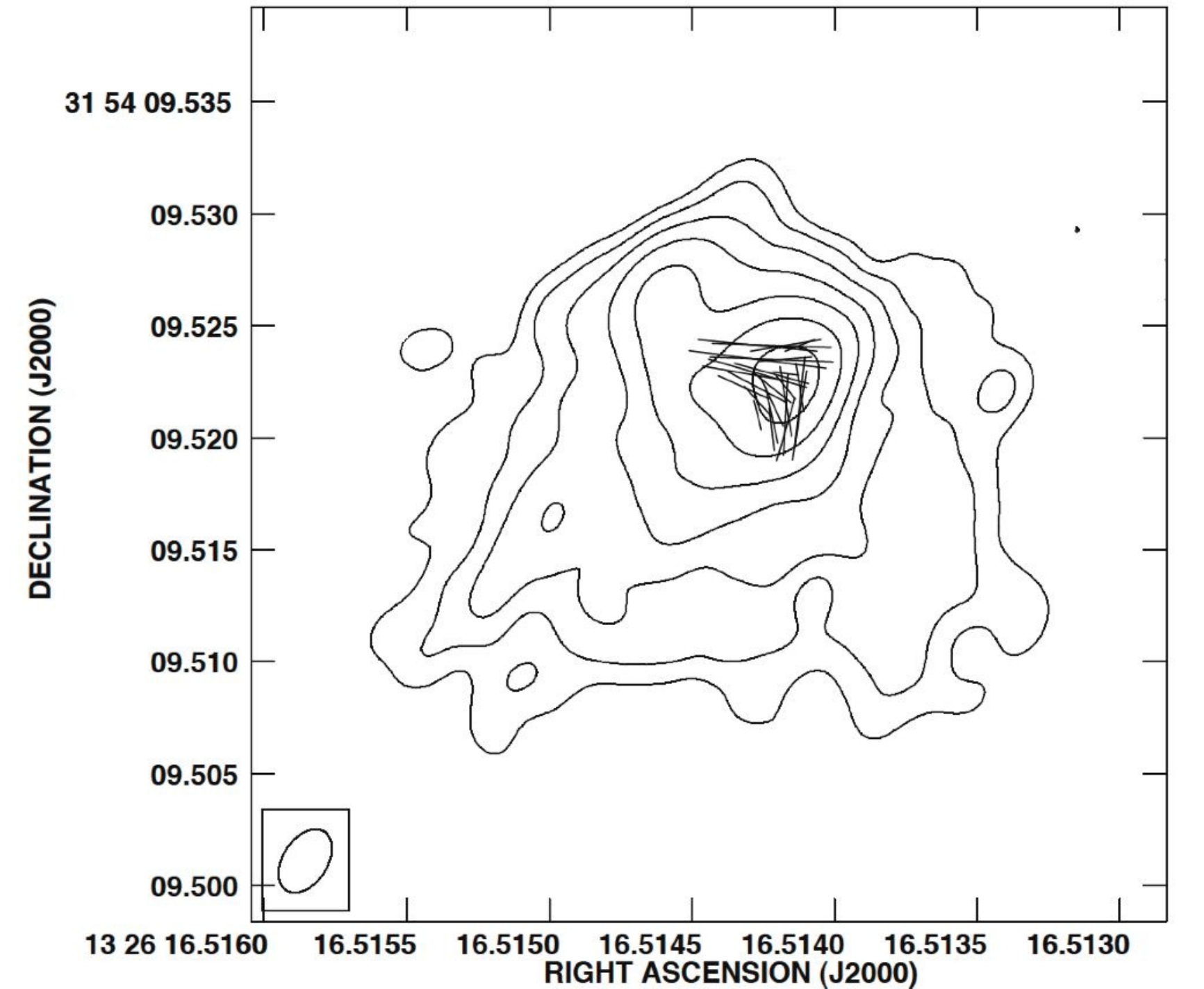}
	}
	\caption{\textbf{Left:} The calculated Rotation Measure map of J13262+3154 generated from 5 and 8 GHz data overlaid onto 5 GHz contours. RMs of $\sim$ $-300$ to 500 rad m$^{-2}$ are observed with a notable sign change across the feature, corresponding to B-Fields of $\sim\pm$0.06-2 $\mu G$ assuming $n_e$ of 0.1 cm$^{-3}$ and an $L$ of 0.3-10 parsecs. \textbf{Right:} The corrected polarisation angle of the magnetic fields $(B)$ plotted on top of  continuum for the north-western lobe. Contour levels begin at 2.0 mJy and increase by factors of $2^\frac{1}{2}$ in both images.}
	\label{fig:J13262RM}
\end{figure*}

\begin{figure*}
	\centering
	 {%
	 \includegraphics[width=0.38\textwidth]{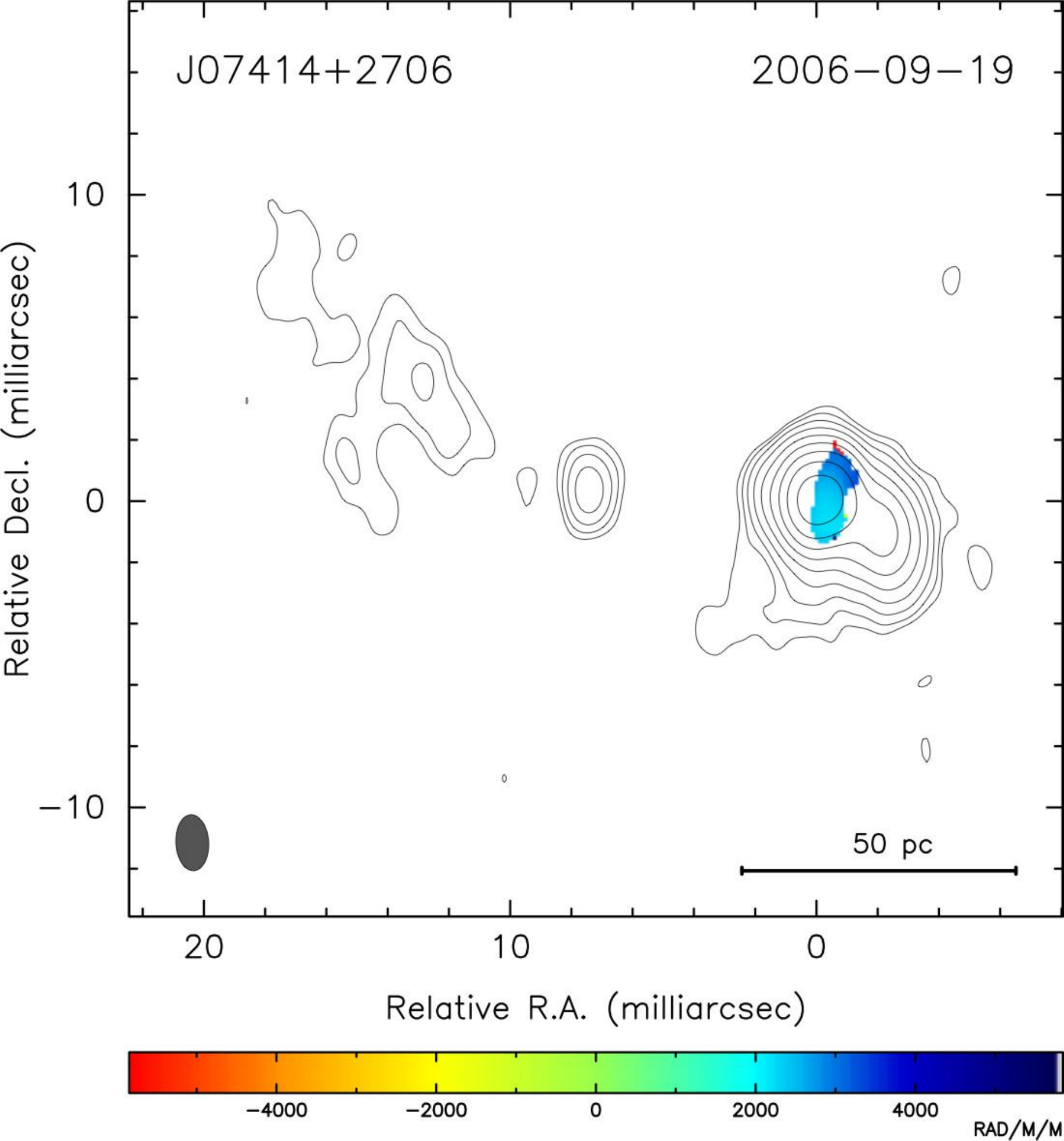}
	}
~\hfill
	\centering
	{%
	\includegraphics[width=0.51\textwidth]{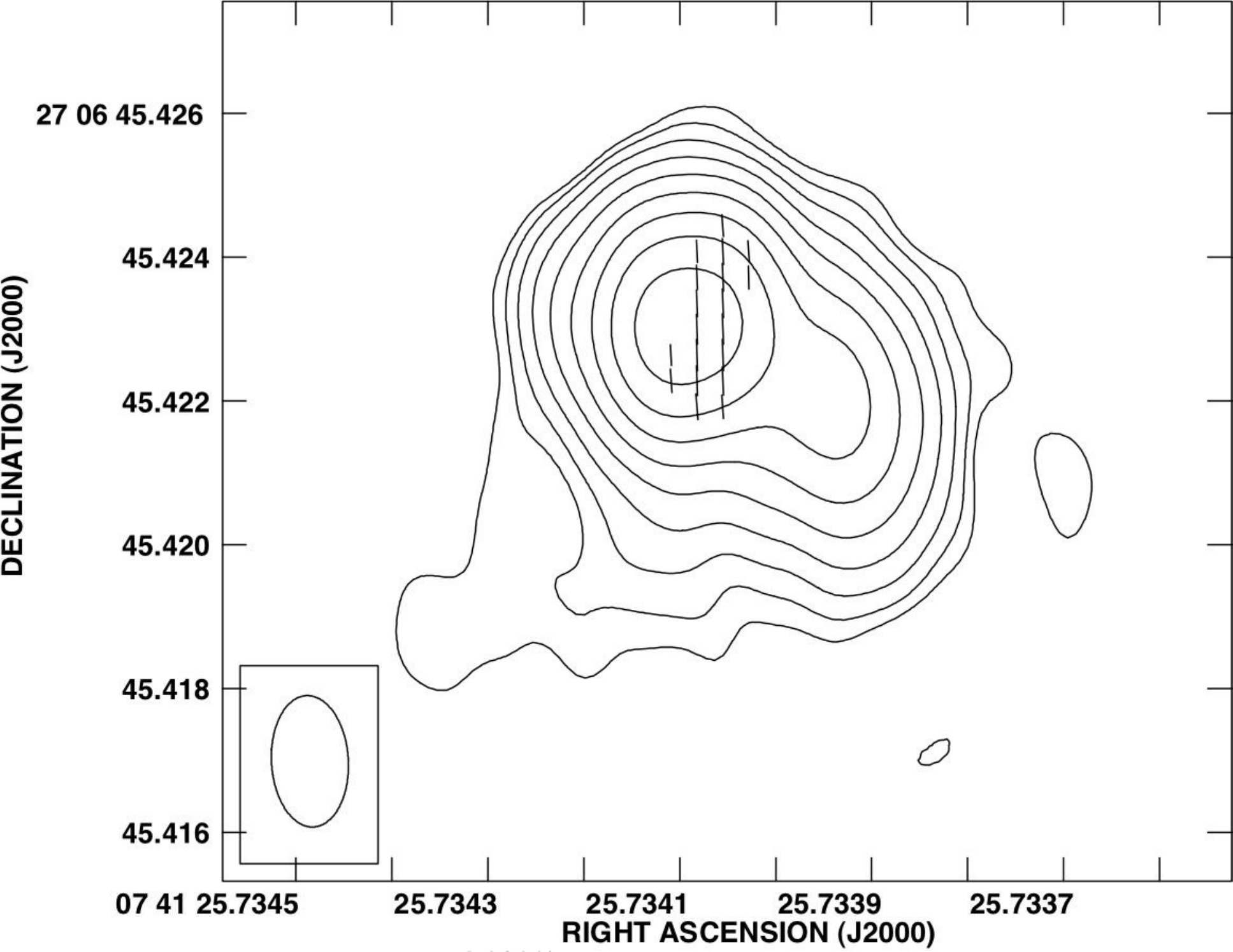}
	}
	\caption{\textbf{Left:} The calculated Rotation Measure map of J07414+2706 generated from 8 and 15 GHz data overlaid onto 8 GHz contours. RMs of $\sim$ 1800 to 4000 rad m$^{-2}$ are observed, corresponding to B-Fields of $\sim\pm$0.36-12 $\mu G$ assuming $n_e$ of 0.1 cm$^{-3}$ and an $L$ of 0.3-10 parsecs.  \textbf{Right:} The corrected polarisation angle of the magnetic fields $(B)$ plotted on top of  continuum for the western lobe Contour levels begin at 0.45 mJy and increase by factors of $2^\frac{1}{2}$ in both images.}
	\label{fig:J07414RM}
\end{figure*}

Polarimetric analysis was performed on all of the confirmed VIPS CSOs, and polarised flux was observed in two of the sources. J07414+2706 exhibited polarised emission at both 5 and 8 GHz and J13262+395 was detected in polarisation at 8 and 15 GHz (see Fig. \ref{fig:J07414RM} and \ref{fig:J13262RM}). Polarised emission from non-CSOs will be presented in a future work, with the notable exception of J16021+3326, whose polarisation  properties were investigated in the process of classifying it as a core-jet \citep{2010ApJ...712..159T}.

\section{Discussion}
\subsection{Definition of a CSO}
\label{definition}

Historically, the definition of a CSO has not been standardised and what one paper considers a requirement, another instead considers a common characteristic. We have adopted a broad definition for a CSO, keeping the term an observation-based definition allowing that this classification might encompass multiple physically distinct species. Even so, we suggest that the CSO category is more focused and relevant classification than the purely spectral classification of Gigahertz peaked spectrum and Compact steep spectrum sources which are highly contaminated with Blazars \citep{2009AN....330..128T}.

We define a compact symmetric object as a source less than 1 kpc in projected extent, where two well defined hotspots (the working surface where the jets push against the surrounding medium) and/or lobes (cavities filled with back flowing material) are observed, which is the simplest and most encompassing definition we can devise. Although the detection of a core is not strictly required, it does provide confirmation of the CSO identification and aids in the understanding the underlying geometry of these systems. The hotspots in these sources are typically discernible as being steep gradients at the ends of the lobes, a characteristic referred to as edge brightening. Several of the candidates in which counter jet/lobe emission was observed were still classified as non-CSOs (J08170+1958, J12035+4632, J12582+5421, J13128+5548 \& J15159+2458) the X band detection of the counter emission of these sources was three sigma or less. We decided these were most likely core-jets oriented at a large enough angle out of the plane of the sky to just detect the de-boosted counter-emission. Lower frequency VLBI observations could reveal them to have structure consistent with CSOs.

\subsection{Confirmed CSOs}

Here each of the 24 confirmed VIPS CSO will be briefly discussed. The properties of these CSOs are reported in Table \ref{tab:cso} and spectral index maps of each source are in Figures \ref{fig:cso_cxspix} \& \ref{fig:cso_xuspix}.

\subsubsection{J07414+2706}

A flat-spectrum compact core is visible in both Figures \ref{fig:cso_cxspix}A and \ref{fig:cso_xuspix}A with lobes to the east and west. The western lobe is appreciably brighter than its counterpart due either to geometry or interaction with the surrounding media. Polarisation is detected in the western lobe in both X and U bands (Fig. \ref{fig:J07414RM}). This source has the highest flux density ratio for the two lobes (19.35), indicating either one side of this source is strongly interacting with its environment or the jets might not be as close to the plane of the sky as is usually assumed in a CSO.  This source is $\sim$200\,pc across at its largest extant.

\subsubsection{J07542+5324}
This source is comprised of lobes extending to the northwest and southeast (Fig. \ref{fig:cso_cxspix}B). While the southern lobe shows flatting of its spectrum towards the centre of the system, there wasn't enough evidence to identify a component to associate with the core of the system. The U band emission from this source was only marginally detected, and didn't allow the generation of an accurate 8-15\,GHz spectral index map. This source is also part of the CSOs Observed In the Northern Sky (COINS; Peck \& Taylor 2000) project, where is was previously confirmed as a CSO.

\subsubsection{J09062+4636}
This source was strongly detected at all three frequencies and is comprised of an unresolved core with weak extension to the north and south (Figs. \ref{fig:cso_cxspix}C and \ref{fig:cso_xuspix}C). We note the lack of edge brightening of the extended emission from this source that is typical of CSOs and is instead characterised with FRI like morphology.  Weak polarisation was detected at 8\,GHz, but no rotation measures could be calculated. This source was confirmed as a CSO by the Compact Radio Sources at Low Redshift (CORALZ; \citealt{2004MNRAS.348..227S,2009A&A...498..641D}) project. 

\subsubsection{J09432+1702}
The brightest feature of this source at 5 GHz appears to be a flat-spectrum core (Figs. \ref{fig:cso_cxspix}D and \ref{fig:cso_xuspix}D). The core ($\alpha\sim-1$ to $-1.5$) is edge brightened on its northeastern side. A small steep-spectrum protrusion to the southeast of the core is similarly brightened on its southeastern side.  This approximately 125\,pc source is first identified as a CSO here in this paper and no other morphological studies have been conducted.

\subsubsection{J10351+5628}
This 188\,pc source contains two steep-spectrum oppositely edge brightened features (Figs. \ref{fig:cso_cxspix}E and \ref{fig:cso_xuspix}E). While no obvious core was observed, the morphology of this double system is convincingly that of a CSO. In Fig. \ref{fig:cso_cxspix}E, there are small ($<$ the beam width) edge effects apparent (positive stripe on the east and negative stripe on the west), which most likely arise from a slight misalignment of the two frequencies due to the complicate spectral nature of the north-eastern lobe. This source was confirmed as a CSO in COINS (Peck \& Taylor 2000). 

\subsubsection{J10426+2949}
The two lobes of this source are comprised of bright, steep-spectrum, resolved components (Figs. \ref{fig:cso_cxspix}F and \ref{fig:cso_xuspix}F). The eastern component is connected to extended emission tracing back towards the west, and ending in a faint, flat-spectrum component tentatively classified as the core. Automated images generated by the  FIRST \citep{1995ApJ...450..559B} and the NVSS \citep{1998AJ....115.1693C} surveys both hint at possible extension to the northwest on kiloparsec scales suggesting a possible earlier epoch of emission.

\subsubsection{J11113+1955}
The two steep-spectrum structures within this source feature opposing brightened edges (Figs. \ref{fig:cso_cxspix}G and \ref{fig:cso_xuspix}G). The morphology of this source is indicative enough to classify it as a CSO even though no core was observed. It is possible the very steep spectrum ($\alpha<-2$) observed in Fig. \ref{fig:cso_xuspix} is an artefact of the 15\,GHz flux being near the detection threshold. This source was confirmed as a CSO in COINS \citep{2000ApJ...534...90P}.

\subsubsection{J11359+4258}
This source is characterised by symmetric, edge brightened, steep-spectrum lobes that are consistent with CSO morphology (Fig. \ref{fig:cso_cxspix}H). While the core is not apparent from these observations, and the redshift for this source is unknown, based on the structure we are classifying this source as a CSO.

\subsubsection{J11488+5924}
This source is characterised by a flat-spectrum core with diffuse lobes extending to the northwest and southeast (Figs. \ref{fig:cso_cxspix}I and \ref{fig:cso_xuspix}I). This is a previously known nearby low-power CSO (Taylor, Wrobel \& Vermeulen 1998; Peck \& Taylor 2000). The apparently missing spectral information in Fig. \ref{fig:cso_cxspix}I is due to the core of this source having a strongly inverted spectral index ($\alpha\sim1-1.5$) between 5 and 8\,GHz, which `runs off' the right side of the spectral index scale whose values were fixed to maximise information content across all sources.  This inversion is not steep enough to substantiate free-free absorption alone, as would be the case if we observed $\alpha\sim4-5$. The low power ($L_{5GHz}\sim1\times10^{23}$) and the lack of edge-brightening of this well studied source could make it the prototype for CSOs with FRI-like properties. This source is confirmed as a CSO in COINS \citep{2000ApJ...534...90P}.

\subsubsection{J11584+2450}
 The flat-spectrum compact core has lobes extending to the north and south which is most easily seen in Fig. \ref{fig:cso_xuspix}J, although Fig. \ref{fig:cso_cxspix}J  highlights the complicated nature of this radio galaxy. These lobes then appear to sweep back and connect with the extended emission to the west. A very unusual CSO which seems to be interacting with its environment. See \cite{2008ApJ...684..153T} for deeper analysis and discussion of this interesting source.

\subsubsection{J11598+5820}
The lobes of this 515\,pc size source are extended, steep-spectrum components with evidence of edge brightening at opposing edges (Fig. \ref{fig:cso_cxspix}K). There is no easily identified core observed but the supporting morphological structure is strong enough evidence to classify it as a CSO.  

\subsubsection{J12043+5202}
This source has a compact, flat-spectrum core with edge brightened, steep-spectrum hotspots to the northwest and southeast (Fig. \ref{fig:cso_cxspix} L). This source is an excellent example of the classic CSO morphology and this paper is the first to identify it as such. This is the only example, in our sample, of a CSO where the brightest hotspot is much closer to the core than it's counterpart (Table \ref{tab:cso}) while all three components (the core and both hotspots) lie on an approximately straight line. If environmental interaction is the cause of the arm ratio, one might reasonably expect to see a deviation in the geometry which we do not observe.

\subsubsection{J12201+2916}
The dominant bright, flat-spectrum component is identified as the core (Figs. \ref{fig:cso_cxspix}M and \ref{fig:cso_xuspix}M). From the north and south ends of the core diffuse jets extend out to the west and east respectively, creating an `S-shaped' symmetry observed in other CSOs (e.g. 2352+495 \citealt{1993AAS...182.5307R};  1946+708 \citealt{2009ApJ...698.1282T}). No edge brightening is visible in this source, and it is not obvious where the jets terminate due to the FR-I morphology. As mentioned in section \ref{vlbi_recovery}, the non-instantaneous spectral energy distribution (SED) (Fig. \ref{fig:SED}) suggests possible extended emission. \cite{2013A&A...550..A76} characterised the source as a compact source with two sides lobes with extensions to the south.

\subsubsection{J12279+3635}
This source extends 453\,pc from end to end. At the eastern edge of the source there is a bright, edge brightened, steep-spectrum component with multiple smaller components tracing back towards the west, likely knots within a jet (Fig. \ref{fig:cso_cxspix} N). An unresolved, flat-spectrum component, potentially the core, is then encountered followed by a weak, diffuse component bracketing the west side. This source was part of COINS \citep{2000ApJ...534...90P} where is was classified as a core-jet. Here, we reclassify this as a CSO based primarily on our identification of the core, and the observed emission in each lobe. The source was also  studied by \cite{2013MRAS..433} where they claim the source has a core 32\,mas from the edge brightening or the jet. We believe they were in fact referring to the western-most knot of the eastern jet instead of the much fainter core.

\subsubsection{J12342+4753}
The reasonably compact, flat- to steep-spectrum($\alpha_{5-8}\sim-0.5$ and $\alpha_{8-15}\sim-1.0$) core has a jet extending out to the northwest and a symmetric jet component to the southeast (Figs. \ref{fig:cso_cxspix}O and \ref{fig:cso_xuspix}O). This is another example of a CSO with FRI morphology in our sample. 

\subsubsection{J12448+4048}
The flat-spectrum, compact core has apparent jet emission to the southwest terminating in a lobe (Figs. \ref{fig:cso_cxspix}P and \ref{fig:cso_xuspix}P). A counter-lobe to the northeast and its luminosity and morphology suggest this lobe is directed away from us. This source was classified as  CSO in COINS \citep{2000ApJ...534...90P}, and was the first quasar (QSO) classified as such.

\subsubsection{J12545+1856}
This compact source exhibits slight extensions to the east and west out of a compact, flat-spectrum centre (Figs. \ref{fig:cso_cxspix}Q and \ref{fig:cso_xuspix}Q). This seems to be a very small (8.4\,pc) CSO just resolved in these observations.  

\subsubsection{J13113+1658}
The centre of this source is flat-spectrum and has jet emission to both the north and south (Figs. \ref{fig:cso_cxspix}R and \ref{fig:cso_xuspix}R). 
This is a COINS source \citep{2000ApJ...534...90P}, but remained a candidate in that sample. Here, we classify this source as a CSO due to the centrally located flat-spectrum emission. Note that this source doesn't display the classic edge brightening of many other CSOs and instead belongs to the FR-I like CSOs discussed in section \ref{sec:morph}.

\subsubsection{J13262+3154}
This source is an archetypical example of a CSO. The compact, flat-spectrum core is centrally located between two large steep-spectrum lobes which each exhibit edge brightening (Fig. \ref{fig:cso_cxspix}S). The extended emission then curves towards the southwest. An exceptional characteristic of this CSO is the detection of polarisation in the northwestern lobe (Fig. \ref{fig:J13262RM}). This source was observed as part of the 2 cm survey, where \cite{1998AJ....115.1295K} described it as having CSO morphology despite their lack of detection of the core.  \citet{2013A&A...555A...4M} found this source to depolarised between 15\,GHz and 8\,GHz with low fractional polarisation.  

\subsubsection{J13354+5844}
We did not perform follow-up observations on this source. This source was observed as part of the CJF survey \citep{1998ASPC..144...17P} and folded into VIPS and classified as a CSO candidate after the follow-up list was constructed. However, \cite{2009AN....330..153S} have images confirming the CSO status of this source and analysis completed by \citet{2011A&A.535..A24} also classified this source as CSO.

\subsubsection{J14136+1509}
This source is a peculiar CSO with a bent morphology. The core is identified as the flat-spectrum ($\alpha_{5-8}\sim0.1$, $\alpha_{8-15}\sim-0.5$) component at roughly the centre of the source (Figs. \ref{fig:cso_cxspix}T and \ref{fig:cso_xuspix}T). A  jet heads southeast, while a steeper spectrum counter-jet bends around to the west.  A spectroscopic redshift of this source has not been measured, so an exact size can not be determined. 

\subsubsection{J14142+4554}
This source was observed as part of the VIPS pilot project \citep{2005ApJS..159...27T} and did not get added to the follow-up observation list. This source has been previously confirmed as a CSO as part of COINS \citep{2005ApJ...622..136G} therefore, for completeness, we include it in the list of VIPS CSOs.

\subsubsection{J16449+2536}
The core of this source is easily identified as the compact flat-spectrum component just east of the bright component in the centre of this source (Figs. \ref{fig:cso_cxspix}U and \ref{fig:cso_xuspix}U). Symmetric S-shaped jets then emerge to the north and south, terminating in edge brightened hotspots separated by $\sim200$pc (Table \ref{tab:cso}). 

\subsubsection{J17003+3830}
The compact flat-spectrum core of this source has lobes both to the east and west (Fig. \ref{fig:cso_cxspix}V). The western lobe is notably brighter and closer to the core either due to geometric effects or interaction with the local environment (Table \ref{tab:cso}). The $\sim30^\circ$ deviation from linearity for the three components is consistent with environmental interaction.

\subsection{Remaining Candidates and Refuted CSOs}
We were unable to either confirm or refute 33 of  the candidates (Fig. \ref{fig:cand_cxspix} \& \ref{fig:cand_xuspix} and Table \ref{tab:nonCSOs}). Most of these sources, 16 out of 29, were not detected at 15 GHz which contributed to the high number of persisting candidates. The refuted CSO candidates are included for completeness in Fig. \ref{fig:refuted_cxspix}, \ref{fig:refuted_xuspix} and Table \ref{tab:notCSOs} but further analysis past invalidating them as CSO candidates is outside the scope of this paper.

\subsection{Morphology \& Luminosity}
\label{sec:morph}
\begin{figure}
        \includegraphics[width=0.55\textwidth,center]{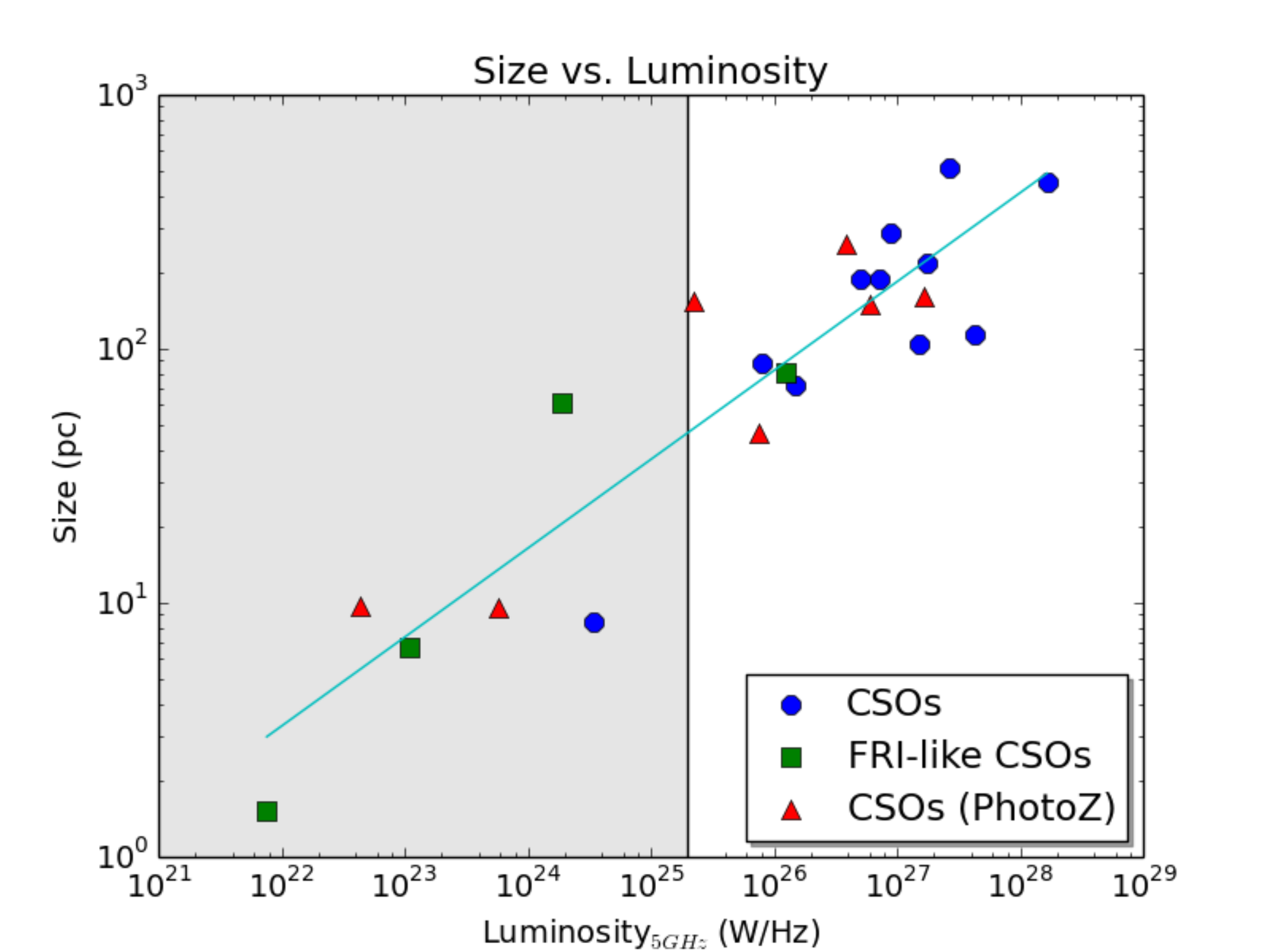}
\caption{Here we plot the size of the CSOs from end to end (as determined by modelfiting the data and looking at the separation between the components describing the edge of each jet) versus the 5 GHz luminosity. The blue circles represent archtypically `edge-brightened' CSOs with reliable redshift data, the green squares represent the `edge-darkened' FRI-like CSOs, and the red triangles the `edge-brightened' CSOs where we had to use SDSS photometric redshift data to calculate the size and luminosity. The gray region of the plot indicates a luminosity less than $2\times10^{25}$ W Hz$^{-1}$, our imposed cut between FR-I and FR-II morphology. A power law, with an index of 0.35, was fit to these data and is plotted in cyan.}
\label{fig:size_lum}
\end{figure}

The connection between the morphology and other properties of radio galaxies was first revealed when \cite{1974MNRAS.167P..31F} took a sub-sample of the 3CR catalogue \citep{1971MNRAS.154..209M} and discovered a distinct luminosity cutoff ($2.5\times10^{25}$ W Hz$^{-1}$ sr$^{-1}$ at 178 MHz with an assumed  H$_{0}$=50 km s$^{-1}$) separating diffuse sources whose highest flux was contained near the core (so-called FR-I sources) and edge-brightened sources that were most luminous where the jets terminated and ran into the surrounding medium (FR-II sources). While the canonical morphology of CSOs includes an edge-brightened lobe structure similar to what is observed in FR-II radio galaxies, several of the CSOs in this sample exhibit the more-diffuse, `wispy' emission typically associated with the lower powered FR-I sources. Interestingly, when we plot the 5 GHz luminosity of each source in Fig. \ref{fig:size_lum} an apparent separation of these morphologies by power is revealed ($\sim2\times10^{25}$ W Hz$^{-1}$), just as for their much larger analogues. Using  Fanaroff \& Riley's 178 MHz cutoff and ours yields a spectral index value, $\alpha$, of $-1.6$ which is consistent with these lobe dominated sources. A more stringent evaluation of these subsets using a two sample  Kolmogorov-Smirnov (K-S) test can not confirm these to be two independent populations, yielding p-value 0.059, which is just above the 0.050 cutoff typically used to constrain significance. This is not surprising given that these populations would overlap in luminosity and we observed only four CSOs with distinct FR-I morphology.

Overall, Figure \ref{fig:size_lum} also shows a strong relationship between the size of a source and its luminosity, with a Spearman correlation coefficient of 0.88 and an associated p-value of $1.36\times10^{-5}$. This could mean that the larger sources we observe are the more powerful ones that were able to expand further, or it could be due to them simply being older and having more time to fill their lobes with high-energy material. Fitting a power law to these data gives the relationship to be $Size \propto Luminosity^{0.35}$. 

\begin{figure}
\includegraphics[width=0.55\textwidth,center]{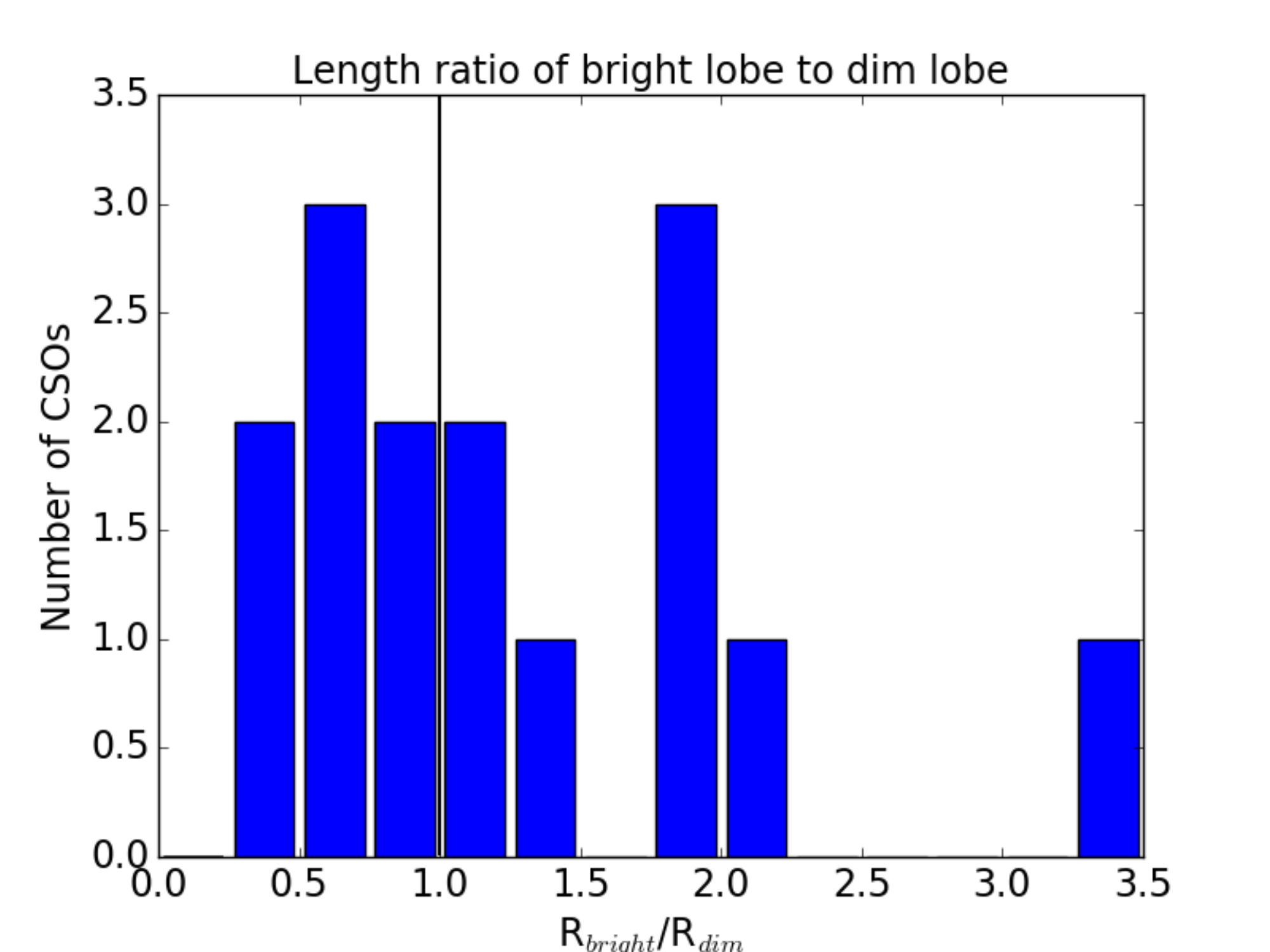}
\caption{A histogram, with bin widths of 0.25, of the ratio of the arm length of the brighter lobe divided by the arm length of the dimmer one. The ratios less than 1 (indicated by the black line) might represent significant interaction with the local environment, causing the foreshortening of the typically longer, bright arm.}
\label{fig:bright}
\end{figure}

It has been proposed that unlike traditional radio galaxies the difference in brightness between the two lobes of a CSO might be due more to interaction with the surrounding medium than simply orientation based boosting/deboosting \citep{2001MNRAS.321...37S, 2006A&A...449...49R}. When observed in large radio galaxies, the brighter lobe is typically further away from the core due to the time of arrival difference caused by one lobe being physically closer to us than the other. To investigate this, we measured the arm lengths of each lobe, $R$,  using the distance from the core to the furthest model component in each lobe. We then plot the ratio of $R_{bright}/R_{dim}$ in Figure \ref{fig:bright}, which will be greater than one for sources following the previously described scenario. While some sources certainly exhibit the expected behaviour, with a maximum ratio of 3.26, 7 of the 15 CSOs have hotspot length ratios less than unity, the smallest of which is 0.46. While projection effects could create such a result, it is more than plausible that this behaviour is due to interactions of these small jets with local inhomogeneities in their environment.

\subsection{CSO Polarisation}
As mentioned in section \ref{polarisation}, two of these CSOs have detections of polarised emission. In both of these sources we observed the polarised flux in the brighter lobe of the source. This is consistent with \cite{2007ApJ...661...78G}, where polarisation results for another two confirmed CSOs were reported, which also detected polarised emission only in the brighter lobe. They also found unusually high flux ratios (10 \& 14) for the lobes of these CSOs,  whereas only one of ours is high (19 and 1.6 in J07414+2706 and J13262+3154 respectively, while the median of the distribution is 1.79). This is consistent with the AGN unification scenario where the emission of most CSOs is depolarised as it passes through the dusty circumnuclear torus on its way to us, but for a few tilted CSOs (where one of the lobes would appear significantly brighter) the emission from the end of the near lobe doesn't pass through the torus on its path towards us.

The three IF pairs that detected stronger polarisation in each source were used to compute rotation measures ($RM$s) as a function of position (Figs. \ref{fig:J07414RM} \& \ref{fig:J13262RM}) wherever polarisation was detected at all three pairs by fitting the change in polarisation angle ($\beta$) to $RM=\frac{\beta}{\lambda^2}$, with values ranging from 1800 to 4000 and  -300 to 500 rad m$^{-2}$ for J07414+2706 and J13262+3154 respectively. The $RM$ is dependent on the magnetic field ($\vec{B}$) and the electron number density ($n_e$) by:
\begin{equation}
 RM=\frac{e^3}{2\pi m_e^2c^4}\int_0^Ln_e\vec{B}\cdot \vec{ds}
\end{equation}
where $L$ is the path length of the Faraday screen, $e$ and $m_e$ are the charge and mass of the electron and $c$ is the speed of light. These rotation measures were then used to calculate the magnetic field polarisation angle corrected for Faraday rotation and are plotted on the right side of each figure.  Following the assumptions of \cite{2007ApJ...661...78G} and the references therein (specifically \citealt{1979ApJ...231L..51S}), if we assume $n_e=1100\pm350$ cm$^{-3}$, and lower and upper limits for $L$ of 0.3 and 10 pc, we calculate magnetic fields of  0.21 to 16 and 0.04 to 2.7  $\mu G$ for J07414+2706 and J13262+3154 respectively.

Here, we point out  the sign change in $RM$ from one side of the polarised feature of J13262+3154 to the other, which could indicate a reversal of the magnetic field as seen in the above equation. However, since the polarised region is not well resolved (less than 3 beams across), it does not satisfy the requirements for a $RM$ gradient outlined in \cite{2010ApJ...722L.183T} and therefore is not strong evidence for a magnetic field reversal in this source.

\subsection{High energy emission from CSOs}

Despite theoretical expectations that CSOs might be isotropic $\gamma$-ray emitters, presumably due to inverse Compton scattering within the lobes of these compact sources (e.g. \citealt{2008ApJ...680..911S}) none of our CSOs have been detected by telescopes such as the \textit{Fermi Gamma-ray Space Telescope}. This is well matched with empirical results from other CSOs, with in fact only one candidate being detected by Fermi to date \citep{2014A&A...562A...4M}. It is still possible these sources are emitting $\gamma$-rays via conventional means and that these are being beamed in the direction of the jets. In fact, it is quite possible that some of the Blazars detected by \textit{Fermi} have a physical extent less than 1 kpc, but this information is impossible to disentangle. 

\subsection{Incidence of CSOs}
Using the number of confirmed CSOs as a lower limit and adding in the remaining candidates for an upper limit, we have 24-57 CSOs within the VIPS catalogue of 1127 sources (2.1\%-5.1\%). While the spectral and flux density prerequisites imposed on the VIPS sample limit what can be surmised about the fraction of CSOs within the entire population of radio-loud AGN, this is the most complete sample to date to investigate the CSO population. Comparing this with other samples and supersets, \cite{2000ApJ...534...90P} report CSO incidences of 11\%, 4.4\% and 2.1\% for PR, PR+CJ, and PR+CJ+VCS, where PR, CJ and VCS are Pearson-Readhead survey \citep{1988ApJ...328..114P}, the Caltech-Jodrell Bank survey \citep{1994ApJS...95..345T} and the VLBA Calibrator Survey \citep{1998ASPC..144..155P} respectively. Direct comparison of these numbers is difficult since each sample had independent selection criteria and the follow-up analysis, particularly of the VCS, was not always rigorous. Peck \& Taylor speculated that the CSO fraction was inversely related to 
the limiting flux density of the survey, with PR having a cutoff of 1.3 Jy at 5 GHz, compared to just 0.1 Jy at 5 GHz for the VCS.  The VIPS survey has a limiting flux density of 0.085 Jy at 8 GHz \citep{2007ApJ...658..203H}, comparable to VCS, yet has a CSO fraction closer to that of PR+CJ.  It seems likely that the VCS survey was deficient in CSOs due to its worse imaging characteristics.  The high CSO fraction in the PR survey can be explained as a fluke of small number statistics.

\begin{figure}
\includegraphics[width=0.55\textwidth,center]{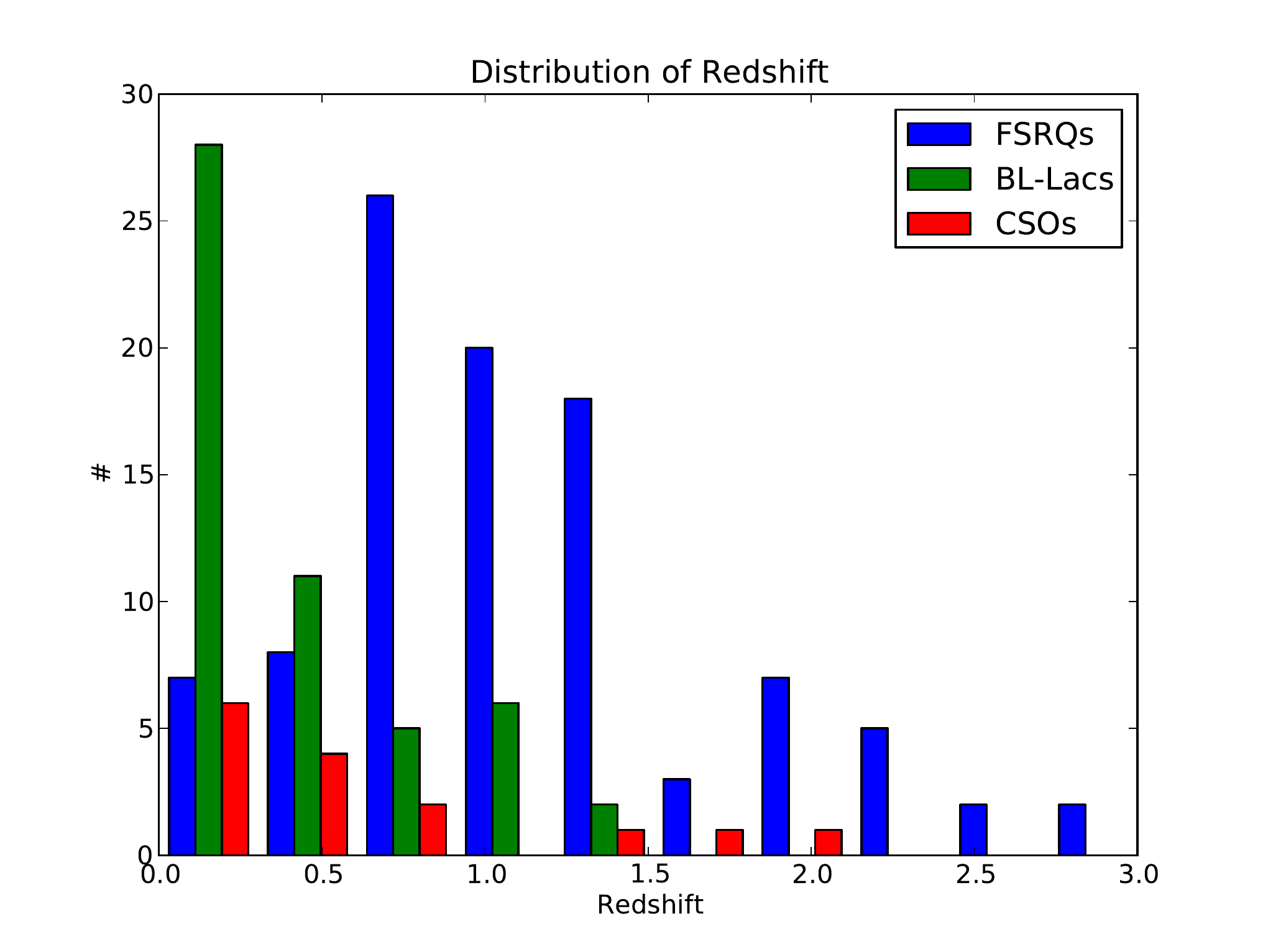}
\caption{Here the redshift ($z$) distribution of the CSOs with reliable redshift data are plotted along with the  distribution of redshifts for FRSQs and Bl-Lac object is plotted for comparison. While the CSOs and FSRQs are clearly different distributions, possibly due to selection effects, the same cannot be said for CSOs and BL-Lacs.}
\label{fig:zdist}
\end{figure}

The redshift distribution of VIPS CSOs (Fig. \ref{fig:zdist}) shows a decay in incidence with redshift. This may be mostly due to selection effects, since although the emission from CSOs is bright it is by definition not strongly beamed towards us and therefore not doppler boosted. When compared to the redshift distribution of \textit{Fermi} detected flat spectrum radio quasars (FSRQs) and BL-Lacertae objects (BL-Lacs), which seem to be representative samples of those types of sources, we see that the VIPS CSOs and \textit{Fermi} FSRQs almost certainly come from independent populations with a two sample K-S test p-value of $3.25\times10^{-12}$. It is important to point out that this could either be due to selection effects or inherent characteristics of the populations. The same cannot be said for CSOs and BL-Lacs, with a K-S test p-value of 0.49, meaning we cannot rule out the possibility these are from a similar population distribution.

\subsection{VIPS SBBH Candidates}
These multi-frequency follow-up observations were additionally searched for either:
\begin{itemize}
\item Two distinct flat- or inverted-spectrum compact cores
\item Jet structure that cannot be easily traced back to a single point of origin
\end{itemize}

We were unable to confirm any new SBBHs with these follow-up observations, and none of the 15 remaining candidates are particularly promising (see Figures \ref{fig:cso_cxspix}-\ref{fig:refuted_xuspix}). Making the assumption that no SBBHs were observed in VIPS, then from the combined VIPS and Caltech-Jodrell Bank flat-spectrum (CJF) sample together only one SBBH (0402+379; Rodriguez et al. 2006) has been detected out of 1279 sources. Reconciling this within the current merger driven paradigm of galaxy evolution minimally requires that our understanding of black hole merger times, particularly the so-called `final parsec problem' of the merger stalling for small radii \citep{2006ApJ...648..976M},  be re-evaluated. This is consistent with \cite{2011MNRAS.410.2113B} result of finding only a single SBBH within a sample of 3114 AGN with VLBI observations, which was also 0402+379. The detailed analysis within that paper of the 1575 sources with redshift data suggest that black hole binaries do not stall indefinitely at any radius $<$ 500 parsecs unless the most pessimistic estimates of merger rates hold true.


\section{Conclusions}
We have identified 24 CSOs within the VIPS sample using the spectral and morphological characteristics of each source, 15 of which were previously unclassified, and are still left with a remaining 33 CSO candidates ($2\%-5\%$ of VIPS). Subsequent 15 GHz observations using higher bandwidth ($\geq$ 2 GHz) should confirm or deny at least half of the remaining candidates. Our sample of CSOs is large enough to observe an FR-I/FR-II distinction in both luminosity and morphology, similar to larger radio galaxies, which has not been previously identified. Correlation is observed within these data between the size and luminosity of CSOs, which could result either from pile-up of emitting material if the larger sources are also older or could indicate the ability of more powerful sources to expand more efficiently within their environments. Further evidence of the significance of environmental interaction with CSO jets is given by $\sim 1/2$ of the CSOs having arm length ratios of the bright lobe divided by the dim lobe less than unity.

Polarisation was detected in two of these CSOs, doubling the number of known polarised CSOs. In all four of these cases, polarisation was only detected in the brighter lobe/jet and it was usually ($3/4$) detected near the working surface of the jet. This, coupled with $3/4$ of the sources having unusually high luminosity ratios for the lobes ($\sim10-19$) is consistent with most of the inherently polarised emission from CSOs being depolarised by the dusty molecular torus surrounding the nucleus, and only the sources that are oriented out of the plane of the sky enough to attain a line of sight to the end of the jets that doesn't pass through the torus being observed to be polarised.

Our lack of new supermassive binary black hole detections is consistent with other comparable work, and indicates either that the `final parsec' problem doesn't exist or that the more pessimistic projections of merger event rates could be correct.

\section*{Acknowledgements}

The authors thank Mike Shaw for assistance with the HET observations and data analysis. Additionally, we thank Craig Walker, Trish Henning, Ylva Pihlstr{\"o}m and Yuri Kovalev for numerous suggestions about how to improve this work, which have produced a more rigorous and robust paper. S.T. would also like to thank Justin Linford for enlightening discussions that improved this work greatly. The authors would also like to thank the anonymous referee, who contributed to the clarity of this work.

The Hobby-Eberly Telescope (HET) is a joint project of the University of Texas at Austin, the Pennsylvania State University,  Stanford University, Ludwig-Maximillians-Universit\"at M\"unchen, and Georg-August-Universit\"at G\"ottingen.  The HET is named in honor of its principal benefactors, William P.~Hobby and Robert E.~Eberly. The Marcario Low-Resolution Spectrograph is named for Mike Marcario of High Lonesome Optics, who fabricated several optics for the instrument but died before its completion;
it is a joint project of the Hobby-Eberly Telescope partnership and the Instituto de Astronom\'{\i}a de la Universidad Nacional Aut\'onoma de M\'exico. This research has made use of the NASA/IPAC Extragalactic Database (NED) which is operated by the Jet Propulsion Laboratory, California Institute of Technology, under contract with the National Aeronautics and Space Administration. The National Radio Astronomy Observatory is a facility of the National Science Foundation operated under cooperative agreement by Associated Universities, Inc.  This research was partially conducted by the Australian Research Council Centre of Excellence for All-sky Astrophysics (CAASTRO), through project number CE110001020.




\begin{landscape}
\begin{table}
\begin{minipage}{\linewidth}
\caption{\textbf{Confirmed VIPS Compact Symmetric Objects}}
{Col. 1: Source Name; Col. 2 Date of observations; Col. 3: Redshift of source; Col. 4: Whether or not a core was observed; Col. 5: Whether or not the source has `FRI-like morphology; Col. 6: Whether or not this is a SBBH Candidate; Col. 7-9: 5, 8, and 15 GHz integrated flux; Col. 10: The extent of the source as determined by model fitting; Col. 11: The calculated luminosity based on 5 GHz flux; Col 12: The ratio of the integrated fluxes of the brighter lobe over the dimmer lobe for each source; Col. 13: The ratio of the length of the brighter jet over the length of the dimmer jet; Col. 14: The angle subtended between the jets;  Col. 15: The object type as per NED where `G' is galaxy, `QSO' is quasar, `VisS' is an object with visible data but no designation and `\--'  only has radio data and no designation.}
\label{tab:cso}
\begin{tabular}{lcccccccccccccc}
\hline \hline
Source & Date & $z$ & Core & FRI & SBBH & $S_{5\,GHz}$ & $S_{8\,GHz}$ & $S_{15\,GHz}$ & Size & $L_{5\,GHz}$ & Flux & Arm & Angle & Object \\
  &  & & & & Cand. & Jy & Jy & Jy & pc & W Hz$^{-1}$ & Ratio & Ratio & deg. & Type \\
\hline  
VIPS J07414+2706 & 2006 Sept. 19 & 0.77$^{1}$ & Y & N & N & 0.578 & 0.418 & 0.270 & 104.8$\pm0.0$ & 1.50E27 &19.35  & 1.02 & 159.2 & VisS\\
VIPS J07542+5324 & 2007 Dec. 31 & 0.84$^{\star}$ & N & N & N & 0.185 & 0.083 & ND & ? & 5.98E26 & 1.33 & ? & ? & \--\\
VIPS J09062+4636 & 2007 Dec. 31 & 0.0848$^{2}$ & Y & Y & N & 0.113 & 0.126 & 0.121 & 61.6$^{\dag}$ & 1.90E24 & ? & ? & ? & G\\
VIPS J09432+1702 & 2008 May 16 & 1.598$^{3}$ & Y & N & N & 0.269 & 0.253 & 0.185 & 113.4 & 4.32E27 & 2.52 & 3.26 & 138.5 & QSO\\
VIPS J10351+5628 & 2007 June 30 & 0.45$^{4}$ & N & N & N & 1.021 & 0.714 & 0.233 & 187.9 & 7.15E26 & 1.32 & ? & ? & G\\
VIPS J10426+2949 & 2008 May 16 & 0.61$^{\star}$ & Y & N & N & 0.264 & 0.122 & 0.045 & ? & 3.86E26 & 1.49 & 1.76 & 175.9 & \--\\
VIPS J11113+1955 & 2008 May 16 & 0.2991$^{5}$ & N & N & N & 0.562 & 0.314 & 0.077 & 71.6 & 1.50E26 & 1.31 & ? & ? & G\\
VIPS J11359+4258 & 2007 Dec. 31 & 1.0$^{\star}$ & N & N & N & 0.332 & 0.155 & ND & ? & 1.65E27 & 1.89 & ? & ? & \--\\
VIPS J11488+5924 & 2007 June 30 & 0.010751$^{6}$ & Y & Y & N & 0.417 & 0.385 & 0.279 & 6.6 & 1.08E23 & 1.70 & 0.51 & 157.7 & G\\
VIPS J11584+2450 & 2006 Sept. 19 & 0.2016$^{7}$ & Y & N & N & 0.723 & 0.441 & 0.202 & 88.3 & 7.95E25 & 2.62 & 0.54 & 164.8 & G\\
VIPS J11598+5820 & 2007 Dec. 31 & 1.278$^{1}$ & N & N & N & 0.286 & 0.103 & ND & 514.8 & 2.63E27 & 1.50 & ? & ? & G\\
VIPS J12043+5202 & 2007 Dec. 31 & 0.01$^{\star}$ & Y & N & N & 0.213 & 0.107 & ND & ? & 4.37E22 & 1.51 & 0.68 & 179.3 & \--\\
VIPS J12201+2916 & 2008 June 26 & 0.002068$^{8}$ & Y & Y & N & 0.388 & 0.349 & 0.236 & 1.5 & 7.61E21 & 1.25 & 2.11 & 151.1 & G\\
VIPS J12279+3635 & 2008 June 26 & 1.973$^{9}$ & N & N & N & 0.610 & 0.296 & ND & 453.1 & 1.66E28 & ? & ? & ? & QSO\\
VIPS J12342+4753 & 2007 Dec. 31 & 0.372349$^{10}$ & Y & Y & N & 0.275 & 0.223 & 0.140 & 80.8 & 1.23E26 & 2.35 & 0.92 & 162.9 & QSO\\
VIPS J12448+4048 & 2008 Jan. 19 & 0.813$^{9}$ & Y & N & N & 0.592 & 0.322 & 0.114 & 217.0 & 1.77E27 & 2.31 & 1.86 & 176.1 & QSO\\
VIPS J12545+1856 & 2008 June 26 & 0.1145$^{11}$ & Y & N & N & 0.107 & 0.102 & 0.095 & 8.4 & 3.43E24 & 1.13 & 0.80 & 159.0 & G\\
VIPS J13113+1658 & 2008 May 16 & 0.03 $^{\star}$ & Y & N & N & 0.303 & 0.202 & 0.092 & ? & 5.77E23 & 5.76 & 1.23 & 172.7 & VisS\\
VIPS J13262+3154 & 2008 May 16 & 0.36801$^{12}$ & Y & N & N & 2.045 & 1.246 & ND & 285.4 & 8.93E26 & 1.59 & 1.84 & 173.5 & G\\
VIPS J13354+5844 & NF$^{\ddag}$ & 0.59$^{\star}$ & ? & ? & ? & 0.637$^{\diamond}$ & - & - & ? & ? & ? & ? & ? & VisS\\
VIPS J14136+1509 & 2006 Nov. 4 & 0.35$^{\star}$ & Y & N & N & 0.195 & 0.139 & 0.092 & ? & 7.51E25 & 5.09 & 0.469 & 142.7 & G\\
VIPS J14142+4554 & NF$^{\ddag}$ & 0.186$^{2}$ & ? & ? & ? & 0.171$^{\diamond}$ & - & - & ? & ? & ? & ? & ? & G\\
VIPS J16449+2536 & 2006 Sept. 19 & 0.588$^{1}$ & Y & N & N & 0.383 & 0.271 & 0.152 & 187.0 & 5.12E26 & 3.69 & 1.31 & 178.8 & VisS\\
VIPS J17003+3830  & 2008 Jan. 19 & 0.23$^{\star}$ & Y & N & N & 0.152 & 0.078 & ND & ? & 2.21E25 & 5.29 & 0.46 & 161.3 & \--\\
\hline
\end{tabular}%
\smallskip

$\dag$ Not measured by component separation\\
$\ddag$ No followup observations performed\\
$\star$ Photometric redshift from SDSS\\
$\diamond$ Flux density value from VIPS\\
(1) This work; (2) Falco, Kochanek \& Munoz 1998; (3) Healey et al. 2008; (4) Aller, Aller \& Hughes 1992; (5) Peck et al. 2000; (6) de Vaucouleurs et al. 1991; (7) Zensus et al. 2002; (8) Ackermann et al. 2011; (9) Xu et al. 1994; (10) Hewett \& Wild 2010; (11) Owen, Ledlow \& Keel 1995; (12) Holt, Tadhunter \& Morganti 2008
\end{minipage}
\end{table}

\end{landscape}

\begin{table*}
\begin{minipage}{\linewidth}

\caption{\textbf{Remaining VIPS CSO Candidates}}
{Col. 1: Source Name; Col. 2 Date of observations; Col. 3: Redshift of source;  Col. 4: Whether or not this is a SBBH Candidate; Col. 5-7: 5, 8, and 15 GHz integrated flux; Col. 8: The extent of the source as determined by modefitting; Col. 9: The calculated luminosity based on 5 GHz flux}
\label{tab:nonCSOs}
\begin{center}

\begin{tabular}{lcccccccc}
\hline \hline
Source & Date & z & SBBH & $S_{5\,GHz}$ & $S_{8\,GHz}$ & $S_{15\,GHz}$ & Size & $L_{5\,GHz}$\\
& & & Cand. & Jy & Jy & Jy & pc& W Hz$^{-1}$\\
\hline
VIPS J07502+3119 & 2008 June 26 & 0.68$^{\star}$ & N & 0.072 & 0.033 & ND & 1675 $\pm$ 2 & 1.38E26\\
VIPS J07530+4231 & NF$^{\ddag}$ & 3.589250$^{1}$ & N & 0.398$^{\diamond}$ & - & - & 101.7 $\pm$ 2.4 & -\\
VIPS J08316+4608 & 2008 Jan. 19 & 0.131138$^{2}$ & Y & 0.091 & 0.067 & ND & 25 $\pm$ 1 & 3.87E24\\
VIPS J08322+1832 & 2006 Nov. 22 & 0.153$^{3}$ & N & 0.484 & 0.369 & 0.226 & 36 $\pm$ 1 & 2.94E25\\
VIPS J08553+5751 & NF$^{\ddag}$ & 0.39$^{\star}$ & N & 0.119 $^{\diamond}$ & - & - & ? & -\\
VIPS J09452+2729 & 2008 Jan. 19 & 0.68$^{\star}$ & N & 0.118 & 0.093 & ND & 76 $\pm$ 2 & 2.26E26\\
VIPS J10066+4836 & 2007 June 30 & 0.7$^{\star}$ & Y & 0.109 & 0.091 & 0.047 & 63 $\pm$ 2 & 2.24E26\\
VIPS J10320+5610 & 2007 Dec. 31 & 1.0$^{\star}$ & Y & 0.111 & 0.060 & ND & 196 $\pm$ 3 & 5.53E26\\
VIPS J11106+4817 & 2007 Dec. 31 & 0.74$^{4}$ & N & 0.140 & 0.079 & ND & 242 $\pm$ 3 & 3.31E26\\
VIPS J11434+1834 & 2008 May 16 & 0.7$^{\star}$ & Y & 0.326 & 0.230 & 0.088 & 118 $\pm$ 2 & 6.70E26\\
VIPS J12018+3919 & 2008 Jan. 19 & 2.37$^{5}$ & N & 0.214 & 0.093 & ND & 427 $\pm$ 3 & 9.13E27\\
VIPS J12105+6422 & 2007 June 30 & 0.9$^{\star}$ & Y & 0.121 & 0.088 & 0.029 & 122 $\pm$ 3 & 4.64E26\\
VIPS J12152+1730 & 2008 May 16 & 0.268$^{6}$ & N & 0.211 & 0.066 & ND & 559 $\pm$ 1 & 4.37E25\\
VIPS J12407+2405 & 2008 May 16 & 0.337$^{6}$ & N & 0.278 & 0.167 & 0.079 & 232 $\pm$ 2 & 6.67E25\\
VIPS J12477+2551 & 2008 Jan. 19 & 0.14$^{\star}$ & Y & 0.075 & 0.058 & ND & 31 $\pm$ 1 & 3.63E24\\
VIPS J13100+3403 & 2007 Dec. 31 & 0.96$^{\star}$ & N & 0.104 & 0.050 & ND & 178 $\pm$ 3 & 4.68E26\\
VIPS J13136+5458 & 2007 Feb. 19 & 0.613$^{7}$ & N & 0.398 & 0.250 & 0.149 & 302 $\pm$ 2 & 5.92E26\\
VIPS J13199+3840 & 2008 Jan. 19 & 0.45$^{\star}$ & Y & 0.117 & 0.077 & ND & 206 $\pm$ 2 & 8.19E25\\
VIPS J13222+2645 & 2008 June 26 & 0.53$^{\star}$ & N & 0.223 & 0.072 & ND & 165 $\pm$ 2 & 2.32E26\\
VIPS J13242+4048 & NF$^{\ddag}$ & 0.496$^{7}$ & N & 0.365$^{\diamond}$ & - & - & 118 $\pm$ 2 & -\\
VIPS J13253+2109 & 2008 May 16 & 0.675$^{6}$ & Y & 0.108 & 0.058 & ND & 110 $\pm$ 2 & 2.03E26\\
VIPS J13576+4353 & 2008 Jan. 19 & 0.646$^{8}$ & Y & 0.469 & 0.285 & 0.123 & 121 $\pm$ 2 & 7.93E26\\
VIPS J13586+4737 & 2007 Dec. 31 & 0.23$^{7}$ & Y & 0.295 & 0.201 & 0.105 & 36 $\pm$ 1 & 1.89E25\\
VIPS J14344+4236 & 2007 Feb. 19 & 0.452$^{6}$ & N & 0.304 & 0.194 & 0.075 & 196 $\pm$ 2 & 2.15E26\\
VIPS J14402+6108 & 2007 June 30 & 0.445$^{6}$ & N & 0.113 & 0.064 & ND & 166 $\pm$ 2 & 7.70E25\\
VIPS J14426+3042 & 2008 June 26 & 1.0$^{\star}$ & Y & 0.144 & 0.064 & ND & 360 $\pm$ 3 & 7.17E26\\
VIPS J15136+2338 & 2008 May 16 & 0.218$^{6}$ & N & 0.751 & 0.475 & 0.362 & 224 $\pm$ 1 & 9.72E25\\
VIPS J15590+5924 & 2007 June 30 & 0.0602$^{9}$ & N & 0.118 & 0.094 & ND & 15 $\pm$ 1 & 1.94E22\\
VIPS J16022+2418 & 2008 Jan. 19 & 1.788871$^{10}$ & Y & 0.181 & 0.131 & 0.060 & 463 $\pm$ 6 & 1.17E27\\
VIPS J16061+5521 & 2007 June 30 & 0.339$^{6}$ & N & 0.077 & 0.043 & ND & 201 $\pm$ 2 & 2.75E25\\
VIPS J16087+1511 & 2008 May 16 & 0.94$^{\star}$ & Y & 0.091 & 0.077 & 0.043 & 123 $\pm$ 2 & 3.89E26\\
VIPS J16092+2641 & 2008 June 26 & 0.473$^{11}$ & N & 1.5678 & 0.878 & 0.275 & 327 $\pm$ 2 & 1.24E27\\
VIPS J17309+3811 & 2008 June 26 & ? & N & 0.169 & 0.099 & ND & ? & - \\
\hline
\end{tabular}%
\end{center}
\smallskip

$\ddag$ No followup observations performed\\
$\star$ Photometric redshift from SDSS\\
$\diamond$ Flux density value from VIPS\\
(1) Abazajian et al. 2004; (2) Abazajian et al. 2003; (3) Adelman-McCarthy et al. 2008; (4) Hook et al. 1996; (5) Fanti et al. 2001; (6) This work; (7) Vermeulen, Taylor \& Readhead 1996; (8) Vermeulen et al. 2003; (9) Falco, Kochanek \& Munoz 1998; (10) Hewett \& Wild 2010; (11) NED 1992
\end{minipage}
\end{table*}%

\begin{table*}
\begin{minipage}{\linewidth}
\caption{\textbf{Refuted VIPS CSO Candidates}}
{Col. 1: Source Name; Col. 2 Date of observations; Col. 3: Redshift of source;  Col. 4: Whether or not this is a SBBH Candidate; Col. 5-7: 5, 8, and 15 GHz integrated flux; Col. 8:  The calculated luminosity based on 5 GHz flux}

\label{tab:notCSOs}

\begin{center}

\begin{tabular}{lccccccc}

\hline \hline
Source & Date & z & SBBH & $S_{5\,GHz}$ & $S_{8\,GHz}$ & $S_{15\,GHz}$ & $L_{5\,GHz}$\\
& & & Cand. & Jy & Jy & Jy & W Hz$^{-1}$\\
\hline
VIPS J07334+5605 & 2007 June 30 & 0.104$^{1}$ & N & 0.084 & 0.046 & 0.034 & 2.15E24\\
VIPS J07369+2604 & 2008 May 16 & 0.997$^{2}$ & N & 0.234 & 0.205 & 0.149 & 1.16E27\\
VIPS J08116+4308 & 2008 Jan. 19 & ? & N & 0.150 & 0.086 & 0.047 & -\\
VIPS J08170+1958 & 2008 May 16 & 0.138$^{3}$ & N & 0.123 & 0.090 & 0.058 & 5.87E24\\
VIPS J08182+6109 & 2007 June 30 & ? & N & 0.103 & 0.065 & 0.018 & -\\
VIPS J08203+5621 & 2007 Dec. 31 & 2.255$^{\star}$ & N & 0.143 & 0.069 & 0.036 & 5.39E27\\
VIPS J08398+4301 & 2008 June 26 & 0.22$^{\star}$  & N & 0.054 & 0.041 & 0.040 & 7.19E24\\
VIPS J09097+4753 & 2007 June 30 & 2.535$^{\star}$ & N & 0.090 & 0.066 & 0.033 & 4.52E27\\
VIPS J09115+1958 & 2008 May 16 & 1.635343$^{4}$ & N & 0.192 & 0.134 & 0.089 & 3.27E27\\
VIPS J09128+4422 & 2008 Jan. 19 & 1.727$^{5}$ & N & 0.164 & 0.129 & 0.087 & 3.21E27\\
VIPS J09267+2758 & 2006 Sept. 19 & 0.22$^{\star}$  & N & 0.128 & 0.123 & 0.071 & 1.69E25\\
VIPS J09292+2536 & 2008 June 26 & 0.53844$^{6}$ & N & 0.108 & 0.086 & 0.110 & 1.17E26\\
VIPS J09355+3633 & 2007 Dec. 31 & 2.858$^{7}$ & N & 0.216 & 0.156 & 0.113 & 1.45E28\\
VIPS J10019+5540 & 2006 Nov. 4 & 0.003723$^{8}$ & N & 0.075 & 0.075 & 0.048 & 2.77E21\\
VIPS J10051+2403 & 2008 June 26 & 0.17$^{\star}$  & N & 0.142 & 0.094 & 0.055 & 1.06E25\\
VIPS J10138+2449 & 2006 Nov. 22 & 1.636$^{6}$ & N & 0.816 & 0.925 & 0.660 & 1.39E28    \\
VIPS J10225+3041 & 2008 June 26 & 1.319802$^{4}$ & N & 0.362 & 0.212 & 0.109 & 3.62E27\\
VIPS J10305+5132 & 2008 Jan. 19 & 0.518450$^{5}$ & N & 0.102 & 0.077 & 0.038 & 1.01E26\\
VIPS J10509+3430 & 2008 May 16 & 2.52$^{9}$ & N & 0.278 & 0.177 & 0.080 & 1.38E28\\
VIPS J10511+5347 & 2007 June 30 & 0.16$^{\star}$ & N & 0.199 & 0.127 & 0.048 & 1.29E25\\
VIPS J10512+4644 & 2008 Jan. 19 & 1.419418$^{4}$ & N & 0.145 & 0.148 & 0.144 & 1.74E27\\
VIPS J10580+4248 & 2008 June 26 & ? & N & 0.113 & 0.057 & 0.045 & -\\
VIPS J11240+2336 & 2008 June 26 & 1.549$^{10}$ & N & 0.278 & 0.257 & 0.280 & 4.14E27\\
VIPS J11285+3243 & 2008 Jan. 19 & 0.369528$^{6}$ & N & 0.115 & 0.071  &0.047 &  5.07E25\\
VIPS J11408+5912 & 2007 June 30 & 0.14$^{\star}$  & N & 0.192 & 0.128 & 0.053 & 9.31E24\\
VIPS J12009+2008 & 2008 June 26 & ? & N & 0.198 & 0.219 & 0.193 & -\\
VIPS J12035+4632 & 2006 Nov. 4 & 0.14$^{\star}$  & N & 0.148 & 0.104 & 0.056 & 7.17E24\\
VIPS J12066+3941 & 2008 Jan. 19 & 1.518589$^{4}$ & N & 0.308 & 0.329 & 0.296 & 4.36E27\\
VIPS J12074+2754 & 2007 Feb. 19 & 2.182064$^{4}$ & N & 0.428 & 0.329 & 0.239 & 1.49E28\\
VIPS J12414+5458 & 2007 June 30 & 0.04$^{\star}$  & N & 0.129 & 0.070 & 0.026 & 4.42E23\\
VIPS J12582+5421 & 2007 June 30 & ? & N & 0.225 & 0.088 & 0.024 & -\\
VIPS J12595+5140 & 2006 Nov. 22 & 0.405$^{\star}$ & N & 0.249 & 0.398 & 0.484 & 1.36E26\\
VIPS J13128+5548 & 2007 June 30 & 0.975$^{\star}$ & N & 0.209 & 0.131 & 0.036 & 9.79E26\\
VIPS J13223+4303 & 2007 Dec. 31 & 0.07$^{\star}$  & N & 0.108 & 0.037 & 0.019 & 1.19E24\\
VIPS J14091+3642 & 2008 June 26 & 0.996$^{11}$ & N & 0.190 & 0.101 & 0.050 & 9.38E26\\
VIPS J14489+5326 & 2007 Dec. 31 & 0.58$^{\star}$  & N & 0.135 & 0.104 & 0.069 & 1.75E26\\
VIPS J14519+6357 & 2006 Nov. 22 & ? & N & 0.193 & 0.163 & 0.058 & -\\
VIPS J15077+5857 & 2007 June 30 & 0.25$^{\star}$  & N & 0.166 & 0.134 & 0.034 & 2.94E25\\
VIPS J15159+2458 &  2006 Sept. 19 & 0.151$^{12}$ & N & 0.116 & 0.089 & ? & 1.28E24\\
VIPS J15186+5002 & 2008 Jan. 19 & 1.513443$^{4}$ & N & 0.073 & 0.055 & 0.034 & 1.02E27\\
VIPS J15451+4751 & 2007 Dec. 31 & 1.277$^{13}$ & N & 0.288 & 0.197 & 0.103 & 2.65E27\\
VIPS J15594+1624 & 2008 May 16 & 0.17$^{\star}$  & N & 0.181 & 0.118 & 0.051 & 1.35E25\\
VIPS J16002+1838 & 2008 June 26 & 2.40466$^{14}$ & N & 0.119 & 0.092 & 0.049 & 5.27E27\\
VIPS J16021+3326 & 2007 Feb. 19 & 1.1$^{15}$ & N & 1.406 & 1.115 & 0.732 & 8.91E27\\
VIPS J16048+1926 & 2006 Nov. 4 & 0.35$^{\star}$  & N & 0.221 & 0.142 & 0.048 & 8.50E25\\
VIPS J16323+2643 & 2008 Jan. 19 & 2.683038$^{4}$ & N & 0.124 & 0.110 & 0.076 & 7.15E27\\
VIPS J16325+3547 & 2006 Nov. 22 & 0.07$^{\star}$  & Y & 0.237 & 0.141 & 0.050 & 2.61E24\\
VIPS J16538+3503 & 2007 Dec. 31 & 0.57$^{\star}$  & N & 0.097 & 0.050 & 0.023 & 1.20E26\\
VIPS J16559+5430 & 2007 June 30 & 1.04$^{17}$ & N & 0.132 & 0.122 & 0.060 & 7.27E26\\
VIPS J17073+4204 & 2007 Dec. 31 & ? & N & 0.123 & 0.052 & 0.022 & -\\
VIPS J17233+3417 & 2008 May 16 & 0.206$^{18}$ & N & 0.218 & 0.178 & 0.086 & 2.48E25\\
VIPS J17246+6055 & 2006 Nov. 4 & 0.33$^{\star}$  & Y & 0.202 & 0.157 & 0.105 & 6.79E25\\
\hline
\end{tabular}%
\end{center}
\smallskip

$\ddag$ No followup observations performed\\
$\star$ Photometric redshift from SDSS\\
$\diamond$ Flux density value from VIPS\\
(1) Marcha et al. 1996; (2) Healey et al. 2008;(3) Glikman et al. 2007; (4) Hewett \& Wild 2010; (5) Abazajian et al. 2004; (6) Adelman-McCarthy et al. 2007; (7) Adelman-McCarthy et al. 2005; (8) Hagiwara, Klockner \& Baan 2004;  (9) Hewitt \& Burbidge 1989; (10) Ackermann et al. 2011; (11) Kunert-Bajraszewska \& Marecki 2007; (12) This work; (13) Vermeulen \& Taylor 1995; (14) Adelman-McCarthy et al. 2008; (15) Snellen et al. 2000; (16) Abazajian et al. 2005; (17) V{\'e}ron-Cetty \& V{\'e}ron 2001; (18) Wills \& Wills 1976
\end{minipage}
\end{table*}%


\end{document}